\documentclass[rmp,aps,amsfonts,amssymb,nofootinbib,twocolumn,amsmath]{revtex4}
\usepackage{graphics}
\usepackage{hyperref} 
\usepackage{epsfig} 
\usepackage{color}

\renewcommand{\k}{{\bf k}} 
\newcommand{\q}{{\bf q}}
\newcommand{\I}{\mathbb{I}}
\newcommand{\T}{\mathcal{T}}  
\newcommand{\GVec}[1]{\mbox{\boldmath$#1$}}

\begin{document}
\title{Weyl and Dirac Semimetals in Three-Dimensional Solids\footnote{\vspace{-1.75cm}{\bf Accepted Reviews of Modern Physics}}}
\author{N.P. Armitage}
 \affiliation{The Institute for Quantum Matter, Department of Physics and Astronomy, The Johns Hopkins University, Baltimore, MD 21218, USA}

\author{E. J. Mele}
\affiliation{Department of Physics and Astronomy, University of Pennsylvania, Philadelphia, PA 19104, USA}

\author{Ashvin Vishwanath}
\affiliation{Department of Physics, University of California, Berkeley,  CA 94720, USA}
\affiliation{Department of Physics, Harvard University, Cambridge, MA 02138, USA}

\date{Jan. 1st, 2018}

\begin{abstract}
Weyl and Dirac semimetals are three dimensional phases of matter with gapless electronic excitations that are protected by topology and symmetry.  As three dimensional analogs of graphene, they have generated much recent interest. Deep connections exist with particle physics models of relativistic chiral fermions, and -- despite their gaplessness -- to solid-state topological and Chern insulators.  Their characteristic electronic properties lead to protected surface states and novel responses to applied electric and magnetic fields. The theoretical foundations of these phases, their proposed realizations in solid state systems,  recent experiments on candidate materials, as well as their relation to other states of matter are reviewed.

\end{abstract}

\maketitle

\tableofcontents

\section{Introduction}
\label{IntroductionSection} 

In 1928 P.A.M. Dirac proposed -- in the first successful reconciliation of special relativity and quantum mechanics -- his now eponymous Dirac equation \cite{Dirac28a}.  Its form resulted from the constraints of relativity that space and time derivatives must appear in the same order in the equations of motion, as well as constraints from the probabilistic interpretation of the wave function that the equations of motion depend only on the first derivative of time.  Dirac's solution used $4 \times 4$ complex matrices that in their modern form are referred to as gamma matrices and a four component wavefunction.   The four components allowed for both positive and negative charge solutions and up and down spin.  This epochal moment in theoretical physics -- originating in these simple considerations -- led to a new understanding of the concept of spin, predicted the existence of antimatter, and was the invention of quantum field theory itself.  A number of variations of the Dirac equation quickly followed.  In 1929, the mathematician Hermann Weyl proposed a simplified version that described massless fermions with a definite chirality (or handedness) \cite{Weyl}.   In 1937, Ettore Majorana found a modification using real numbers, which described a neutral particle that was its own anti-particle \cite{Majorana37a,Elliott15a}.  These developments have found vast application in modern particle physics.  The Dirac equation is now the fundamental equation describing relativistic electrons and the Majorana equations are a candidate to describe neutrinos.  The Dirac equation is also a key concept leading to topological phenomena such as zero modes and anomalies in quantum field theories.  Unfortunately in the nearly 90 intervening years no candidate Weyl fermions have been observed as fundamental particles in high-energy particle physics experiments.

In condensed matter physics, where one is interested in energy scales much smaller than the rest mass of the electron, it would appear at first blush that a non-relativistic description, perhaps with minor corrections, would suffice and that Dirac physics would not play an important role. However, the propagation of even slow electrons through the periodic potential of a crystal, leads to a dressing of the electronic states. In certain instances this results in an effective low energy description that once again resembles the Dirac equation. Perhaps the best known example of this phenomena is in graphene, where a linear in momentum dispersion relation is captured by the massless two dimensional Dirac equation. These 2D carbon sheets provide a condensed-matter analogue of (2+1)-dimensional quantum electrodynamics (QED) \cite{Semenoff84a} and with the contemporary isolation of single layer graphene sheets has resulted in a large body of work on their electronic properties \cite{Novoselov04a,Novoselov05a,Zhang05a,Geim12a,Neto09a}.  Recent work has found evidence for Majorana bound states in 1D superconducting wires \cite{Mourik12a,Elliott15a}.  In this review we will mainly be concerned with analogous physics in \textit{three dimensional crystals} with linearly dispersing fermionic excitations that are describe by the massless 3D Weyl and Dirac equations.  These solid state realizations offer a platform where predictions made by relativistic theories can be tested, but at the same time entirely new properties that only exist in a condensed matter context emerge, such as Fermi arc surface states at the boundary of the material.   Moreover, the fact that the strict symmetries of free space do not necessarily hold in a lattice, means that new fermion types with no counterpart in high-energy physics \cite{Soluyanov15a,Bradlyn16a,Yong2015a,Wieder16a}, can emerge (e.g. Type II Weyl, Spin-1 Weyl, Double Dirac etc.).

Let us further review of history of three dimensional Weyl equation \cite{Weyl}, its relation to Dirac and Majorana fermions, and their manifestation in condensed matter systems.  In 1929 shortly after Dirac wrote down his equation for the electron which involved these complex $4\times 4$ matrices, Weyl pointed out a simplified relativistic equation utilizing just the $2\times2$ complex Pauli matrices $\sigma_n$. This simplification required the fermions to be massless. Weyl fermions are associated with a chirality or handedness, and a pair of opposite chirality Weyl fermions can be combined to obtain a Dirac fermion. It had been believed that neutrinos might be Weyl fermions.  However with the discovery of a nonvanishing neutrino mass \cite{McDonald16a,Kajita16a}, there are no fundamental particles currently believed to be massless Weyl fermions.  As mentioned above Majorana also found a modification of the Dirac equation that used real numbers and described a neutral particle that was its own anti-particle \cite{Majorana37a,Elliott15a}.

In a seemingly unrelated line of reasoning, the conditions under which degeneracies occur in electronic band structures, was investigated by Herring in 1937 \cite{Herring37a}. It was noted that even in the absence of any symmetry one could obtain accidental two-fold degeneracies of bands in a three dimensional solid. The dispersion in the vicinity of these band touching points is generically linear and resembles the Weyl equation, modulo the lack of strict Lorentz invariance. Remarkably several of the defining physical properties of Weyl fermions, such as the so-called chiral anomaly, continue to hold in this nonrelativistic condensed matter context. The chiral anomaly discussed by Adler-Bell-Jackiw \cite{Adler69a,Bell69a}, is an example of a quantum anomaly which in its simplest incarnation demonstrates that a single Weyl fermion coupled to an electromagnetic field results in the nonconservation of electric charge. To evade this unphysical consequence, the net chirality of a set of Weyl fermions must vanish in a lattice realization, an example of the fermion doubling theorem. However, even in this setting it was realized that the chiral anomaly can have a nontrivial effects as pointed out in a prescient paper \cite{Nielsen83a}, which cemented the link between band touchings in three dimensional crystals and chiral fermions.  These band touchings of Herring were  named ``Weyl nodes" in \cite{Wan11a}. The electrodynamic properties of these and other band touching points had been studied by \onlinecite{Abrikosov71a}.

Topological consequences of Weyl nodes began to be explored with the realization that Berry curvature plays a key role in determining the  Hall effect \cite{Karplus54a,TKKN}, and the Weyl nodes are related to ``diabolic points" discussed by Berry as sources of Berry flux \cite{Berry85}. The fact that diabolic points are monopoles of Berry curvature and could influence the Hall effect in ferromagnets was emphasized in \cite{Fang92, Nagaosa10a}.  In a different context, realizations of Weyl nodes in  superfluids and superconductors have been discussed \cite{Volovik87a,Murakami02a}.  In particular Volovik  \cite{Volovik87a} pointed out that the $A$ phase of superfluid He$_3$ realizes nodes in the pairing function leading to a realization of Weyl fermions.
 
The prediction and discovery of topological insulators (TIs) in two and three dimensions  \cite{Haldane88a,KaneMele2005,Fu07a,Fu07b,Bernevig06a, Moore07a,Roy09a,Hsieh08a,Xia09a,HZhang2009,Chen09a}, has led to an explosion of activity in the study of  topological aspects of band structures \cite{Hasan10, Qi11}. While initially confined to the study of band insulators, where topological properties could be sharply delineated due to the presence of an energy gap, interesting connections to gapless states have begun to emerge. For one, the surfaces of topological insulators in 3D feature a gapless Dirac dispersion, analogous to the two dimensional semimetal graphene, but with important differences in the number of nodal points.  The transition between topological and trivial phases proceeds through a gapless state -  for example a 3D topological to trivial insulator transition, in the presence of both time reversal and inversion symmetry proceeds through the 3D Dirac dispersion \cite{Murakami07a}.  If inversion symmetry is lost, then the critical point expands into a gapless phase with Weyl nodes, that migrate across the Brillouin zone and annihilate with an opposite chirality partner, leading to the change in topology.   

A direct manifestation of the topological aspects of Weyl fermions appeared with the realization that Weyl nodes lead to exotic surface states in the form of Fermi arcs \cite{Wan11a}. The name `Weyl semimetal' (WSM) was introduced to describe a phase where the chemical potential is near the Weyl nodes and a potential realization of such a state in a family of materials, the pyrochlore iridates, was proposed along with the prediction of a  special all-in, all-out magnetic ordering pattern \cite{Wan11a}. Subsequent proposed realization of Weyl semimetals in magnetic systems include the spinel HgCr$_2$Se$_4$ (a double Weyl) \cite{Xu11},  heterostructures of ferromagnets and topological insulators \cite{Burkov11a}, and Hg$_{1-x-y}$Cd$_x$Mn$_y$Te films \cite{bulmash2014prediction}.  Although a clearcut demonstration of a magnetic WSM remains outstanding, the search for inversion breaking Weyl systems as envisaged in  \cite{Murakami07a,Halasz12a} reached fruition with the prediction and discovery of TaAs as a WSM \cite{Weng2015a,Huang15a,Lv15a,Lv15b,Xu15a} (and other members in this material class \cite{Xu15b,Xu15c,xu2015experimental}), and the observation of Fermi arc surface states attached to the bulk Weyl points.

For the case of 3D material systems described by the massless Dirac equation, the possibility of stable $four$-fold degenerate Dirac points was raised by \onlinecite{Abrikosov71a} and much more recently by \onlinecite{Young2012,WangA3Bi2012}.  Unlike the case of Weyl points, this degeneracy is not topologically protected since its net Chern number is zero and residual momentum-conserving terms in the Hamiltonian can potentially mix these terms and gap the electronic spectrum.  However in particular situations this mixing can be forbidden by space group symmetries in which case the nodes remain intact as \textit{symmetry-protected} degeneracies.  For instance this can occur at a phase transition between TI and non-TI phases in a crystal system that preserves inversion \cite{Murakami07a,Murakami2007b}.  However, as pointed out by \onlinecite{Young2012,WangA3Bi2012,Steinberg2014} a DSM can also appear as a robust electronic phase that is stable over a range of Hamiltonian control parameters.   Such systems are called Dirac semimetals (DSM).   A number of material realizations of such symmetry protected DSMs have been discovered \cite{Liu14b,Xu15a,Liu14a,Borisenko14a,Neupane14a}. 

There are a number of excellent reviews on related topics in this general area. Basic concepts related to Weyl semimetals were reviewed in \onlinecite{Turner13a} as well as in the very clear set of lectures by \onlinecite{WittenLectures15}. Transport properties of Weyl semimetals were reviewed in \cite{Hosur13a} and \cite{Burkov15a}, while the extensive contributions of $ab$ $initio$ techniques to the discovery of topological materials (including Weyl and Dirac states) were reviewed in  \onlinecite{Bansil16a,Weng16a}.  \onlinecite{Kharzeev14a} reviews connections of Weyl fermions in solid-state physics to the chiral anomaly in quantum chromodynamics.  A number of other shorter reviews on specific aspects of topological semimetals have appeared recently \cite{Witczak14a,Syzranov17a,Yan16a,Burkov16a,Jia16,Hasan15a,Smejkal17a,Hasan17a,burkov2017weyl} and connections to related systems like nodal superconductors have been reviewed in \onlinecite{Wehling14a,Vafek14a,Schnyder15}. In the present review, we attempt to summarize the theoretical, materials, and experimental situation of 3D Dirac and Weyl semimetals with an emphasis on general features independent of specific material systems.  There have been many interesting developments in this area in the last few years, in regards to the theoretical proposals, the development of new materials, and the study of experimental phenomena.  Although we have been attentive to matters of priority, the experimental data we have chosen to include is not necessarily the first that was shown to demonstrate some phenomenon, but is our estimation of that most illustrative of an effect.  Unfortunately even in the relatively well-defined scope of the current topic, the literature is vast and we cannot hope to cover all work. Important omissions are regrettable but inevitable.

\section{Properties of Weyl semimetals}
\label{WeylSection}

\subsection{Background} 

Here we first review two seemingly unrelated topics that originated in the early days of quantum mechanics -  the problem of level repulsion and accidental degeneracies and the relativistic wave equations for fermions. We will see that these provide complementary perspectives on Weyl semimetals and are unified by the identification of their topological aspects. Towards this end we will review topological invariants of insulators in the third part of this section.

\subsubsection{Accidental degeneracies} As a starting point, consider the basic question of when accidental degeneracies arise in an energy spectrum \cite{vonNeumannWigner}. We focus on a pair of energy levels and ask if one can bring these levels into degeneracy by tuning Hamiltonian parameters. The energy levels (up to an overall constant) are determined by the most general $2 \times 2$ Hamiltonian:  $ H =  f_1 \sigma_x+ f_2 \sigma_y + f_3 \sigma_z$, with an energy splitting between the levels $\Delta E = 2\sqrt{f_1^2+f_2^2+f_3^2}$.  In general in the absence of any symmetry, this cannot be accomplished by tuning just one parameter;  degeneracy requires tuning all three terms to give zero simultaneously.  If we focus on real Hamiltonians with time reversal symmetry, we can exclude the imaginary Pauli matrix. Then, a pair of levels can be brought into coincidence typically by tuning two parameters, since we can typically solve two equations $\epsilon_x=0$ and  $ \epsilon_z=0$ with two variables. But in the absence of any such symmetry, we need to tune three independent parameters to achieve a degeneracy.  As mentioned above, these points of degeneracy in the extended two or three parameter space are termed diabolic points and have been discussed in the context of Berry's phase \cite{Berry85} and not surprisingly will be associated with a topological property as we discuss below.

\subsubsection{Weyl and Dirac fermions}

The Dirac equation in $d$ spatial dimension and effective speed of light $c=1$ is
\begin{equation}
(i\gamma^\mu\partial_\mu -m)\psi=0
\end{equation}
where $\mu=0,\,1\dots,\,d$ label time and space dimensions, and the $d+1$ gamma matrices satisfy the anti-commutation relation $\{\gamma^\mu,\,\gamma^\nu\}=0$ for  $\mu\neq \nu$ and $(\gamma^0)^2=-(\gamma^i)^2= \mathbb{I}$,
where $i=1,\dots,\,d$.  $\mathbb{I}$ is the $2\times 2$ unit matrix.
The minimal sized matrices that satisfy this property depend of course on the dimension, and are $2^{k+1}\times 2^{k+1}$ dimensional matrices in both spatial dimensions  $d=2k+1 \, {\rm and} \, 2k+2$.

Weyl noticed that this equation can be further simplified in certain cases in odd spatial dimensions \cite{Weyl}. For simplicity consider $d=1$. Then one needs only two anti-commuting matrices eg. the $2 \times 2$ Pauli matrices eg. $\gamma^0 = \sigma_z$ and $\gamma^1=i\sigma_y$. Therefore the Dirac equation in $1+1$ dimension involves a two component spinor and can be written as: $i\partial_t\psi = (\gamma^0\gamma^1  {p}  +m\gamma^0)\psi$, where $p=-i\partial_x$. If one were describing a {\em massless} particle $m=0$, this equation can be further simplified by simply picking eigenstates of the Hermitian matrix $\gamma_5 = \gamma^0\gamma^1=\sigma^x$. If $\gamma_5\psi_\pm=\pm \psi_\pm$. One then has the 1D Weyl equation:
\begin{equation}
i\partial_t\psi_\pm = \pm {p} \psi_\pm
\end{equation}
The resulting dispersion is simply $E_\pm(p) = \pm p$ which denotes a right (left) moving particle, which are termed chiral or Weyl fermions. Analogous dispersions arise at the one dimensional edge of an integer quantum Hall state, but are not allowed in an isolated one dimensional system where chiral fermions must appear in opposite pairs.  The fermion mass term interconverts opposite chiralities. We will see that an analogous situation prevails in 3+1 dimensions and indeed the analogy with 1D fermions will be a theme that we will repeatedly return to.

In any odd spatial dimension $d=2k+1$ one can form the Hermitian matrix $\gamma_5 = i^k \gamma^0\gamma^1\dots\gamma^d$. This is guaranteed to commute with the `velocity' matrices $\gamma^0\gamma^i$, which can be simultaneously diagonalized along with the massless Dirac equation. At the same time it differs from the identity matrix since it anticommutes with $\gamma^0$. In even spatial dimensions, the latter property no longer holds, since all the gamma matrices are utilized, and their product is just the identity.

Let us now specialize to $d=3$. The gamma matrices are, as Dirac found, now $4\times 4$ matrices, and can be represented as $\gamma^0 = \mathbb{I}\otimes \tau_x$, $\gamma^i = \sigma^i\otimes i\tau_y$ and $\gamma^5= -\mathbb{I}\otimes \tau_z$. Again, if we identify chiral components: $\gamma_5\psi_\pm=\pm \psi_\pm$, where $\psi_\pm$ are effectively two component vectors, we have for the massless Dirac equation:
\begin{eqnarray}
\nonumber
i\partial_t\psi_\pm &=& H_\pm \psi_\pm \nonumber \\
H_\pm &=& \mp {\vec{p}\cdot\vec{\sigma}}
\label{Weyl1}
\end{eqnarray}
Thus Weyl fermions propagate parallel (or antiparallel) to their spin, which defines their chirality. We will see that a single chirality of Weyl fermions cannot be realized in 3D, but momentum separated pairs can arise.  These are the Weyl semimetals.

\subsubsection{Topological invariants for band insulators}
Band theory describes the electronic states within a crystal in terms of one particle Bloch wave functions $|u_n(\k)\rangle$ that are defined within the unit cell and are labelled by a crystal momentum $\k$ and band index $n$. The Berry phase of the Bloch wavefunctions within a single band $n$ is captured by the line integral of the Berry connection ${\mathcal{A}}_n(\k) = -i \langle u_n(\k)|\nabla_\k|u_n(\k)\rangle$, or equivalently the surface integral of the Berry flux: ${\mathcal F}^{ab}_n(\k) = \partial_{k_a}{\mathcal A}^b_n-\partial_{k_b}{\mathcal A}^a_n$.  For a two dimensional insulator, the Berry flux for each isolated band ${\mathcal F}^{xy}_n = {\mathcal F}_n$ is effectively a single component object, and the net Berry flux is quantized to integers values since:
\begin{equation}
 \int \frac{d^2\k}{2\pi}{\mathcal F}_n(\k) =N_n
\end{equation}
The quantized Hall conductance is obtained by summing over all occupied bands. In a three dimensional crystal, the Berry flux behaves like a dual magnetic field $\epsilon^{abc}{\mathcal B}_c(\k) = {\mathcal F}^{ab}(\k) $, switching the roles of position and momentum (with suppressed band index $n$). The  semiclassical equations of motion for an electron now take the following symmetric form \cite{Xiao10}
\begin{eqnarray}
{\bf \dot{r}} &=& {\bf v} - {\bf \dot{p}} \times {\mathcal B}\\
{\bf \dot{p}} & = & e{\bf E} +e {\bf \dot{r}}\times {\bf B} 
\end{eqnarray}
where  ${\bf v}$ is an appropriately defined renormalized band velocity and ${\bf E},\, {\bf B}$ are externally applied electric and magnetic fields. Note, the Berry flux restores the symmetry $r\leftrightarrow p$ of these equations of motion, which is otherwise broken by the Lorentz force. However there is an important difference between  ${\bf B}$ and ${\mathcal B}$. Unlike the physical magnetic field, the Berry field ${\mathcal B}$  is allowed to have magnetic monopoles. We will see that these precisely correspond to the Weyl points in the band structure.

\subsection{Topological aspects of Weyl semimetals}
\label{WSMmodels}

In Weyl semimetals,  the conduction and valence bands coincide in energy over some region of the Brillouin zone. Furthermore, this band touching is stable at least to small variations of parameters. A key input in determining conditions for such band touchings is the degeneracy of bands, which in turn is determined by symmetry. If spin rotation symmetry is assumed, e.g. by ignoring spin-orbit coupling, the bands are doubly degenerate. Alternatively, doubly degenerate bands also arise if both time reversal $\mathcal T$  and inversion symmetry $\mathcal P$ ($r\rightarrow-r$) are simultaneously present or their combined $ {\mathcal PT}$ symmetry is. Then, under the operation $\tilde{\mathcal  T}  = {\mathcal PT}$, crystal momenta are invariant and moreover ${\tilde{\mathcal  T}}^2=-1$, which ensures double degeneracy. On the other hand, if only $\mathcal T$ is present, bands are generally nondegenerate since crystal momentum is reversed under its action. Only at the time-reversal invariant momenta (TRIM) where $\k \equiv -\k$, is a Kramers degeneracy present. Similarly, if time reversal is broken, and only inversion is present, the bands are typically nondegenerate.  As discussed above, the conditions for a pair of such nondegenerate bands to touch can be captured by discussing just a pair of levels whose effective Hamiltonian can generically be expanded as:  $H(\k) = f_0(\k)\mathbb{I} + f_1(\k)\sigma_x + f_2(\k)\sigma_y + f_3(\k)\sigma_z $. To bring the bands in coincidence we need to adjust all three coefficients $f_1=f_2=f_3=0$ simultaneously, which, by the arguments of the previous section requires that we have three independent variables, i.e. that we are in three spatial dimensions. As we can then expect band touchings without any special fine tuning, we can readily argue that the existence of Weyl nodes is stable to small perturbations of Hamiltonian parameters. The location of the Weyl nodes can be geometrically visualized as follows. We consider the real function $f_1(\k)$ and ask where it vanishes in momentum space;  typically this will be a 2D surface that separates positive and negative values of the function. If we demand a simultaneous zero of $f_2(\k),\,f_3(\k)$, this specifies the intersection of three independent surfaces, which will typically occur at a point. Now, consider a perturbation that changes the functions $f_a$ by a small amount. This will also move the zero surfaces, and their points of intersection by a small amount, but the intersection will persist, just at a different crystal momentum.   The Weyl nodes cannot be removed by any small perturbation, and may only disappear by annihilation with another Weyl node.  Below, we will describe a topological perspective that makes this fact obvious.

\begin{figure}[t!]
\includegraphics[width=\columnwidth]{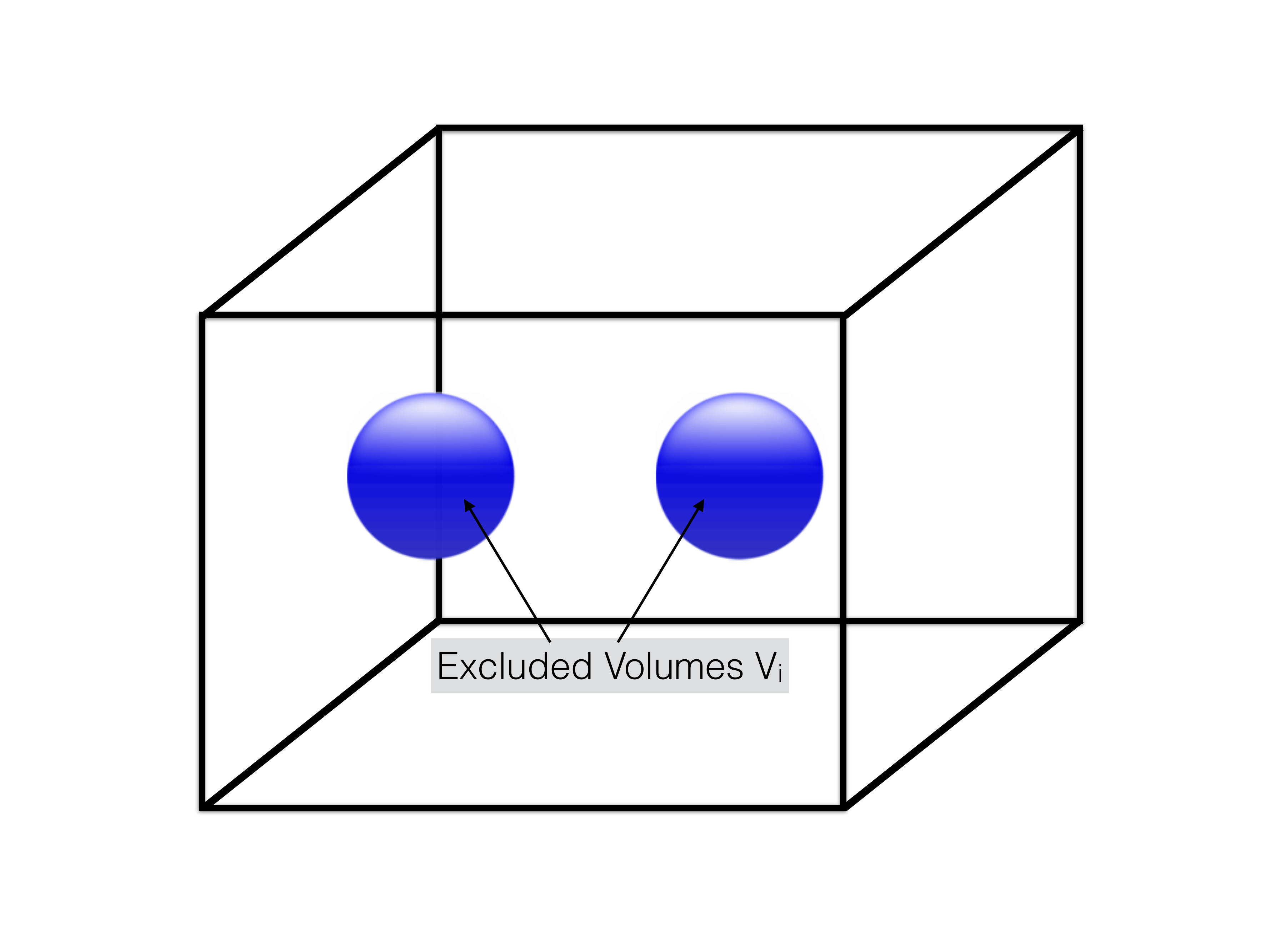}\vspace{-.1in}
\caption{Net chirality of Weyl nodes must be zero, which is a consequence of the fact that the net Berry flux integrated over the Brillouin Zone (a closed volume) must vanish.}
\label{fig:BZ}
\vspace{-.1in}
\end{figure}

\begin{figure*}[ht]
\includegraphics[width=12.7cm]{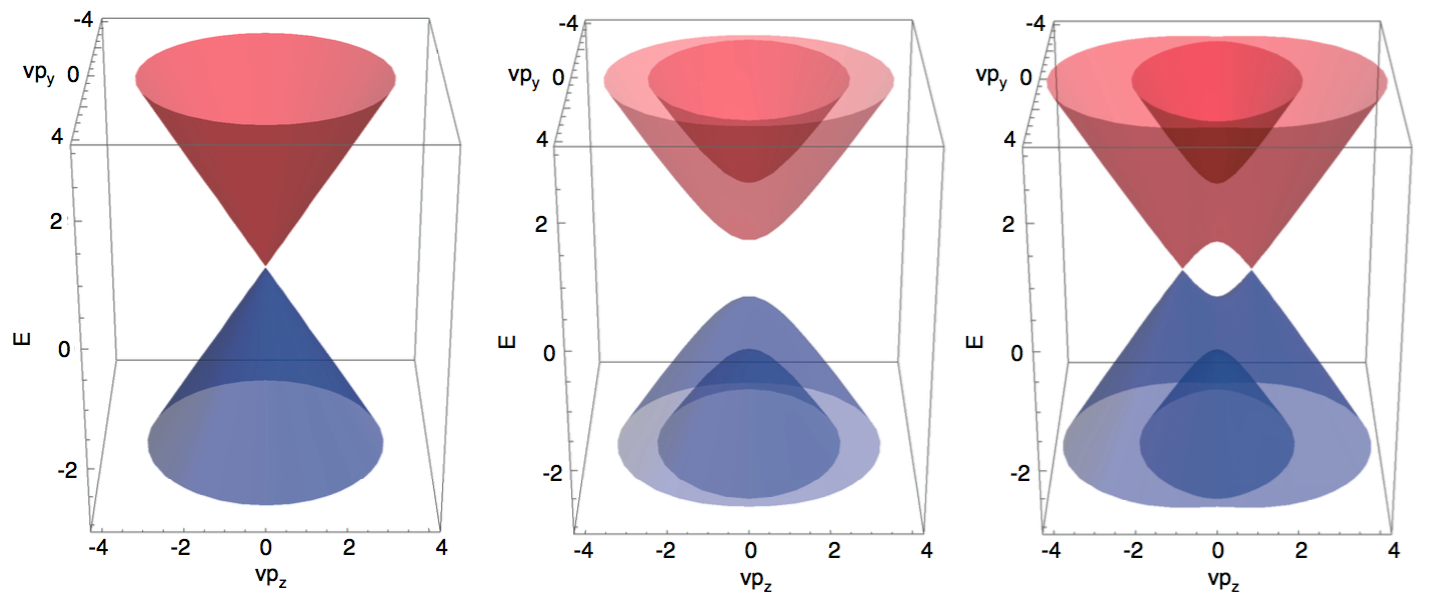}
\includegraphics[width=4.7cm]{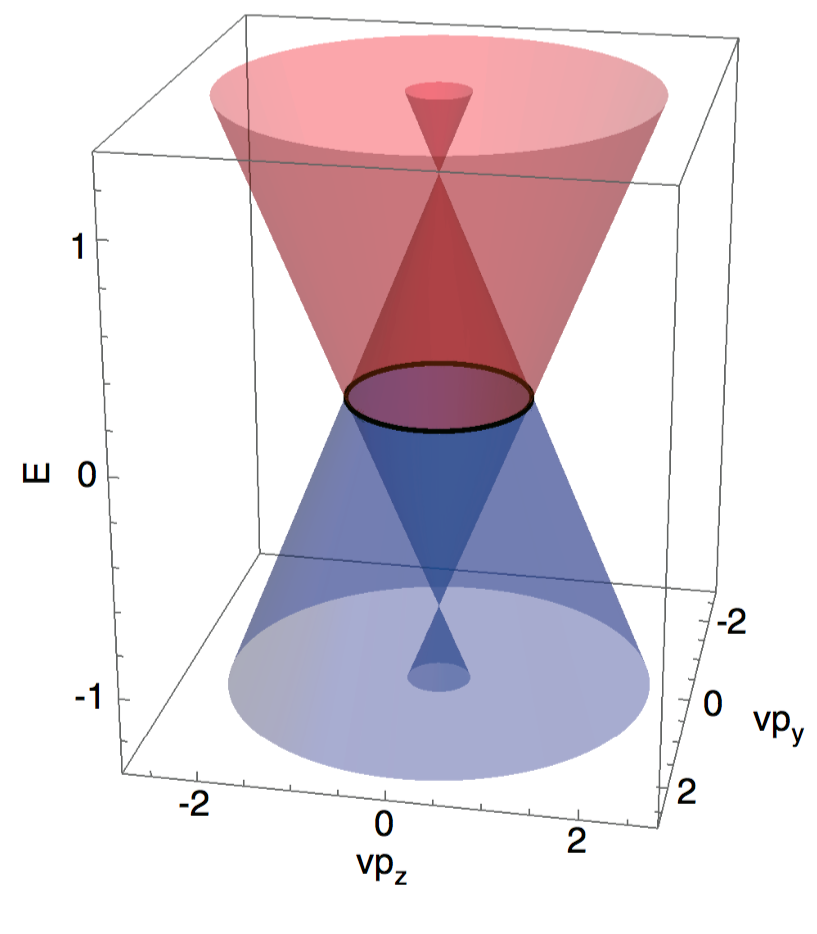}\vspace{-.1in}
\caption{ (left to right) Energy spectra of $\varepsilon_{s\mu}(0,p_y,p_z)$
        for the Dirac semimetal  ($m=b=b'=0$), magnetic semiconductor ($m=1, b=0.5, b'=0$), Weyl semimetal  ($m=0.5, b=1, b'=0$), and line node semimetal   ($m=0, b=0, b'=1$) for the Hamiltonian Eq. \ref{4band}.  From \cite{Koshino16a}. }
\label{KoshinoFig}
\end{figure*}

Based on this reasoning it may appear that all we need to do to realize a WSM is to find a 3D crystal with nondegenerate bands by breaking appropriate symmetries.  While indeed Weyl nodes are quite natural, typically one also imposes an additional requirement - that they be close to the Fermi energy, so this also requires that we find candidates for which $f_0(\k)$ is nearly zero.  We can further discuss the generic dispersion near the band touching point, by expanding the Hamiltonian about $\k = \delta \k +\k_0$. This gives
\begin{equation}
H(\k) \sim f_0(\k_0) \I+ {\bf v}_0 \cdot \delta\k \;\I + \sum_{a=x,y,z}{\bf v}_a \cdot \delta\k\; \sigma^a
\label{WeylHamiltonian}
\end{equation}
where ${\bf v}_\mu = \nabla_k f_\mu (\k)\big|_{\k=\k_0}$ (with $\mu =0,\,\dots ,\,3$) are effective velocities  which are typically nonvanishing in the absence of  additional symmetries. Note, if we revert to the special limit where ${\bf v}_0 =0$ and ${\bf v}_a = v_0 \hat{a}$ ($a=1,\,\dots 3$), we obtain the Weyl equation (\ref{Weyl1}). We therefore refer to these band touchings as  {\em Weyl nodes}.  While this makes a connection to Weyl fermions with a fixed chirality ($C={\rm sign }\left ( {\bf v}_x \cdot {\bf v}_y\times {\bf v}_z \right )$) it remains unclear why Weyl nodes should come in opposite chirality pairs.  To realize this we need a topological characterization of Weyl nodes which is furnished by calculating the Berry flux on a  surface surrounding the Weyl point.

Furthermore, we can check that the Berry flux piercing any surface enclosing the point $\k_0$ is exactly $2\pi C$ , where $C$ is the chirality e.g. Weyl points are monopoles of Berry flux.  If we consider the sphere surrounding a Weyl point, and consider its 2D band structure, it has a nonvanishing Chern number $C=\pm 1$. However, if we expand this surface so that it covers the entire Brillouin zone, then by periodicity, it is actually equivalent to a point, and must have net Chern number zero. Therefore, the net Chern number of all Weyl points in the Brillouin zone must vanish. This can be seen from the Fig. \ref{fig:BZ} where we isolate band touchings within the volumes ${\mathcal V}_i$. The integral of $\nabla_{\k}\cdot {\mathcal B}(\k) =0$ over this volume vanishes, but can be expressed as an integral over the surfaces of the excluded volumes $\sum_i \oint_{\partial V_i}  {\mathcal B}(\k)\cdot d{\mathcal S}_k=-2\pi \sum_i C_i$ which must vanish. In the continuum, one can define a single Weyl node, since the momentum space is no longer compact. However, in lattice model realizations of Weyl fermions the net chirality must vanish.  This also shows that Weyl nodes can only be eliminated by distortions to the Hamiltonian in a pairwise fashion e.g. by annihilation with another Weyl node of opposite chirality. Note also, the Berry flux must be an integer multiple of $2\pi$ which allows, for example, band touching with $C=\pm 2$, which corresponds to ``double Weyl" nodes, which do not have a linear dispersion in all directions.  

\begin{figure*}[ht]
\includegraphics[width=6.7cm]{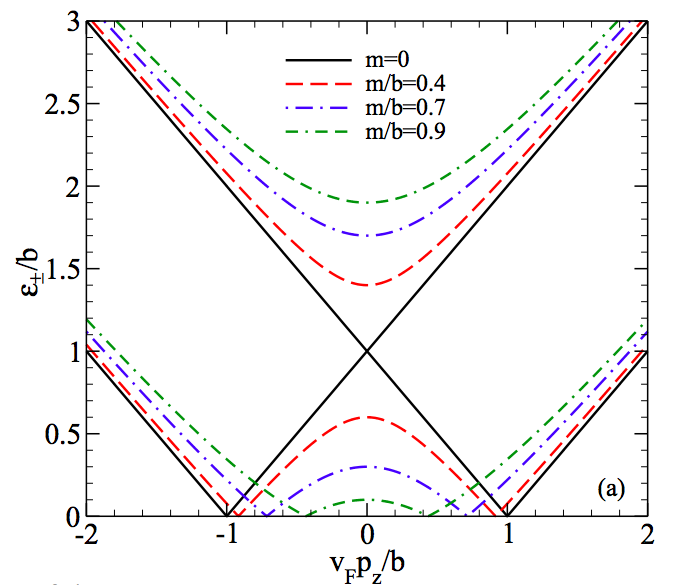}
\includegraphics[width=6.4cm]{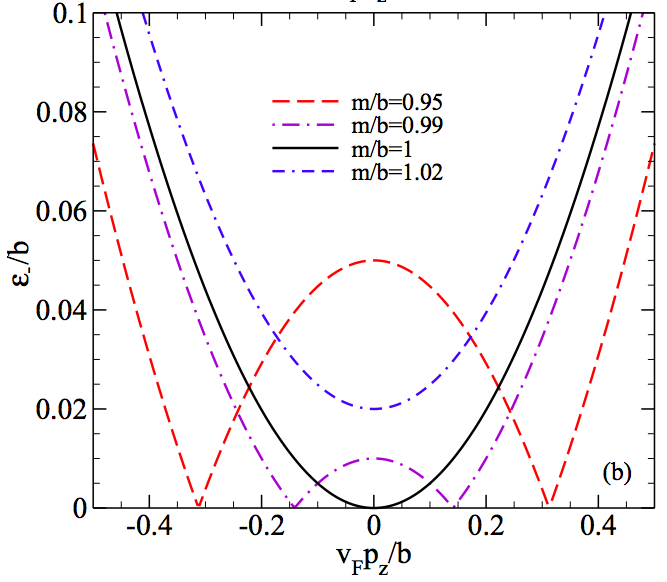}
\includegraphics[width=4.5cm]{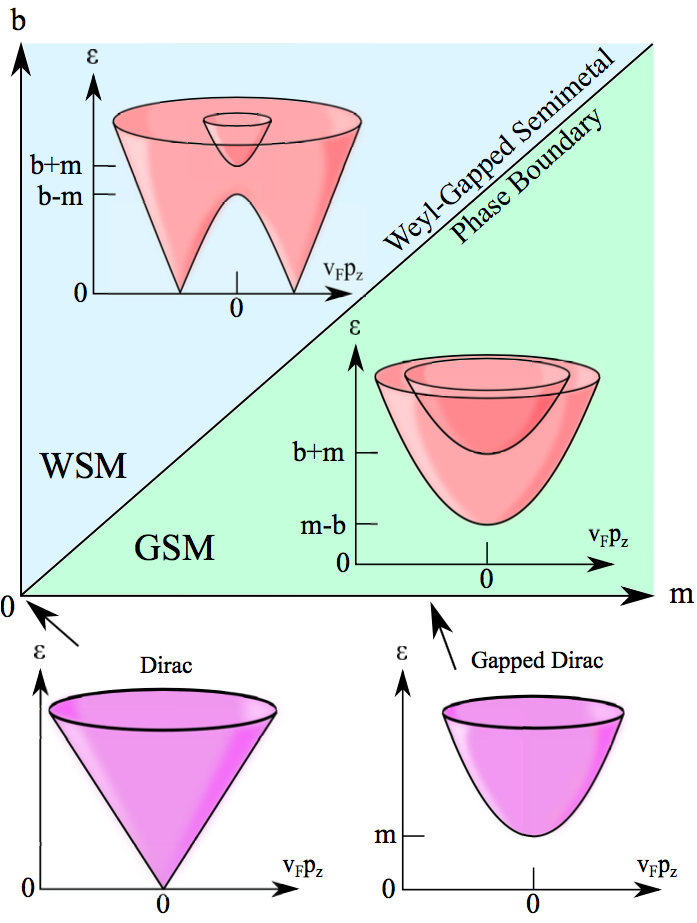}\vspace{-.1in}
\caption{ (left) Band structure from Eq. \ref{4band} for values of $m/b$ for the $s = +$ and $\mu = \pm$  bands for increasing $m/b$ in the WSM phase. For finite $m$ the $\mu = +$ band is gapped, while $\mu = -$ contains two Weyl nodes. (center) The $s=+$ and $\mu=-$ band near the phase transition at $m/b = 1$.  (right)  Phase diagram of Eq. \ref{4band}.  At $m /b < 1$, the system is a WSM, while $m/b > 1$, a gapped semimetal exists. Along $b = 0$, a degenerate massive DSM is observed.  At $m = b = 0$, massless degenerate Dirac fermions exist.  From \cite{Tabert16a}. }
\label{Tabert16}
\end{figure*}

To build intuition and make the possibilities more explicit in the space of $4\times 4$ Hamiltonians \cite{Burkov11b}, we can consider a simple continuum system with two orbitals plus spin, which describes the cases of WSMs, ``line node" semimetals\footnote{As pointed out elsewhere \cite{Burkov11b}, this term is an oxymoron.   Yet it persists.}, as well as conventional gapped magnetic semiconductors.  Expanding around the $\Gamma$ point, we consider a $4\times 4$ Hamiltonian matrix,
\begin{align}
H &= v\tau_x (\GVec{\sigma}\cdot \k) + m\tau_z + b\sigma_z + b' \tau_z \sigma_x
\nonumber\\
&=
\begin{pmatrix}
   m \mathbb{I} + b\sigma_z +b' \sigma_x & v \GVec{\sigma}\cdot\k \\
   v \GVec{\sigma}\cdot\k & -m \mathbb{I} + b\sigma_z - b'\sigma_x
 \end{pmatrix},
\label{4band}
\end{align}
where $\k=(k_x,k_y,k_z)$ is the momentum, and the $\tau_n$'s are Pauli matrices for the pseudospin orbital degrees of freedom.    Here $m$ is a mass parameter, and $b$ and $b'$ are Zeeman fields that physically can correspond to magnetic field in the $z$ and $x$ directions respectively.  A number of interesting and relevant cases can be obtained as a function of $m$, $b$, and $b'$.   For $b'$ equal to zero, one obtains the eigenvalues
\begin{equation}
\varepsilon_{s\mu}(\k) = s \sqrt{m^2+b^2+v^2 k^2 + \mu  2b\sqrt{v^2 k_z^2 + m^2}},
\end{equation}
where $k=|\k|$,  and $s=\pm1$ and $\mu=\pm1$.  The  spectrum for $\varepsilon_{s\mu}(0,k_y,k_z)$ is plotted 
in Fig. \ref{KoshinoFig} for  cases (far left) $m=b=0$, which corresponds to a Dirac semimetal 
composed of a pair of degenerate linear bands, which touch at $\k=0$,  (center left) $|m|>|b|$ describes a gapped magnetic semiconductor, where the energy bands are gapped in the range  $|E| < |m|-|b|$, and (center right) $|b|>|m|$ that represents the WSM where the middle bands  touch at a pair of isolated point-nodes $\k = (0,0,\pm \sqrt{b^2-m^2}/v)$.  Further plots from \cite{Tabert16a} are shown in Fig. \ref{Tabert16} that reflects different regimes in the $m$ and $b$ parameter space.

For the case of $m=b=0$, but $b'$ finite, one obtains the eigenvalues
\begin{equation}
\varepsilon_{s\mu}(\k) = s \sqrt{v^2 k_x^2 + \left[v\sqrt{k_y^2+k_z^2} + \mu b' \right]^2},
\label{E_line}
\end{equation}
where the zero-energy contour becomes a circle at $k_x=0$ and  $\sqrt{k_y^2+k_z^2}=b' /v$ as shown in Fig. \ref{KoshinoFig} (far right).  The spectrum is immediately gapped for $k_x$ away from 0.

Although idealized here as  $\mathcal{T}$  breaking fields in this continuum model, $b$ and $b'$ are representative of the symmetry breaking perturbations that may be encountered in lattices, which may require different considerations.  See for example \onlinecite{Carter12a}.  It is also important to note that in general it is impossible to apply laboratory magnetic fields large enough to generate Zeeman splittings of the Weyl nodes that are substantial fractions of the Brillouin zone.  Therefore, the fields $b$ and $b'$ above should be considered as effective internal fields.

\begin{figure}[t]
\includegraphics[width=0.85\columnwidth]{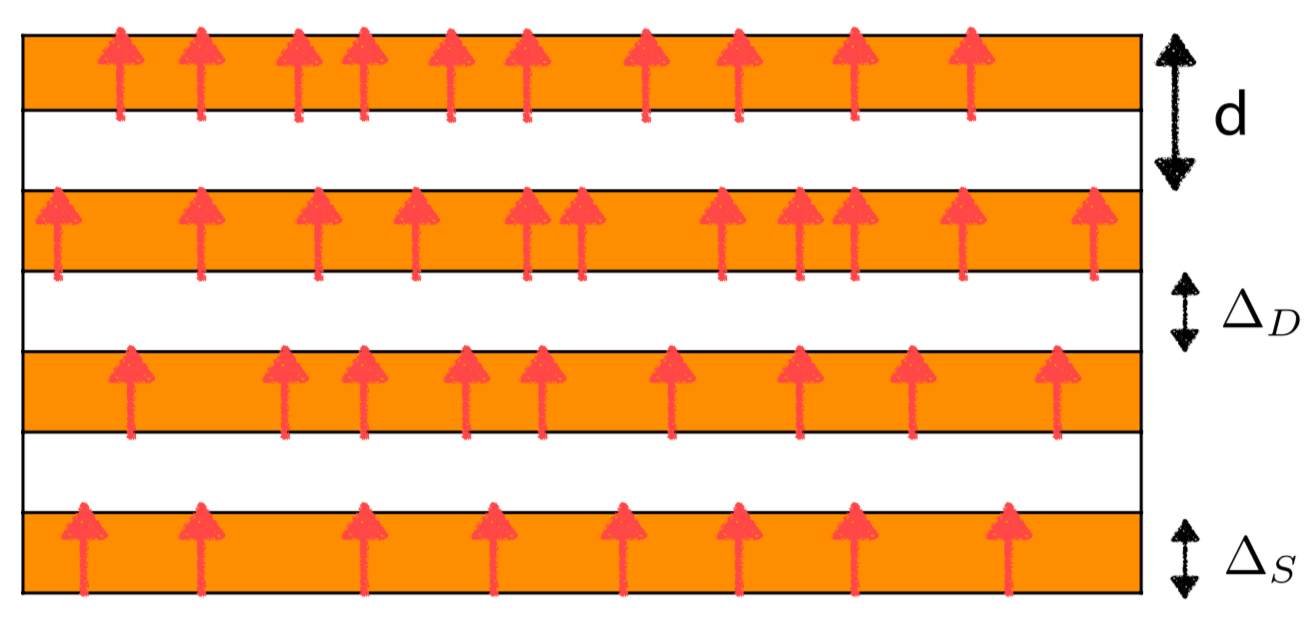}\vspace{-.1in}
\caption{Cartoon of a heterostructure model of a Weyl semimetal of topological and normal insulators. Doped magnetic impurities are shown by arrows.  $d$ is the real space periodicity of the lattice.  $\Delta_S$ and $\Delta_D$ are tunneling between topological surface states on the same topological insulator layer and between different layers respectively.  From \cite{Burkov15a}.}
\label{BurkovLayers}
\vspace{-.1in}
\end{figure}

To gain further intuition into the physics, it is very instructive to consider a real space model of a WSM.   As a simplest realistic model for a ${\mathcal T}$ broken WSM \onlinecite{Burkov11a} considered a repeating structure of period $d$ of normal insulators and TIs in which ${\mathcal T}$ was broken through a Zeeman field.  As shown in Fig. \ref{BurkovLayers}, one can model such system with two different tunneling matrix elements for tunneling between surface states on the same layer ($\Delta_S$) and between surface states on neighboring layers ($\Delta_D$).   With no magnetism this system shows a topological--normal band inversion transition as a function of the relative strength of $\Delta_S$ and $\Delta_D$.  The multilayer is a bulk 3D TI when $\Delta_D > \Delta_S $ and a normal insulator for $\Delta_D < \Delta_S $.  If the layers are magnetized giving a spin splitting $b$ one finds a WSM phase for values of $\Delta_S$ near $\Delta_D$.   The Weyl nodes are found at 
\begin{equation}
k_z^{\pm} = \frac{\pi}{d}  \pm  \frac{1}{d} \mathrm{arccos} \Big(  \frac{\Delta_S^2 + \Delta_D^2 - b^2}{2 \Delta_S \Delta_D} \Big).
\label{WeylNodePosition}
\end{equation}
One can see that that as a function of  $\Delta_S$, $\Delta_D$, and $b$ the Weyl nodes move through the Brillouin zone (BZ) and can annihilate at the BZ edge for a critical value of $b$ giving a fully magnetized state and as discussed below a quantized anomalous Hall conductivity.  A related construction is possible for ${\mathcal P}$ breaking WSMs \cite{Halasz12a}.   These layered models not only suggest a possible route towards creating new WSM states but also provide an alternate viewpoint on WSMs as a state of matter with a periodically inverted and uninverted ``local band gap".   With this perspective, aspects like the presence of Fermi arcs and the anomalous Hall effect follow naturally.  Related schemes for building up 3D topological semimetals by stacking one dimensional primitives based on the Aubry-Andre-Harper model have also been studied \cite{Ganeshan15a}.

 \subsubsection{Weyl semimetals with broken ${\mathcal T}$ symmetry}
The simplest setting to discuss a WSM is to assume broken time reversal symmetry, but to preserve inversion. This allows for the minimal number of Weyl nodes, i.e. two with opposite chirality. Inversion symmetry guarantees they are at the same energy and furthermore provides a simple criterion to diagnose the existence of Weyl points based on the parity eigenvalues at the time reversal invariant momenta (TRIMS).

 Let us discuss this in the context of the following toy model. We envision a magnetically ordered system so the bands have no spin degeneracy, but with a pair of orbitals on each site of a simple cubic lattice. Further assume that the orbitals have opposite parity (e.g.  $s,\,p$ orbitals), so $\tau_z$, which is diagonal in the orbital basis, is required in the definition of inversion symmetry: $H(\k) \rightarrow \tau^zH(-\k)\tau_z$.  The Hamiltonian is
 \begin{eqnarray}
 H(\k) &=& t_z(2-\cos k_x a -\cos k_y a  + \gamma - \cos k_z a) \tau_z \nonumber \\ &&+ t_x (\sin k_x a) \tau_x + t_y (\sin k_y a) \tau_y.
 \label{Weyl2}
 \end{eqnarray}
 For $-1<\gamma <1$ we have a pair of Weyl nodes at location $\pm\k_0 = (0,0,\pm k_0)$ where $ \cos k_0 =  \gamma$. The low energy excitations are obtained by approximating $H_\pm(\k) \approx H(\pm \k_0 +\q)$ where we assume small $|\q| \ll k_0$. Then, $H_\pm = \sum_a v^\pm_a q_a\tau_a$ where $v^\pm = (t_x,\,t_y,\,t_z\sin k_0)$.

Note, at the 8 TRIM momenta $(n_x,\,n_y,\, n_z)  \frac{\pi}{a}$ where $n_a=0,\,1$, only the first term in the Hamiltonian is active, and if $\gamma>1$ the parity eigenvalues of all the TRIMs are the same and the bands are not inverted.  However, at  $\gamma=0$, the parity eigenvalue of the $\Gamma$ point shows the bands are inverted and it is readily shown that this immediately implies Weyl nodes, i.e. an odd number of inverted parity eigenvalues is a diagnostic of Weyl physics \cite{Turner12,Hughes11a,Wang16c}. At the same time let us compare the Chern numbers $\Omega(k_z)$ of two planes in momentum space $k_z=0$ and $k_z= \frac{\pi}{a} $.  Then the Chern number vanishes at $\Omega(k_z= \frac{\pi}{a})=0$, but $\Omega(k_z=0)=1$. Starting at  $\gamma =  -1$,   Weyl nodes form at the BZ boundaries and move towards each other before annihilating at the zone center at $\gamma = 1$.   As $\gamma \rightarrow 1$, the entire Brillouin zone is filled with unit Chern number along the $k_z$ direction, and a three dimensional version of the integer quantum Hall state is realized \cite{Halperin87}.  Therefore the WSM appears as a transitional state between a trivial insulator and a TI.

When the chemical potential is at $E_F=0$, the Fermi surface consists solely of two points $\pm \k_0$. On increasing $E_F$, two nearly spherical Fermi surfaces appear around the Weyl points and a metal exists. The Fermi surfaces are closed two dimensional manifolds within the Brillouin zone.  One can therefore define the total Berry flux penetrating each, which by general arguments is required to be an integer, and in the present case is quantized to $\pm 1$, which is a particular feature characteristic of a Weyl {\em metal}. When $E_F>E^* = t_z(1-\gamma)$ the Fermi surfaces merge through a Lifshitz transition and  the net Chern number on a Fermi surface vanishes. At this point, one would cease to call this phase a Weyl metal. This discussion highlights the importance of the Weyl nodes being sufficiently close to the chemical potential as compared to $E^*$. Ideally, we would like the chemical potential to be tuned to the location of the Weyl nodes just from stoichiometry, as occurs for ideal graphene.

\subsubsection{Weyl semimetals with broken ${\mathcal P}$ symmetry}
If $\T$ is preserved then inversion symmetry must be broken to realize a WSM. A key difference from the case of the $\T$ broken WSM is that the total number of Weyl points must now be a multiple of four. This occurs since under time reversal a Weyl node at $\k_0$ is converted into a Weyl node at $-\k_0$ with the {\em same} chirality. Since the net chirality must vanish, there must be another pair with the opposite chirality.  Although potentially more complicated with their greater number of nodes, such WSMs may be more experimentally compatible as external fringe fields from a ferromagnet may be problematic for angle-resolved photoemission spectroscopy's (ARPES) momentum resolution.   Additionally, without the complications of magnetism in principle some properties of the system should be simpler under strong magnetic field.

A useful perspective on $\T$ symmetric WSMs is to view them as the transition between a 3D topological insulator and a trivial insulator. When a 3D TI possesses both time reversal symmetry and inversion there is a simple `parity' criterion to diagnose its band topology \cite{Fu07b}.  To achieve a transition between topological and trivial states, Kramers doublets with opposite parities must cross each other at one of the TRIMS.  Since these pairs of states have opposite parity eigenvalues, level repulsion is absent and they can be made to cross by tuning just one parameter.  At the transition point, a four fold degeneracy occurs at the TRIM, leading generically to a Dirac dispersion. Hence the transition between a trivial and topological insulator, in the presence of inversion, proceeds via a Dirac point. However, on breaking inversion, the Kramers doublets at the TRIM points can no longer cross each other while adjusting just a single tuning parameter. How then does the transition proceed? The key observation is that the bands are now nondegenerate away from the TRIMs, and by our previous counting, can be brought into coincidence by tuning the crystal momenta. Tuning an additional parameter to drive the transition involves moving the Weyl nodes towards each other and annihilating them as described by \onlinecite{Murakami08}.  For example when an inversion symmetry breaking staggered potential is applied to the Fu-Kane-Mele model of the 3D TI, the transition between weak and strong TI becomes a WSM phase. It has been recently argued that the band closing transition of a semiconductor lacking inversion symmetry always proceeds through a gapless phase, consisting either of Weyl points or nodal lines \cite{Murakami16a}.

\begin{figure}[t!]
\includegraphics[width=\columnwidth]{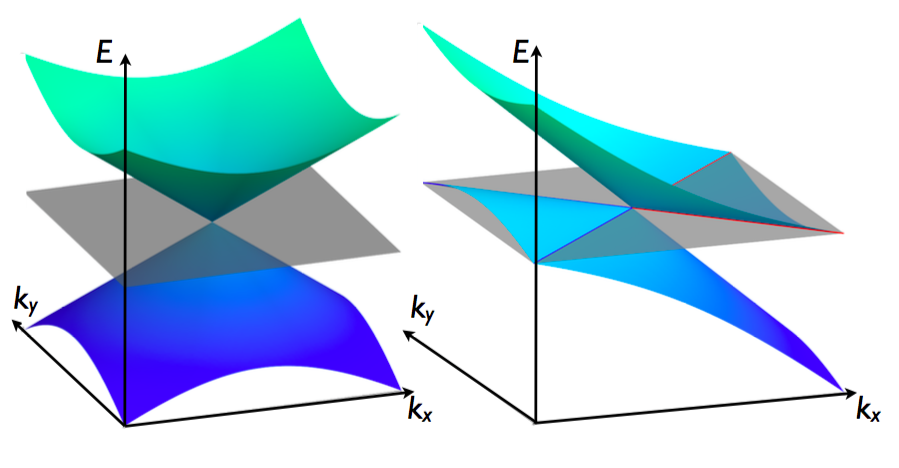}\vspace{-.1in}
\caption{Left: Conventional Type I Weyl point with point-like Fermi surface. Right: Type II Weyl point is the touching point between electron and hole pockets.   Red and blue (highlighted) iso-energy contours (red and blue) denote the Fermi surface coming from electron and hole pockets with chemical potential tuned to the touching point.}
\label{WeylII}
\end{figure}

The velocity parameter ${\mathbf{v}}_0$ in Eq. \ref{WeylHamiltonian} introduces an overall tilt of the Weyl cone. Such a term is forbidden by Lorentz symmetry for the Weyl Hamiltonian in vacuum but it can generically appear in a linearized long wavelength theory near an isolated twofold band crossing in a crystal \cite{Wan11a,Soluyanov15a}.  Small ${\mathbf{v}}_0$ simply induces a crystal field anisotropy into the band dispersion near a Weyl point. However sufficiently large ${\mathbf{v}}_0$  produces a qualitatively new momentum space geometry wherein the constant energy surfaces are open rather than closed  and the resulting electron and hole pocket contact at a point as shown in Fig. \ref{WeylII}.  This new semimetallic phase has been termed a ``Type II" Weyl semimetal (or structured Weyl semimetal \cite{Yong2015a}), in contrast to a ``Type I" semimetal with closed constant energy surfaces.  Although Type I and Type II WSM's cannot be smoothly deformed into each other, they share electronic behavior that derive from the presence of an isolated band contact point in their bulk spectra.  Interestingly the topological character of the Weyl point is still fully controlled by the last term in Eq. \ref{WeylHamiltonian} and persists even for Type II Weyl semimetals.  Thus Type II Weyl semimetals support surface Fermi arcs that terminate on the surface projections of their band contact points which are the signature of the topological nature of the semimetallic state.  In addition to arcs, \onlinecite{mccormick2017minimal} have shown that Type II WSMs support an additional class of surface states they call ``track states".  These are closed contours that are degenerate with the arcs but do not share their topological properties.  They can be generated when the connectivity of Weyl nodes changes as one tunes the parameters in a system with multiple sets of Weyl points.

Type II systems are also expected to support a variant of the chiral anomaly when the magnetic field direction is well aligned with the tilt direction, have a density of states different than the usual form, possess novel quantum oscillations due to momentum space Klein tunneling, and and a modified anomalous Hall conductivity \cite{Soluyanov15a,OBrien16a,Udagawa16a,Zyuzin16a}.    It is proposed that tilting of the cones has a strong effect on the transport Fano factor $F$ (the ratio of shot noise power and current) \cite{Trescher15a}. Type I and Type II nodes of opposite chirality can be merged and annihilate.  And it is likely that some materials can undergo Type I to Type II transitions with doping or under pressure.  Claims for a Type II state have been made recently in MoTe$_2$ \cite{deng2016experimental,Liang16c,Jiang17a,Tamai16a,Huang16a}, WTe$_2$ \cite{Wang16e}, their alloy Mo$_{x}$W$_{1-x}$Te$_2$ \cite{belopolski2016discovery,belopolski2016fermi} and TaIrTe$_4$ \cite{Koepernik16a,Haubold16a,Belopolski16b}, although the evidence in the case of WTe$_2$ is controversial as discussed below and in \cite{bruno2016observation}.

\subsection{Physical consequences of topology}
We have seen that there are topological aspects of WSMs which is most simply stated in terms of them being monopoles of Berry curvature.  Here we will explore some of the consequences of that topology. From experience with topological insulators and quantum Hall states, we are used to two different manifestations of topology. The first is to look for nontrivial surface states, and the second is to study the response to an applied electric and/or magnetic field. We will follow these general guidelines in this case and will not be disappointed.  Indeed WSMs have special surface states called Fermi arcs and an unusual response to electric and magnetic fields due to the above discussed chiral anomaly.
 
\subsubsection{Fermi arc surface states}
\label{FermiArcSS}

\begin{figure}[t!]
\includegraphics[width=0.45\columnwidth]{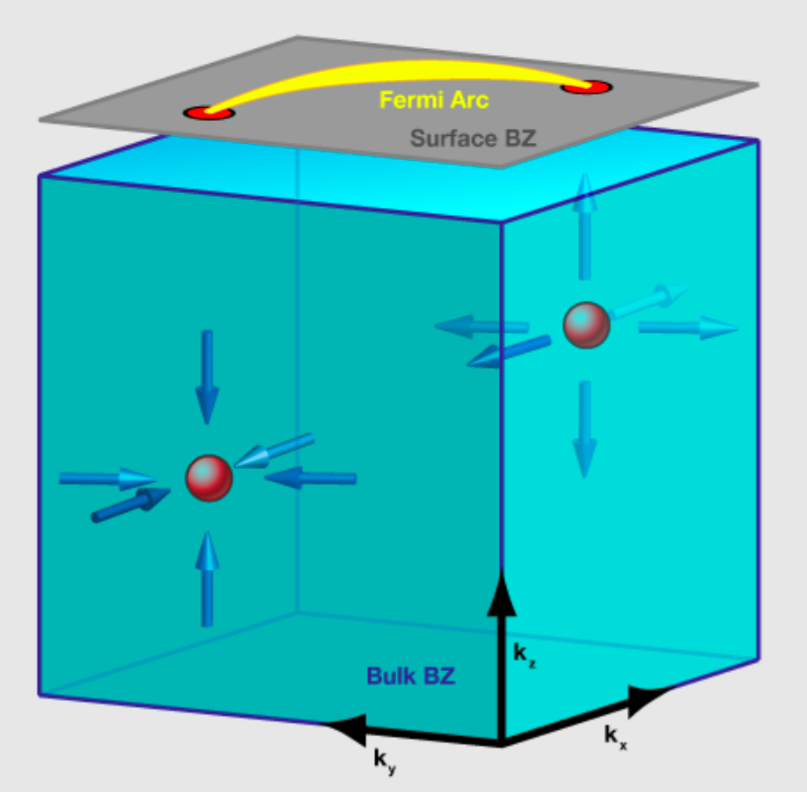}
\includegraphics[width=0.50\columnwidth]{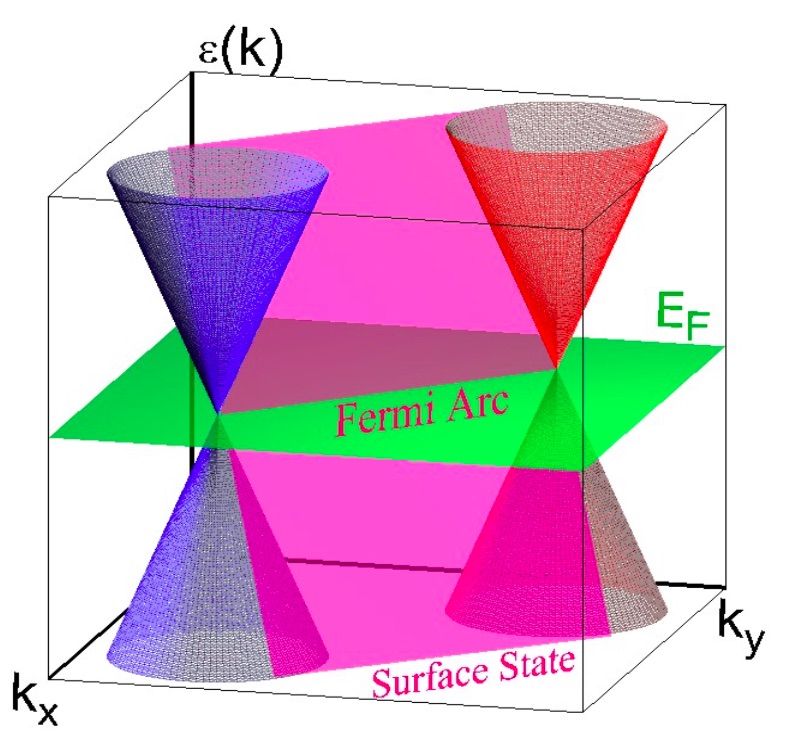}
\includegraphics[width=0.95\columnwidth]{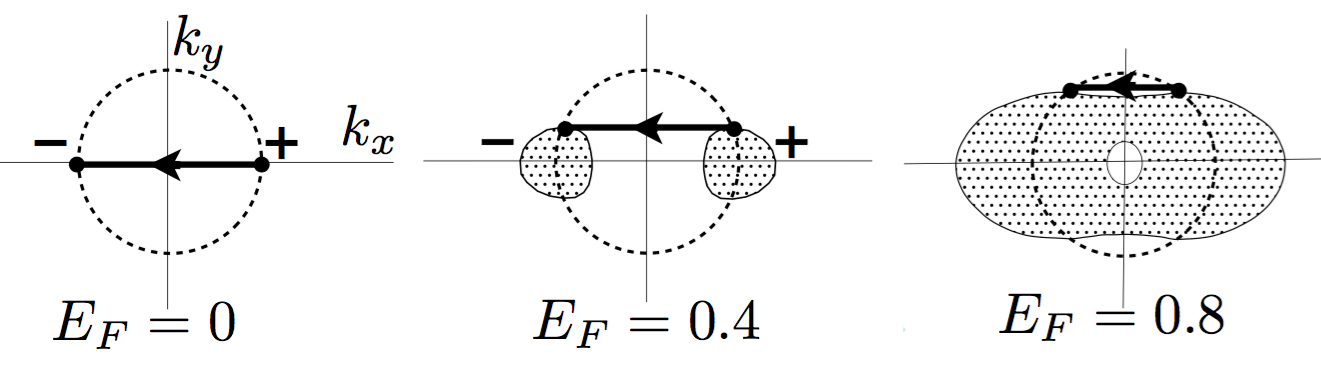}
\vspace{-.1in}
\caption{ (top left) Chern number, Weyl points and surface Fermi arcs.  (top right) Connection of surface states to bulk Weyl points.  (bottom) Evolution of Fermi arc with chemical potential in a particular microscopic model on raising the chemical potential from the nodal energy ($E=0$).  Fermi arcs are tangent to the bulk Fermi surface projections, and may persist even after they merge into a trivial bulk Fermi surface.   From \cite{Balents11a,Wan11a,Haldane14}.  }
\label{fig:FermiArcs}
\vspace{-.1in}
\end{figure}

Surface states are usually associated with band insulators.  Well-defined surface states can exist within the bulk band gap and are typically exponentially localized near the surface. How can we define surface states when the bulk is gapless, as in WSMs? For this we need to further assume translational invariance, so we label surface states by crystal momenta within the 2D surface Brillouin zone (sBZ). Then, we only require that there are regions of the sBZ that are free of bulk states at the same energy. Indeed if we consider the idealized limit of a pair of Weyl nodes at the chemical potential ($E_F=0$) at momenta $\pm\k^*_0$ in the sBZ, one can define surface states at the same energy at all momenta except at the projection of the Weyl points onto the sBZ (Fig. \ref{fig:FermiArcs} top left). At those two points, surface states can leak into the bulk even at $E_F=0$ and are not well defined. If one considers other energies, the momentum region occupied by bulk states grows as shown at the bottom of Fig. \ref{fig:FermiArcs}. The presence of these bulk states allow for surface states that are impossible to realize in both strictly 2D but also on the surface of any three dimensional insulator, where there is a finite energy gap throughout the entire Brillouin Zone.

We can now discuss the nature of the surface states that arise in WSMs, which, at $E_F=0$ are Fermi arcs that terminate at $\pm\k^*_0$. These are a direct consequence of the fact that Weyl nodes are sources and sinks of Berry flux. Hence, if we consider a pair of planes at $k_z=0$ and $k_z= \frac{\pi}{a}$ in the model given in Eq. \ref{Weyl2}. Since they enclose a Weyl node, or Berry monopole, there must be a difference in Berry flux piercing these two planes that accounts for this source. Indeed, in model of Eq. \ref{Weyl2} we see that $k_z=0$ ($k_z=  \frac{\pi}{a}$) has Chern number $C=1$  ($C=0$). In fact, any plane $-k_0<k_z<k_0$ will have Chern number $C=1$, so each of the of 2D Hamiltonians $H_{k_z}(k_x,k_y)$, represents a 2D Chern insulator. If we consider a surface perpendicular to the $x$ direction, we can still label states by $k_z,\,k_y$. The 2D Chern insulators $H_{k_z}$  will each have a chiral edge mode that will disperse as $\epsilon \sim vk_y$ near the Fermi energy as shown also in Fig. \ref{fig:FieldEffect}c.  In the simplest model, $v$ is independent of $k_z$ as long as it is between the Weyl nodes. The Fermi energy $E_F=0$ crosses these states at $k_y=0$ for all $-k_0<k_z<k_0$, leading to a Fermi arc that ends at the Weyl node projections on the sBZ,  and in this particular model, is a straight line.  An interesting alternative continuum derivation of the Fermi arc surface states is contained in the lecture notes of \onlinecite{WittenLectures15}, where boundary conditions are formulated to characterize scattering of Weyl electrons from the boundary of the solid.

On changing the chemical potential away from the Weyl nodes, the Fermi arc is displaced by virtue of its finite velocity. The surface states all disperse in the same direction, and inherit the chiral property of the Chern insulator edge states. At the same time, the bulk Fermi surface now encloses a nonvanishing volume, and their projection onto the sBZ is now a pair of filled discs that encloses the Weyl node momenta. How are the Fermi arc surface states attached to the projection of the bulk Fermi surface?  In the top right of Fig. \ref{fig:FermiArcs}, a plot of both surface (pink) and bulk bands projected to the sBZ is shown, and sections of this dispersion at two energies resemble the two left figures at the bottom of Fig. \ref{fig:FermiArcs}.  In a conventional 2D electron dispersion traversing a band around a closed iso-energetic contour in momentum space returns one to the starting momentum.  In contrast, in a WSM system, on following a closed contour around an end point of the Fermi arc one moves between the valence and conduction bands. A useful analogy is to the Riemann surface generated by a multi-valued function \cite{Fang16}. Therefore, such a band structure, although impossible in 2D, is allowed as a surface state since the surface states can be absorbed by bulk bands on moving away from the Weyl nodes in energy.

In \onlinecite{Haldane14} it was argued that the Fermi arc surface states must be tangent to the bulk Fermi surfaces projected onto the sBZ. This follows from the fact that the surface states must convert seamlessly into the bulk states as they approach their termination points.  Putting this differently, the evanescent depth of the surface state wavefunction grows until at the point of projection onto the bulk states, the surface states merge with the bulk states.  They should inherit the velocity of the bulk states, which implies they must be attached tangentially to the bulk Fermi surface projections as shown in Fig. \ref{fig:FermiArcs}.  Surface states calculated in the model of \cite{Haldane14}  also show that Fermi arcs could continue to exist above the Lifshitz transition, when the Fermi surfaces surrounding the two Weyl points merge.  However, Fermi arc surface states bridging disconnected Fermi surfaces imply that they carry nontrivial Chern number. This suggests an experimental diagonstic to determine the existence of a Weyl metal.  Consider a closed $k$-space curve at constant energy in the surface Brillouin zone, and determine the electronic states intersected by it. If an odd number of surface states are encountered, and no bulk states, then one is required to have a nontrivial Chern number on the bulk Fermi surface enclosed by the curve \cite{Lv15a,Belopolski16a} and hence we can define this as a ``Weyl metal". The surface states intersected need to be counted in a sign sensitive fashion, with $\pm1$ depending on whether their velocity is along or opposite to the direction of traversal of the contour.  This quantity is related to the total Chern number of Fermi surfaces enclosed by the contour.

A useful alternate viewpoint on Fermi arc surface states is to imagine growing the three dimensional bulk beginning with a thin slab. Initially, the opposite surfaces are close to one another, and viewed as a 2D system, this should have a conventional closed Fermi surface. As the separation between the opposite faces increases, opposite halves of the Fermi surface migrate to opposite surfaces, leading to the Fermi arcs. This is analogous to obtaining a single Dirac node on the surface of a topological insulator by starting with a pair of Dirac nodes in 2D and gradually separating them to opposite surfaces \cite{Wu13a}. One important distinction for the Fermi arc case is that the surface states must become extended into the bulk at the termination point of the Fermi arc. One can utilize this viewpoint to construct models of WSMs with any given surface Fermi arc dispersion as in Ref. \cite{Hosur12b}. A more mathematical perspective on Fermi arc surface states was described in Ref. \cite{Mathai2017}. 

The most direct observation of Fermi arc surface states has been achieved through ARPES and more recently scanning tunneling microscopy (STM) studies on the WSM candidate TaAs, which are reviewed below. Another standard probe of Fermi surfaces is quantum oscillations, which can also be used to study Fermi arc surface states.  However as the corresponding theory involves both Fermi arcs and chiral Landau levels, stitched together in a  consistent fashion,  we will discuss this below in a separate Section \ref{Sec:AnomalyArcs}.

\subsubsection{The chiral anomaly}
\label{ChiralAnomaly}

In a WSM with a pair of Weyl nodes of opposite chirality, the number of electrons in the vicinity of each is modified in the presence of electric and magnetic fields via the equation
\begin{equation}
\frac{dn^{3D} _{R/L}}{dt} = \pm \frac{e^2}{h^2}{\bf E\cdot B}
\label{3DChiralAnomalyN}
\end{equation}
where we have inserted the superscript to remind us that we are dealing with 3D Weyl fermions. Therefore, even in the presence of spatially uniform fields, which may be oriented in an arbitrary direction relative to the separation of the Weyl nodes, the density of electrons at an individual node is not conserved. In particular, this immediately tells us that a single Weyl node, or any set with an unbalanced chirality, is problematic, since it will lead to non-conservation of electric charge. However, if the chirality is balanced, as happens for any lattice realization, the opposite Weyl nodes act as sources and sinks of electrons, leading to nodal (or valley) polarizations, while preserving the total charge. To give some intuition for how this arises, let us first consider the one dimensional analog (the chiral anomaly appears in any odd spatial dimension). The number density at a pair of one dimensional Weyl nodes will correspondingly obey:
\begin{equation}
\frac{d(n^{1D} _{R/L})}{dt} = \pm \frac{e}{h}{E}
\label{1DChiralAnomalyN}
\end{equation}
This is readily derived from the semiclassical equation of motion which will accelerate electrons along the field $\dot{k}=eE/\hbar$. When this change of momentum equals the spacing between momentum states $2\pi/L$, an extra electron is added (removed) from the right moving (left moving) Weyl point leading to the equation above. The key ingredient of course is the fact that in a condensed matter context the left and right Weyl points are not really distinct entities in a lattice model, rather they are connected beneath (and above) the Fermi level, so one cannot clearly separate electrons associated with one group or the other, at arbitrarily high energies. In fact the one dimensional chiral anomaly is an essential ingredient in generating electrical conductivity. If we associate a scattering mechanism that relaxes any density imbalance between the two nodes with a rate $1/\tau_a$, then we can write the following modified rate equation:
\begin{equation}
\frac{d(n^{1D} _{R} - n^{1D} _{L} )}{dt} = 2 \frac{e}{h}{E}   - \frac{n^{1D} _{R} - n^{1D} _{L} }{\tau_a}
\label{1DChiral}
\end{equation}
In steady state this leads to a current: ${\mathcal I}  = 2e^2 E l/h$, where the scattering length $l = v_F\tau_a$. A very similar calculation can be applied to metals in 2D or 3D,  where the different points of the Fermi surface can be regarded as one dimensional chiral fermions propagating along the local Fermi velocity, and the shift of the Fermi surface in an electric field being just the manifestation of the 1D chiral anomaly. However, the three dimensional chiral anomaly is rather distinct and requires the application of {\em both} electric and magnetic fields.
\begin{figure}[t!]
\includegraphics[width=\columnwidth]{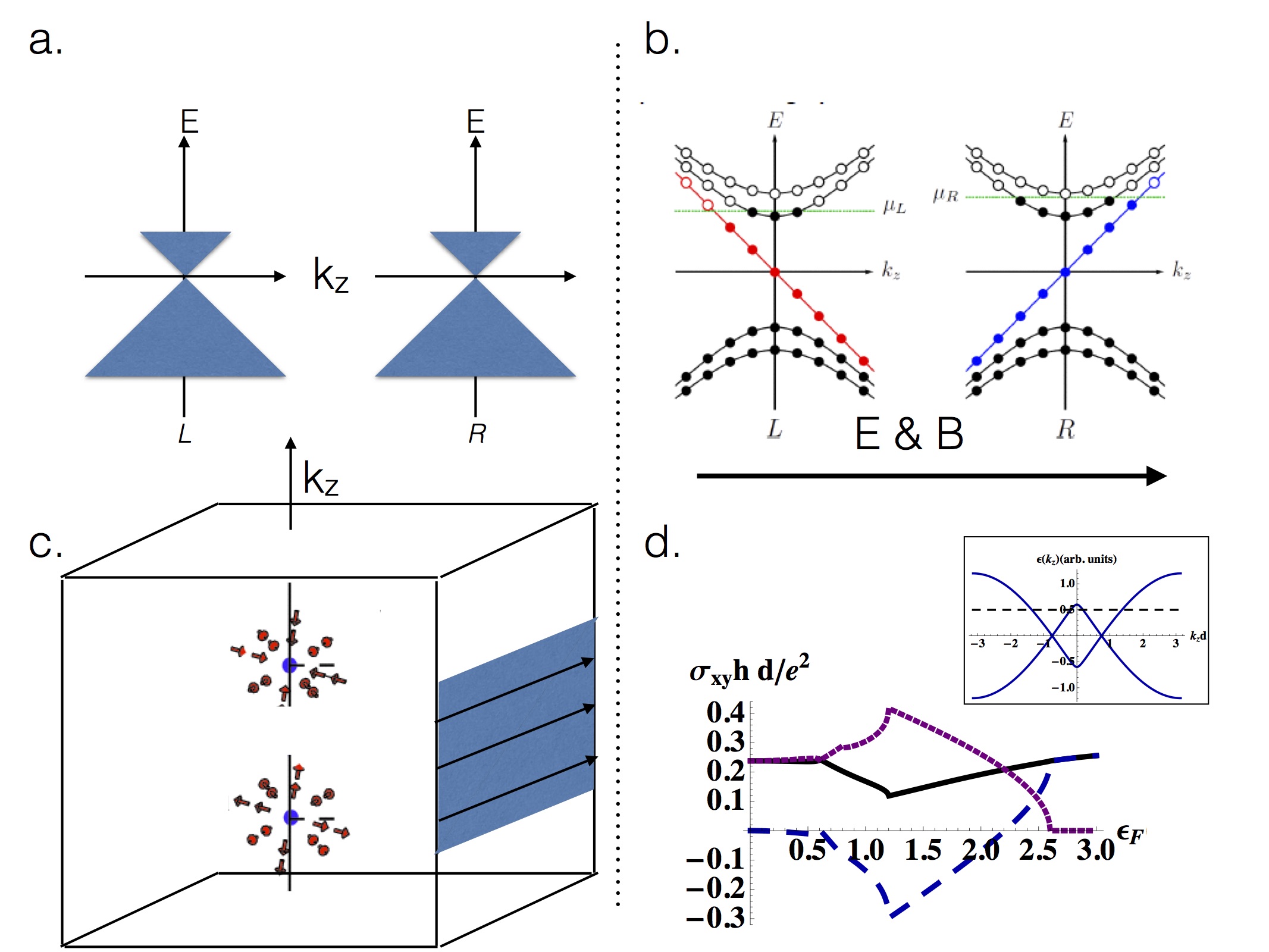}\vspace{-.1in}
\caption{ (a) Opposite Weyl nodes in a T-breaking WSM in the absence of fields. (b) Spectrum in a magnetic field along the `z' axis displaying Landau levels that disperse along the field. The zeroth Landau levels are chiral. In addition, an electric field along `z' generates valley imbalance. (c) Anomalous Hall effect from chiral Fermi arc surface states whose magnitude is determined by the Weyl node separation in momentum space, when the chemical potential is at the Weyl nodes. (d) On moving the chemical potential away from the Weyl nodes, the anomalous Hall conductivity changes but only marginally in the model considered in \cite{Burkov14} up to the chemical potential when the Fermi surfaces enclosing the two Weyl points merge. }
\label{fig:FieldEffect}
\vspace{-.1in}
\end{figure}
A simple way to understand the 3D chiral anomaly is to first consider the effect of the magnetic field in the clean system, which leads to Landau levels that disperse only along the field direction. The zeroth Landau can be shown to be chiral, i.e. it propagates only along or opposite to the field direction, with reversed velocities at the two opposite chirality Weyl nodes. Consider a single isotropic Weyl node with chirality $C =\pm1$, that is minimally coupled to an external magnetic field ${\bf B} = B \hat{z}$.
\begin{equation}
H_C = C v_F({\bf p} -e{\bf A})\cdot {\bf \sigma}
\end{equation}
Labeling the conserved momentum along the field $p_B = {\bf p\cdot B}$ , we can set this to zero, where we recover the problem of a single 2D Dirac node in a field, which is known to have the spectrum $\epsilon_n = \frac{v_F \hbar}{l_B} {\rm sgn} (n)\sqrt{|n|}$ . In particular the zeroth Landau level is at zero energy and is polarized along the $\hat{B}$ direction with eigenvalue of ${\bf \sigma}\cdot \hat{B}$ being $\sigma_B=+1$. This corresponds, for the case of graphene, to the sublattice-valley polarization of the zeroth Landau level. However, in the present context it has the following remarkable consequence. Reintroducing the dispersion along the field 
\begin{equation}
H^{{\rm n=0}}_C =C v_F p_B\sigma_B,
\label{Eq:3DChiral}
\end{equation}
we see that for polarizing the spin $\sigma_B=+1$ implies a one way propagation of electrons along the magnetic field for $C=+1$ and the opposite propagation at the opposite Weyl node ($C=-1$). The $n \neq 0$ Landau levels in contrast display a conventional dispersion as shown in Fig. \ref{fig:FieldEffect}b. 
Therefore we can relate the problem of a WSM in 3D in a magnetic field, to an effectively one dimensional problem where the electrons propagate purely along the magnetic field lines, forming chiral one dimensional channels.  Thus, we can utilize the 1D chiral anomaly formula Eq. \ref{1DChiralAnomalyN} with the electric field applied along the magnetic field $E = {\bf E}\cdot \hat{B}$.  Finally we convert the result into a three dimensional  density $n^{3D}_{R/L} = \frac{1}{\mathcal A} n^{1D}_{R/L}$, utilizing the fact that the one dimensional channels have a cross sectional area occupied by a magnetic flux quantum (${\mathcal A} = \phi_0/B$).  This gives us the result advertised at the beginning of this section Eq. \ref{3DChiralAnomalyN}, and previously identified in Ref. \cite{Nielsen83a}. Before we turn to experimental consequences of the chiral anomaly in the solid state context, let us discuss a closely related effect - the Chiral Magnetic Effect (CME).  

Consider a WSM which has an effective chemical potential difference ($\Delta \epsilon$) between the two Weyl nodes and a magnetic field applied in the direction connecting them \cite{Zyuzin12a,Chernodub14a}.  A naive application of the above arguments would suggest that there is a current along the magnetic field arising from the unequal occupation of left and right moving chiral modes, giving a current of 
\begin{equation} 
j_c = \frac{e^2}{h^2}B\Delta \epsilon.
\label{chiralcurrent} 
\end{equation}
\noindent If such a current exists, it cannot be an equilbrium dc transport current as no voltage is applied. Moreover it also cannot be a magnetization current ($j_{\rm mag}=\nabla \times M$) since this would imply the transverse components $M_\perp \propto A_\perp$ \cite{Levitov85}. The latter violates gauge invariance since it is a physical quantity that directly depends on the vector potential. Indeed in any equilibrium situation the current must vanish when all contributions from filled electronic states are taken into account \cite{Kohn64,VafizehFranz}. However, in a {\em nonequilibrium} setting the current can be nonvanishing.  For instance, if an electric field oscillates at a frequency $\omega$ that is faster than the internode relaxation rate, then a chemical potential difference between nodes can be induced and an oscillating chiral current can occur \cite{Burkov15a,Ma15a,Zhong16,Zhou13}.  Note that there is a similar, but ultimately different effect that can occur in chiral metals in response to time varying magnetic fields that has been called the gyrotropic magnetic effect \cite{Ma15a,Zhong16}.   It is governed by the intrinsic magnetic moment of the Bloch states on the Fermi surface and is distinct from the CME.

A related nonequilibrium situation can occur when a density difference of electrons in the two opposite Weyl nodes is created by the chiral anomaly that can pump charge between nodes in the presence of parallel electric and magnetic fields.  It leads to different effective chemical potentials for the Weyl nodes that can lead to observable consequences of the CME mentioned above  \cite{SonSpivak,Aji12a,Burkov15a}. The density difference is determined by the balance between the chiral pumping and the rate of inter-Weyl node scattering ($1/\tau_a$), which results in a finite steady state density difference between nodes proportional to $ {\bf E\cdot B}$.   The natural assumption here is that intranode scattering is much faster than the internode scattering.  This gives an effective chiral chemical potential difference between nodes, that when combined with the CME (Eq. \ref{chiralcurrent}), gives a chiral current $j_c \propto B  {\bf E\cdot B}  \tau_a$.   This contribution to the dc effect can be seen to arise from what is effectively two successive uses of the ${\bf E\cdot B}$ form, the first that establishes the chiral chemical potential difference between nodes and the second that gives a current.  The complete expression is given below but this simplified treatment exemplifies some of the key features.  The magneto-conductivity tensor is quadratic in magnetic field $\delta \sigma_{ab} \propto B_a B_b$ e.g. it has a quadratic dependence on magnetic field that is anisotropic and maximal for transport along the field direction.  Thus, along the magnetic field direction the conductivity is modified from its zero field value as \cite{Burkov15a,SonSpivak} 

\begin{equation}
\sigma(B)=\sigma_0 + \frac{e^4B^2\tau_a}{4\pi^4g(\epsilon_F)}.
\end{equation}
Therefore the magnetoconductivity is predicted to positive (and magnetoresistance negative).  It is remarkable that this expression can be arrived at through both the quantum limit calculation above and one done in the framework of semiclassical kinetics \cite{SonSpivak} e.g.  Landau levels are not required.  Although such an effect has been proposed to be used as a smoking gun signature of a WSM, as discussed below there are dominating experimental artifacts that may obscure such a dependence.  What is particularly diagnostic of the effect is a strong dependence on intervalley scattering $\tau_a$. A strongly disordered WSM with mixing between opposite Weyl nodes should not exhibit any transport signature of an isolated Weyl node - indeed in the limit of small $\tau_a$ the magnetoresistance is also small. However, as the intervalley scattering time increases and the Weyl nature becomes more pronounced, the chiral contribution will dominate and may lead to large negative magnetoresistance. In addition to the experimental issues, one potential intrinsic complication is that the chiral current can also be relaxed by reaching the surface where it can be converted into electrons at the opposite Weyl node by sliding along the Fermi arc surface state. Indeed as discussed by \onlinecite{Ominato16} this is the origin of chiral current relaxation in the absence of direct internode scattering, but leads to unusual scaling with system size. However, the presence of any bulk scattering mechanism between opposite Weyl nodes will eventually dominate in the large volume limit since the Fermi arc relaxation mechanism is a surface effect. Analogous phenomena in thermoelectric transport have also been predicted \cite{Lundgren14a,  Spivak16, Lucas16}  and explored experimentally \cite{Hirschberger16}.

\subsubsection{Anomalous Hall effect}
The simplest manifestation of Weyl physics arises from the anomalous Hall effect \cite{Yang11a}. Of course this requires explicit $\mathcal{T}$ breaking. However, we note that it may also be excluded even in magnetic WSMs where, for example, cubic symmetry is preserved.    Consider the simplest example of a pair of Weyl nodes in a magnetic system, separated along the $z$ direction by a crystal wavevector ${\bf q} = 2k_0{\hat z}$, where we have directed the vector from positive to negative chirality Weyl nodes.   As discussed in Sec. \ref{FermiArcSS}, in such a situation one can view each $k_x,k_y$ plane for $-k_0<k_z<k_0$ as a 2D Chern insulator.  Each 2D Chern insulators  will have a chiral edge mode near the Fermi energy each contributing a Hall conductance $\frac{e^2}{h}$ as shown in Fig. \ref{fig:FieldEffect}c.  Therefore the anomalous Hall effect is particularly simple when the chemical potential is at the Weyl nodes and for generic positions of the pair of Weyl nodes one has:

\begin{equation}
\sigma_{ab} = \epsilon_{abc} \frac{e^2}{2\pi h}q^c
\label{Eq:sigma}
\end{equation}
assuming  ${\bf q} = 2k_0{\hat z}$, this reduces to $\sigma_{xy} = \frac{e^2}{2\pi h}2k_0$. 
Note, $\bf q$ is only defined modulo reciprocal lattice vectors ${\bf G}$. This is physically related to the property that the Hall conductance calculated by this formula are quantized modulo Hall conductances arising from filled bands which may lead to three dimensional quantum Hall states \cite{Halperin87} which are also characterized by a reciprocal lattice vector ${\bf G}$.  On moving the chemical potential away in energy, it was argued that the change of anomalous Hall conductance can be small \cite{Burkov14} until the Fermi surfaces surrounding the opposite Weyl points merge (Fig. \ref{fig:FieldEffect}d).  In a crystal with cubic symmetry, where $\mathcal T$ breaking Weyl nodes may appear as proposed in the pyrochlore iridates \cite{Wan11a,Krempa12,WangMillis16}, symmetry enforces vanishing of the anomalous Hall effect, due to the absence of a preferred axis. However, on applying a uniaxial strain that lowers the symmetry, an anomalous Hall signal should appear, proportional to the degree of symmetry breaking for small strain. This was proposed as a probe of cubic magnetic Weyl semimetals in \cite{Yang11a}.   Note that, although newly appreciated, this may be a very common mechanism for generating an anomalous Hall effect.  Two Fermi pockets are predicted to surround isolated Weyl points in bcc iron and are believed to give a major contribution to its anomalous Hall effect \cite{Gosalbez15a}.  It has also been appreciated that competing interactions in these materials can stabilize other interesting magnetic states that generically support AHE \cite{Goswami17a}.

\subsubsection{Axion electrodynamics of Weyl semimetals}

Both CME and the anomalous Hall effect in WSMs can be represented compactly by the addition of the so-called axion term to the electromagnetic Lagrangian \cite{Zyuzin12a,Son12a,Goswami13a,Vazifeh13a,zyuzin2012weyl,grushin2012consequences}.  The action is
\begin{equation}\label{s1}     
S_\theta=  \frac{1}{ 2 \pi} \frac{ e^2}{ h }  \int dt d {\bf r} \theta({\bf r},t) {\bf E}\cdot {\bf B}.
\end{equation}
This is similar to the formalism used in describing the electromagnetic response of topological insulators \cite{Wu16a,Qi08a,Essin09a}.  For TIs the effect is felt only when there the spatial or temporal gradients of $\theta$ are are finite e.g. at surfaces where in the presence of $\mathcal{T}$ breaking field it generates a half quantum Hall effect from a single surface.  In the present case the $\theta$ term has time and bulk position dependence and in its simplest form it is
\begin{equation}\label{s1}     
\theta({\bf r} ,t) =2({\bf k}_0 \cdot {\bf r}  -b_0 t)
\end{equation}
where $\k_0 $ is the position of the Weyl nodes and $2b_0$ is proportional to the chemical potential difference between Weyl nodes.  Unlike related terms in topological insulators, it leads to observable effects in the bulk of the material.   Minimizing the action in the standard fashion (See SI of \onlinecite{Wu16a} for example) leads to the following equations of motion for the charge density and current.   They are

\begin{eqnarray}
\rho =  \frac{1}{ 2 \pi}  \frac{e^2}{h}  2\k_0   \cdot {\bf B} \label{r1}, \\
{\bf J} =  \frac{1}{ 2 \pi}  \frac{e^2}{  h}( 2\k_0   \times  {\bf E} - 2b_0   {\bf B}  ). \label{r2}
\end{eqnarray}
Eq. \ref{r1} and the first term in Eq. \ref{r2} represent the anomalous Hall effect that is expected  to occur in a Weyl semimetals with broken $ \mathcal{T}$ and is equivalent to Eq. \ref{Eq:sigma}.  The second term in Eq.\ (\ref{r2}) describes the CME discussed above whereby a current is proportional to the applied magnetic field $\bf{B}$ and is equivalent to  Eq. \ref{chiralcurrent} where the energy difference $\Delta \epsilon = 2\hbar b_0$. The existence of this CME term may present a conundrum as discussed above.   Such a term proportional to magnetic field with no voltage applied cannot represent an equilibrium current.  Generally Eq. \ref{r2} needs to be supplemented by an equation describing the relaxation of the chiral charge \cite{burkov2017weyl}.  However, Eq. \ref{r2}  remains valid for dynamics fast compared to the internode scattering time, while the CME vanishes in the dc limit in equlibrium.  There are a host of other -- particularly optical -- effects that have been predicted on the basis of this physics \cite{ZhouPlasma15,Kargarian15a,Hosur14a,Cortijo16a}.   For a detailed discussion of the CME in both condensed matter and particle physics contexts please see \onlinecite{gynther2011holographic,burkov2017weyl}.

\subsubsection{Interplay between chiral anomaly and surface Fermi arcs}
\label{Sec:AnomalyArcs}
\begin{figure}[t!]
\includegraphics[width=\columnwidth]{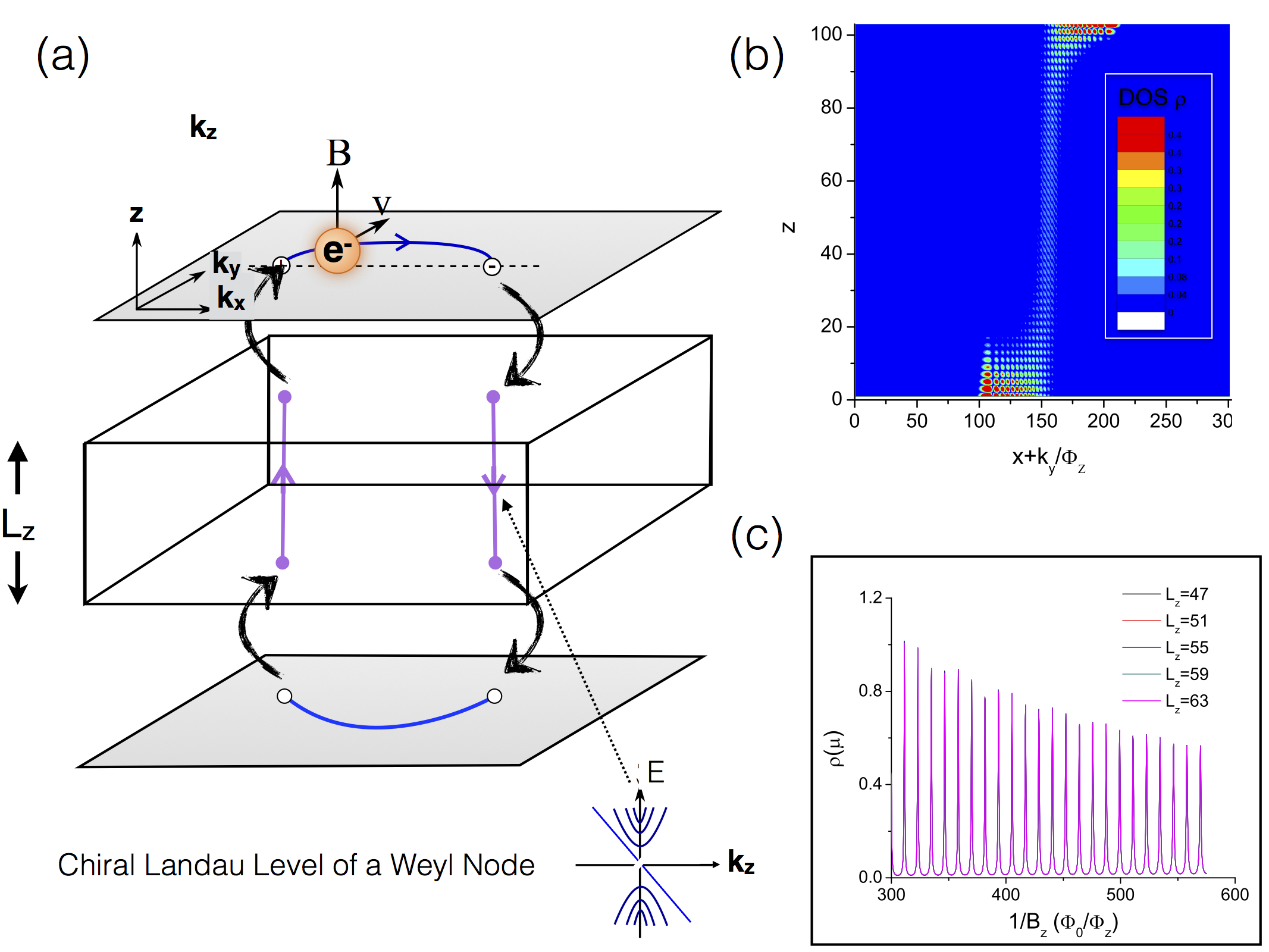}\vspace{-.1in}
\caption{ Weyl semimetal slab in an applied magnetic field. (a)  Weyl cyclotron orbits depicted in hybrid real space ($z$) and momentum space ($k_x,k_y$).  Electrons slide along the surface Fermi arcs and are absorbed by the chiral bulk Landau level which propagates them to the opposite surface from \cite{Potter2014}. Numerically calculated  (b) wave function of Weyl orbits, showing their hybrid surface-bulk character and (c) quantum oscillations in density of states from Weyl orbits \cite{Zhang16}. }
\label{fig:QO}
\vspace{-.1in}
\end{figure}

 Quantum oscillation experiments, which involve measuring the variation of a physical property such as magnetization or conductivity as a function of applied magnetic field, is a sensitive probe of Fermi surface geometry.  It is natural to expect that the unusual Fermi arc surface states of Weyl semimetals will display nontrivial quantum oscillations signatures. Indeed this expectation is consistent with recent theoretical studies  \cite{Potter2014,Zhang16,Gorbar16} described below, which predict a semiclassical trajectory, ``Weyl orbits",  that weave together surfaces and bulk states.
 
 Consider the simplest $\mathcal T$ broken Weyl semimetal with a pair of Weyl nodes displaced by $\k_0$ along the $k_x$ direction. Applying a magnetic field along the $z$ direction as shown in Fig. \ref{fig:QO}, leads to  Lorentz force acting on the surface electrons that makes them slide along the Fermi arc.  For a conventional Fermi surface, the cyclotron motion leads to a closed path which can then be quantized leading to oscillations. However, in the case of Fermi arcs, the electron at the tip of the arc has nowhere to go on the surface. Instead, we would expect it to tunnel into the bulk. Indeed, studying the previously obtained bulk spectrum in the presence of a magnetic field, the chiral Landau levels of Eq. \ref{Eq:3DChiral} are precisely the bulk modes that can absorb the electron and convey it to the bottom surface where it proceeds to rotate along the opposite Fermi arc and returns to the top surface along the oppositely propagating chiral Landau level of the other Weyl node. These trajectories are the ``Weyl orbits" and to describe their properties let us assume for simplicity that we are in the quantum limit for the bulk, so there is no contribution from bulk Fermi surfaces (which are anyway easily distinguished from surface oscillations which only depend on the perpendicular component of field). A numerically calculated quantum oscillation trace is shown in Fig.  \ref{fig:QO}c, which results from the surface-bulk hybrid orbit \cite{Zhang16}. The peaks in the quantum oscillation occur at magnetic fields $B_n$ set by the equation: 
 \begin{equation}
 \frac1{B_n}  = \frac{e}{S_k} \left [ 2\pi (n +\gamma) \hat{z}\cdot \hat{B} -L_z(\k_0 \cdot \hat{B}+\frac{2\mu}{v_\parallel} )\right ]
 \label{ZhangEquation}
 \end{equation}   
 where $S_k$ is the area enclosed between the top and bottom surface Fermi arcs, and $\mu$ is the chemical potential measured from the Weyl nodes. This is further simplified if the field is parallel to the $z$ axis, when we can write: $B_n^{-1} =  \frac{2\pi e}{S_k} (n+\gamma) +\Phi[L_z]$, where the frequency of quantum oscillation is set by $S_k$, with a thickness dependent phase offset $\Phi = - L_z \frac{2\mu}{v_z} $. This is simply the phase accumulated on traversing the bulk - indeed this expression can be obtained most simply by considering the phase accumulated by the semiclassical trajectories described above and quantizing it using the Bohr-Sommerfeld condition. While a simple estimate can be made using energy-time quantization as in \cite{Potter2014}, the complete expression above is obtained from a phase space quantization \cite{Zhang16}.
 
The unusual nature of these Weyl orbits is the fact that they behave both like surface states (the oscillations depend on the vertical component of the magnetic field) while at the same time the thickness dependence appears in the phase offset. Note, thus far we have assumed a perfectly clean system - in the presence of impurities, scattering in the bulk will lead to an exponential suppression of the quantum oscillation signal which will also be thickness dependent. Different scattering processes can degrade the signal - in the weak field limit scattering between the chiral Landau level mode and other nonchiral modes (as in Fig. \ref{fig:FieldEffect}b) can arise even from forward scattering. In the quantum limit, backscattering requires scattering by the large wavevector $\k_0$, but even in the absence of such scattering, differing disorder induced path lengths can lead to interference and contribute to suppression of the quantum oscillation signal \cite{Zhang16}. Experimental investigation  of Fermi arc quantum oscillations are described in Section \ref{QMTransport}.
 
\begin{figure}[t!]
\includegraphics[width=\columnwidth]{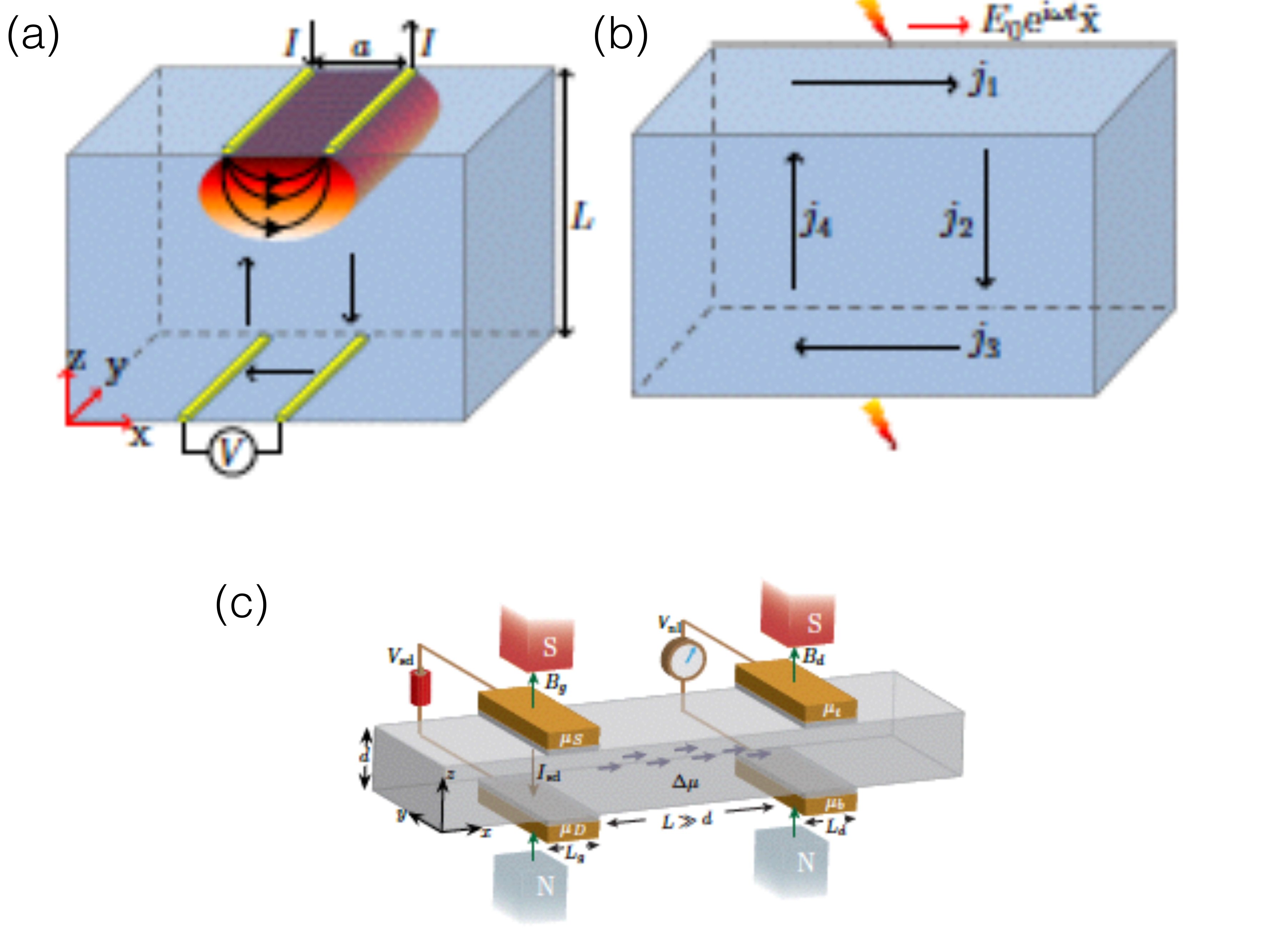}\vspace{-.1in}
\caption{Theoretical proposals of nonlocal transport in WSMs.  Weyl cyclotron orbits lead to (a) a voltage difference on the lower pair of contacts when current is injected between contacts on the top surface and (b) resonances in transmission of electromagnetic waves at frequencies controlled by the magnetic field from Ref. \cite{Baum15} (c) An alternate proposal for nonlocal transport utilizing the choral anomaly. A source-drain current $I_{\text{sd}}$ is injected into a WSM slab of thickness $d$ via tunneling contacts of thickness $L_g$.  In the presence of a local generation magnetic field $B_g$, a valley imbalance $\Delta\mu$ is created via the chiral anomaly and diffuses a distance $L\gg d$ away. If a `detection' field $B_d$ is applied, the valley imbalance can be converted into a potential difference $V_{\text{nl}}$ between top and bottom contacts of size $L_d$.  From Ref. \cite{Parameswaran14a}. }
\label{fig:nonlocal}
\vspace{-.1in}
\end{figure}

 The sensitivity of quantum oscillations to disorder stems from the requirement for coherent electron motion over the periodic trajectory. On the other hand  forward scattering between Landau level modes does not affect the current which continues to propagate along the same direction. This led to the prediction of effects \cite{Baum15} that depend on the current pattern related to the semiclassical orbits, which can show robust and unusual signatures as shown in Fig. \ref{fig:nonlocal}a,b. For example, resonant transmission of electromagnetic waves through the slab at frequencies determined by the intersurface cyclotron orbits are predicted which are estimated to be in the microwave or THz range for experimentally relevant parameters. Nonlocal dc transport that relies on the intersurface cyclotron orbits was also proposed in the same work. A different approach to generating nonlocal voltage based on the chiral anomaly appeared in  \cite{Parameswaran14a} (see Fig. \ref{fig:nonlocal}c) and a preliminary experimental report claiming to observe this effect has appeared in \cite{CZhang15} on the DSM system Cd$_3$As$_2$.
 
\subsection{Disorder effects on Weyl semimetals} 
In a WSM where all Weyl nodes are exactly at the chemical potential, and other bands are removed in energy, the density of states vanishes. An interesting question is the evolution of the density of states as the system is disordered. In the analogous problem of disordered graphene in 2D, a finite density of states immediately appears even at weak disorder \cite{Neto09a}. Analogous calculations for 3D semimetals show a vanishing density of states persisting up to a finite value of disorder strength \cite{Fradkin86,Goswami11a,Hosur12a,Pixley16a} beyond which a metallic state is expected. The critical properties of this interesting semimetal-metal transition have been discussed in several recent works \cite{Altland2016a,Kobayashi2014a,Syzranov15,Syzranov16,Sbierski15,Pixley15,Shapourian16,Bera16,Louvet16}  which used different  numerical and renormalization group based analytical approaches. For example, a relatively simple limit to study the problem was introduced in \cite{Louvet16} by mapping the problem to a Gross-Neveu-Yukawa theory in 4-$\epsilon$ dimensions which can be studied by conventional RG techniques and yields a continuous transition between a disordered semimetal and a diffusive metal.   On the other hand rare events that are not accounted for in perturbative RG approaches can  have a singular effect  by inducing  a small density of states even at weak disorder \cite{Nandkishore14,Pixley16}  that ultimately rounds off the transition at the longest scales.  This implies there is no sharp distinction between the semimetal and metallic regions, nevertheless there is a wide range of length scales where the system is controlled by the previously discussed critical point, before being eventually rounded off at the largest scales \cite{Pixley2017a} by flow to the metallic fixed point.   These issues have reviewed recently in Ref. \cite{Syzranov17a}.  It has been proposed that optical conductivity may be a useful probe in studying the critical properties of the disorder-driven phase transition in Weyl semimetals \cite{roy2016universal}.

\section{Dirac semimetals in three dimensions}
\label{DiracSection}

As discussed above, Weyl points can occur in three dimensional materials only when either time reversal or inversion symmetries are broken.  When inversion symmetry is present a Weyl node at ${\mathbf{k}}$ must be accompanied by a partner node at $- {\mathbf{k}}$ at the same energy that carries the {\it opposite} topological charge \cite{Wan11a,Burkov11a}. Conversely, time reversal symmetry requires that nodes at these momenta are time reversed partners which carry the {\it same} topological charge \cite{Halasz12a}.  Since the net topological charge enclosed within Brillouin is zero, this latter situation further requires the existence of two additional compensating partner Weyl nodes \cite{Murakami07a,Halasz12a}. The presence of both inversion and time reversal symmetries excludes the possibility of a two fold degeneracy at a Weyl point in the spectrum.

Nevertheless when both symmetries are present energetically degenerate Weyl nodes carrying opposite charges can be stabilized at the {\it same} crystal momentum.  This produces a composite point singularity hosting a fourfold degeneracy. This degeneracy is not topologically protected since its net Chern number is zero and residual momentum-conserving terms in the Hamiltonian projected into the degenerate subspace can potentially mix these states and gap the electronic spectrum.  However in special situations this mixing can be forbidden by space group symmetries in which case the nodes remain intact as \textit{symmetry-protected} degeneracies. This is of fundamental interest since the stable merger of two low energy Weyl nodes provides a solid state realization of the 3+1 dimensional Dirac vacuum and materials that support this degeneracy are called {\it Dirac semimetals} (DSMs).  This can occur at a quantum critical point where a three dimensional Hamiltonian is parametrically fine tuned to the bulk gap closure that separates conventional and $\mathbb{Z}_2$ topological insulating states \cite{Murakami07a,Murakami2007b}.  These topological semimetals are sometimes described as ``three dimensional graphenes" although this moniker is inappropriate because unlike the situation in graphene \cite{KaneMele2005} the Fermi surface point of a DSM is a symmetry protected degeneracy in the presence of (possibly strong) spin-orbit interactions.  In graphene this degeneracy is removed by spin-orbit coupling and its gapped phase is the prototype quantum spin Hall insulator \cite{KaneMele2005}.

One can demonstrate how this arises in a simple model \cite{Fu07a} which analyzes the spectrum of a spin-orbit coupled tight binding bands on the diamond lattice.    In this model one isotropic ``$s$'' orbital with two spin polarizations are assigned to each of two sites in the primitive cell.  The Hamiltonian for this system is

\begin{eqnarray}\label{modelHam}
{\cal H} &=& t \sum_{\langle ij \rangle, s}  \, c^\dag_{i,s}  c_{j,s}  \nonumber\\
&+& i \, \frac{ \lambda_{\rm so}}{a^2} \sum_{\langle \langle ij \rangle \rangle;s,s'} \, c^\dag_{i,s} \left( \hat{\boldsymbol\sigma}  \cdot {\mathbf{d}}^{(1)}_{ij} \times {\mathbf{d}}^{(2)}_{ij} \right) c_{j,s'}
\end{eqnarray}

\noindent with a scalar coupling strength $t$ between nearest neighbor sites $\langle i,j \rangle$ and spin-orbit coupling strength $\lambda_{\rm so}$ between second neighbor sites $\langle \langle i,j \rangle \rangle$ bridged by successive nearest neighbor hops along the bond vectors ${\mathbf{d}}^{(1)}_{ij}$ and ${\mathbf{d}}^{(2)}_{ij}$ coupled to operators $\hat{\boldsymbol\sigma}$ that act on the spin degree of freedom. When $t$ is isotropic (the same value on each nearest neighbor bond) the spectrum supports a point of fourfold degeneracy at $E=0$ at each of the three distinct $X$ points located on  centers of the Brillouin zone faces. The fourfold degeneracy is lifted at linear order in ${\mathbf{k}}$ producing a pair of doubly degenerate linear dispersing bands.  Uniaxial strain breaks the cubic symmetry and can gap this spectrum. For example, under a compressive strain along a body diagonal the hopping amplitudes depend on the bond orientations $t_{111}>t_{1 \bar 1 \bar 1}=t_{\bar 1 1 \bar 1}=t_{\bar 1 \bar 1 1}$, which opens a gap at half filling to create a strong topological insulator. In the complementary situation where tensile strain reduces $t_{111} < t_{1 \bar 1 \bar 1}$ the degeneracy is again lifted but the gapped state is instead a weak topological insulator composed of weakly coupled $(111)$ bilayers in two dimensional quantum spin Hall states. A conventional insulator can also be created near this state, although it is not perturbatively accessible from it since this requires introducing a staggered on-site scalar potential exceeding the nonzero spin-orbit scale $\lambda_{\rm so}$.

Material realizations of a DSM at a quantum critical point occur in normal-topological insulator transitions tuned by composition \cite{Sato11a,Novak2015,Xu11a,Zeljkovic2015,Brahlek12a,Wu13a,Salehi16a} and by strain \cite{YoungQCP2011}.  However, a DSM can also appear as a robust electronic phase that is \textit{stable} over a range of Hamiltonian control parameters.  There are at least two different ways of accomplishing this. (Class I) One can exclude the possibility of mass terms appearing in a band-inverted Bloch Hamiltonian ${\cal H}({\mathbf{k}})$ for ${\mathbf{k}}$ lying  along a symmetry axis \cite{WangA3Bi2012,WangCd3As22013}. We refer to this as the ``band inversion" mechanism. (Class II) One can search for space groups that support small groups with four dimensional irreducible representations (FDIR) at discrete high symmetry momenta ${\mathbf{k}}_n$. We refer to this as the ``symmetry enforced" mechanism \cite{Young2012,Steinberg2014,Zaheer2014}.  In the band inversion mechanism the Dirac semimetal is not truly a symmetry-protected state since it actually contains a {\it pair} of Dirac points (DP) and one may continuously tune parameters to un-invert the bands without changing the space group. This eliminates the two DPs by their merger and pairwise annihilation. However, in the symmetry enforced mechanism the appearance of the DP is an unavoidable consequence of the space group of the material.\footnote{ \onlinecite{Gibson15a} mention a 3rd mechanism, whereby 2D graphene-like layers are stacked in such a fashion as to give minimal 3D coupling and only small gaps.   As even graphene itself has a small gap due to SOC, irrespective of 3D couplings these systems will always have small gaps and are not strictly speaking Dirac systems.  Therefore such materials are not discussed in this review.} We discuss these in more detail below.

\subsection{Dirac semimetal from band inversion}{\label{BIDSM}}

 The band inversion mechanism provides perhaps the most direct route to formation of a Dirac semimetal.  The energy eigenvalues in the $n$-th band are related by time reversal symmetry $E_{n,\uparrow}({\mathbf{k}}) = E_{n,\downarrow}(-{\mathbf{k}})$  and by inversion symmetry $E_{n,\sigma}({\mathbf{k}}) = E_{n,\sigma}(-{\mathbf{k}})$. The combined operation of both symmetries  requires that $E_{n,\uparrow}({\mathbf{k}}) = E_{n,\downarrow}({\mathbf{k}})$ so that each band remains doubly degenerate {\it locally} at every ${\mathbf{k}}$. A Dirac node can occur if two such branches undergo an accidental band crossing at a point. Since the small group is trivial at a low symmetry ${\mathbf{k}}$ point in the Brillouin zone, the intersection of a pair of doubly degenerate bands is generically prevented by an avoided crossing.  However when ${\mathbf{k}}$ lies along a symmetry line, lattice symmetries intervene by constraining the possible interactions within this multiplet. For example, if the crossing states transform according to different irreducible representations of the group of the symmetry line their hybridization is prevented and a fourfold degeneracy at this point of intersection is symmetry protected.

  \begin{figure}
 \includegraphics[width=7cm]{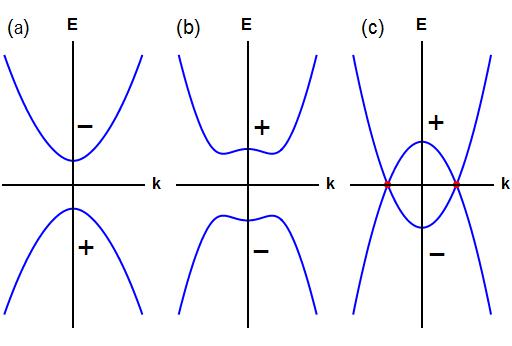}
 \caption{Development of a Dirac semimetal in an inverted band structure. The band inversion transition reverses the parities (${\pm}$) of the ${\mathbf{k}}=0$ eigenstates in the uninverted (a) and inverted (b) level orderings. In (b) the inverted bands are two fold degenerate and undergo an avoided crossing at ${\mathbf{k}} \neq 0$ which gaps the spectrum. In (c) the mixing is forbidden along a symmetry line by the different rotational symmetries of the intersecting bands. This leaves two points each with a fourfold point degeneracies at ${\mathbf{k}} = \pm {\mathbf{k}}_D$ along the symmetry line that is lifted to linear order in ${\mathbf{k}}-{\mathbf{k}}_D$. Uninverting the bands produces a pairwise annihilation of the Dirac points and the system reverts to the conventional insulating state as shown in (a).} \label{DiracFigure}
 \end{figure}

 Fig. \ref{DiracFigure} illustrates how this situation can arise naturally near a band inversion transition. The uninverted (a) and inverted (b) band structures reverse the parities and band curvatures of their ${\mathbf{k}}=0$ eigenstates. Generally, these states are allowed to mix at ${\mathbf{k}} \neq 0$ as shown in panel (b) which produces an avoided crossing and fully gaps the state with a ``Mexican-hat" dispersion (b).  However if these states transform along a symmetry direction according to different irreducible representations of the group of the symmetry line, the spectrum retains a gap closure on the symmetry line as shown in Fig. \ref{DiracFigure}(c). Note that this mechanism generically produces {\it pairs} of fourfold degenerate points along this line.  If one tunes parameters to uninvert the bands these two Dirac points merge and annihilate and the system reverts to a fully gapped state shown in panel (a).   Generally Dirac systems very sensitive to symmetry breaking terms.  First-principles calculations show that even a 1$\%$ compression in the $y$ direction opens an approximately 6 meV energy gap in Na$_3$Bi\cite{WangA3Bi2012}.

The band inversion mechanism can be understood more quantitatively by adopting a four state Hamiltonian for a system near a band inversion transition \cite{YangNagaosa2014,Gao2016}

\begin{eqnarray}\label{fourstategeneral}
{\cal H}({\mathbf{k}}) = \sum_{ij} \, a_{ij} ({\mathbf{k}}) \, \sigma_i \otimes \tau_j
\end{eqnarray}

\noindent where $\boldsymbol\sigma$ and $\boldsymbol\tau$ are Pauli matrices that act in the spin and orbital spaces respectively. When ${\mathbf{k}}$ lies along an n-fold symmetry axis the local Hamiltonian ${\cal H}({\mathbf{k}})$ commutes with the n-fold rotations ${\cal C}_n$ so one can work in a basis where the eigenstates are labelled by rotational quantum numbers $J_z$.  In this basis, and for a momentum $k_z$ along this symmetry line, the Hamiltonian can be written as a sum of commuting terms

\begin{eqnarray}\label{fourstateham}
{\cal H} (k_z) = c_0 + c_1 \tau_3 + c_2 \sigma_3 \tau_3
\end{eqnarray}

\noindent where $c_n(k_z,m)$ are real functions of $k_z$ and a mass parameter $m$ that describes the band inversion.  The gap in this model is $\Delta E = 2 \, {\rm min} (|c_1 \pm c_2|)$ and since its eigenvalues appear in degenerate pairs either $c_1$ or $c_2$ are automatically zero. The constraint $\Delta E=0$ then defines a {\it parametric curve} in the $(m,k_z)$ plane on which the system is gapless and supports a fourfold degeneracy. Any solution on this curve defines a Dirac semimetal that is stable to small variations of the band inversion parameter $m \rightarrow m+\delta m$.

\begin{figure*}
\includegraphics[width=11cm]{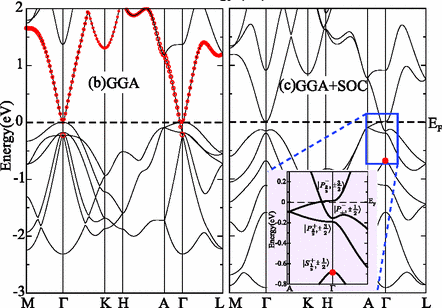}
\caption{Formation of a Dirac semimetal by the band inversion mechanism in ${\rm Na_3Bi}$. Band structure calculations without (left) and with (right) spin-orbit coupling both show a band inversion transition at the zone center illustrated by tracking the ${\rm Na}(3s)$ character of the eigenstates (red (bold) circles). An expanded view of the dispersion along the $\Gamma - A$ direction (inset) shows the symmetry-protected intersection of two two fold degenerate branches with distinct $J_z$ rotational eigenvalues at a Dirac point along the symmetry axis. (Adapted from \onlinecite{WangA3Bi2012}.)}\label{Na3Bi}
\end{figure*}

 Band inversion is predicted to be the mechanism for Dirac semimetal states in the alkali pnictides ${\rm A}_3{\rm B}$ (where ${\rm A=(Na,K,Rb)}$ \cite{WangA3Bi2012,LiuNa3Bi2014} and ${\rm B=(As,Sb,Bi)}$) and in ${\rm Cd}_3 {\rm As}_2$ \cite{Wang2013,LiuCd3As22014,Neupane14a}.  In both families of compounds the low energy physics is controlled by a single band inversion occurring near the $\Gamma$-point of the Brillouin zone.  The band structure in the prototypical case of ${\rm Na}_3{\rm Bi}$  shown in Fig. \ref{Na3Bi}  been calculated using density functional theory \cite{WangA3Bi2012}. The results can be usefully mapped onto a four-state model in the form of Eq. \ref{fourstategeneral} by studying the low momentum symmetry-allowed couplings between the four spin-orbitals involved in the band inversion.  Using the $\Gamma$ point state vectors as a basis these orbitals can be indexed by their parities and transformations under rotations about a symmetry axis. For ${\rm Na}_3{\rm Bi}$ in space group $P6_3/mmc$ (${\rm D^4_{6h}}$) the low energy basis functions can be constructed from bonding and antibonding combinations of the Na $2s$ and  the crystal field split Bi $6p$ orbitals.
Crucially in this four component basis their transformations under rotations about the $c$ axis span  four {\it different} $J_z$ eigenvalues: $ \{|S^+_{\frac{1}{2}}, 1/2 \rangle, |P^-_{\frac{3}{2}},  3/2 \rangle,  |P^-_{\frac{3}{2}}, -3/2 \rangle, |S^+_{\frac{1}{2}}, -1/2 \rangle \}$.  For momentum $k_z$ along the $\Gamma - A$ symmetry line the states are split with a mass term
\begin{eqnarray}
{\cal M}(k_z) = M_o - M_1 k_z^2
\end{eqnarray}
where $M_o M_1 > 0$ describes a band inverted state. This reveals a pair of gap closure points at $k_z = \pm \sqrt{M_o/M_1}$ which are protected by the symmetry under $c$-axis rotations.  For ${\rm Na}_3{\rm Bi} $,  $M_o \approx -0.087 \, {\rm eV}$ and $M_1 \approx -10.64 \, {\rm eV  \cdot \AA^2}$ giving a pair of DP's at momenta symmetrically shifted with respect to the $\Gamma$ point by $k_z = \pm .090 {\rm \AA^{-1}}$ which is approximately one quarter of the way to the zone boundary at $A$ \cite{WangA3Bi2012}.  This DSM is a stable phase over a range of Hamiltonian parameters that preserve the band inversion.

Similar physics occurs for the DSM in ${\rm Cd_3As_2}$ \cite{Wang2013,LiuCd3As22014,Borisenko14a} although there the situation is further complicated by the presence of 1/4 Cd site vacancies which can be ordered at room temperature to form crystals with very large unit cells.   There is a type I structure, which is tetragonal with $P4_2/nmc $ symmetry and a type II structure, which is a body centered tetragonal crystal  with $I4_1/acd $ symmetry. The latter is energetically favored and is inversion symmetric. The vacancy ordering has been imaged in \onlinecite{butler2017observation}.  Despite their enlarged unit cells both structures should feature the same level inversion at the $\Gamma$ point which reverses the conventional energy ordering of the $\Gamma_6$ and $\Gamma_7$ states which reside mainly on the Cd $5s$ and As $4p$ states. The low energy physics is again represented by a minimal four band model spanned by basis functions: $ \{|S_{\frac{1}{2}}, 1/2 \rangle, |P_{\frac{3}{2}},  3/2 \rangle,  |P_{\frac{3}{2}}, -3/2 \rangle, |S_{\frac{1}{2}}, -1/2 \rangle \}$.  This inverted band structure supports two fourfold-degenerate gap closures along the $\Gamma-Z$ direction slightly displaced from the zone center $k_z = \pm k_{\rm D} \sim \pm .03 \, {\AA^{-1}}$ with degeneracy protected by a $C_{4}$ rotational symmetry about the $c$ axis\footnote{The precise crystal structure of ${\rm Cd_3As_2}$ has been a matter of debate.   It was believed that it had the non-centrosymmetric  $I4_1cd$ structure proposed by \onlinecite{steigmann1968crystal}.   However, recent single-crystal X-ray diffraction studies \cite{ali2014crystal} show that the structure possesses an inversion center and is  $I4_1/acd$.  The locations of the Dirac nodes is difficult to predict reliably, since it depends on the magnitude of a small band inversion which is sensitive to the calculated lattice constant as well as to the type and degree of vacancy ordering on the cation sublattice \cite{Aubin1977,Cd3As2Breakdown2015,Cd2As2bands1977,Cd3As2vacancies1984}. As an extreme example of this sensitivity, density functional calculations for the putative ${\rm Cd_3As_2}$ structure  with {\it no} cation vacancies predicts a normal energy level ordering and therefore no Dirac semimetallic state \cite{Wang2013}.}.

A related route to a DSM is the lifting of fourfold degeneracy on a Dirac line node by spin-orbit coupling.  This can occur in a space group that hosts a ${\mathbf{k}}$-space curve on which the bandstructure is fourfold degenerate in the absence of spin-orbit coupling. When a spin-orbit potential is introduced these degeneracies are pairwise lifted at generic ${\mathbf{k}}$ points but can persist where the nodal line intersects a symmetry axes. ${\rm Cu_3PdN}$ in an antiperovskite structure has been proposed as a material that exemplifies this type of multi Dirac material, with three pairs of Dirac points appearing along three Cartesian symmetry directions in the Brillouin zone \cite{Yu2015}.   A similar multi-Dirac point bandstructure is predicted for the ${\rm Ca_3PbO}$ family of materials in a cubic inverse perovskite structure \cite{Ogata2011}. These Dirac points also rely on a band inversion and can be removed by ``shrinking" the parent nodal line to a point so that the Dirac points pairwise annihilate.

The space of candidate DSMs produced by band inversion is enlarged by considering ternary compounds \cite{Ogata2011,Du2015,Weng16a,Gibson15a,Sklyadneva2016}. In principle this allows one to develop criteria for choosing chemically optimized DSMs.  Key considerations are the orbital character of states that permit a Dirac degeneracy along a symmetry line, a stoichiometry where the Dirac states do not overlap the Fermi surface from other nontopological bands and the chemical stability of the material.  This has been illustrated in a survey of materials in the ZrBeSi family which crystallize in the same space group as ${\rm Na_3Bi}$  and exist in a family of materials sufficiently large to allow a separation of candidate Dirac and non-Dirac phases \cite{Du2015,Gibson15a} based on cation electronegativity differences \cite{Gibson15a}.   Such materials considerations will be investigated in more detail below in Sec. \ref{MaterialsSection}.

The minimal models describing the Dirac point physics in these materials are similar in their structure to the four band models frequently used to describe topological insulators in 2D quantum wells \cite{Bernevig06a} and in layered 3D materials in the ${\rm Bi_2 Se_3}$ family \cite{HZhang2009}. In those cases the minimal four band models also describe band inversion in a manifold of spin-orbit and crystal field-split basis states of opposite parity. Crucially, for the TIs this manifold is spanned by orbitals with $J_z=\pm 1/2$ only  so that pairs of states with common rotational eigenvalues are allowed to hybridize so when the band structure is inverted the system remains fully gapped. By contrast for the DSM their basis states carry different rotational eigenvalues and the degeneracy is symmetry protected.
 Nonetheless in this latter case lowering of the symmetry by in-plane strain or by spatially modulated potentials can mix states within the degenerate manifold and revert to a  gapped phase \cite{Yu2015,Ortix2014}.  A related phenomenon can occur for thin films with the rotational symmetry axis aligned with the surface normal which allows intervalley scattering between partner DP's at $\pm k_z$  as found in calculations for thin films in the ${\rm A_3Bi}$ family \cite{TopologicalTuning2014}.

\subsection{Symmetry-enforced Dirac semimetals}{\label{SEDSM}}

Although DSMs produced by the band inversion mechanism are generally stable to some range of Hamiltonian parameters, the presence of such a state is not necessarily assured as one may tune such a system through a band inversion transition and remove these Dirac singularities without changing the symmetry of the Hamiltonian. One is therefore motivated to ask whether space groups exist that {\it require} unremovable Dirac singularities in their band structures and further whether the band filling in possible material realizations allow the chemical potential to reside at or near these singular points. We refer to this class as ``symmetry-enforced" Dirac semimetals. Space groups that allow such point degeneracies have been studied \cite{Zak1999} and identified for specific crystal structures both with and without spin-orbit coupling in two \cite{Young2015,Damljanovic2016} and in three dimensions \cite{Steinberg2014,Young2012,Manes2012}.  Material realizations that also satisfy the band filling constraint have been proposed \cite{Young2015,Steinberg2014,Young2012,Gibson15a}.

The search for candidate Hamiltonians satisfying the first requirement can be carried out systematically by identifying three dimensional space groups $G$  that contain four dimensional irreducible representations (FDIR's) in their small groups $G_{\mathbf{k}}$ at specified momenta ${\mathbf{k}}$.  Interestingly this possibility can be excluded for any of the symmorphic space group in three dimensions. In these space groups FDIR's appear only in their {\it double groups} and then only in the double groups for crystals with cubic symmetry. However for symmorphic lattices with cubic symmetry a FDIR must reside on a threefold symmetry axis.  The basis for the FDIR can be indexed by quantum numbers spanning the set $J_z = \{\pm 3/2, \pm 1/2\}$  which in the presence of threefold symmetry requires a nonvanishing Berry's flux through a closed ${\mathbf{k}}$ space surface surrounding the point of degeneracy. Therefore a FDIR on a threefold axis cannot describe the stable merger of two Weyl points with {\it opposite} handedness.  Instead it describes a merger of two Weyl points carrying the same topological charge, a situation that has been dubbed a ``multi-Weyl" semimetal \cite{Bernevig2bandPRL2012}.

The space of candidate momenta that can support FDIR's is enlarged by considering the nonsymmorphic space groups containing lattice symmetries such as glide planes and screw axes which combine point operations $R_i$ and nonprimitive translations $\boldsymbol\tau_i$: $g_i = \{R_i|\boldsymbol\tau_i\}$.  The action of any such operation on a Bloch state $\psi_{\mathbf{k}}$ can be represented as the product of a unitary operator ${\cal U}_{\mathbf{k}}(R_i)$ acting in the state space and an overall phase factor due to the displacement in the manner
\begin{eqnarray}
\{R_i|\boldsymbol\tau_i\} \psi_{\mathbf{k}} = e^{-i {\mathbf{k}} \cdot \boldsymbol\tau_i } {\cal U}_{\mathbf{k}}(R_i) \psi_{\mathbf{k}} ({\mathbf{r}})
\end{eqnarray}
Then the multiplication rule
\begin{eqnarray}
\{R_1 | \boldsymbol\tau_i \}\{R_2 | \boldsymbol\tau_2 \} =  \{R_1 R_2 | R_1 \boldsymbol\tau_2 + \boldsymbol\tau_1\}
\end{eqnarray}
gives a product rule for the ${\cal U}$'s
\begin{eqnarray}\label{factorsystem}
{\cal U}_{\mathbf{k}}(R_1 R_2) = e^{i(R_1^{-1} {\mathbf{k}} - {\mathbf{k}}) \cdot {\mathbf{r}} }  {\cal U}_{\mathbf{k}}(R_1) {\cal U}_{\mathbf{k}} (R_2)
\end{eqnarray}

\noindent If $R_1$ and $R_2$ are operations in the small group $G_{\mathbf{k}}$ then the shift of the wavevector is a reciprocal lattice vector: $\Delta {\mathbf{k}} = R_1^{-1} {\mathbf{k}} - {\mathbf{k}} \in \{ {\mathbf{G}} \}$. If the shift $\Delta {\mathbf{k}}$  in Eq. \ref{factorsystem} is nonzero it defines a nontrivial factor system for a projective representation of $G_{\mathbf{k}}$ \cite{Hamermesh1962}.  Note that when $\Delta {\mathbf{k}}=0$  the phase factor is unity and this reduces to the regular representation of the small group.  This occurs automatically at the zone center and also at generic low symmetry points in the Brillouin zone.  As noted above, the regular representations of space groups admit FDIR's only in the special case of the cubic groups with the FDIR occurring along a threefold symmetry axis \cite{BandC1972}, in which case they carry a nonzero Berry's flux and do not describe Dirac points.  However on the faces of the Brillouin zone where $\Delta {\mathbf{k}} \in \{{\mathbf{G}}\} \neq 0$  the factor system is nontrivial and identifies a projective representation of $G_{\mathbf{k}}$ that can allow symmetry enforced FDIR's. This possibility is thereby excluded for any point in the interior of the Brillouin zone.

One concludes that a necessary condition for a symmetry-enforced DSM is the presence of a nonsymmorphic space group, which hosts a small group $G_{\mathbf{k}_n}$ at zone boundary points ${\mathbf{k}}_n$ that host FDIR's.   This is not a sufficient condition, since one needs to additionally verify that the degeneracy is broken to linear order in momentum ${\mathbf{k}}-{\mathbf{k}}_n$ near the point of degeneracy. To guarantee this, the symmetric Kronecker product of the FDIR  with itself must contain the vector representation of $G_{\mathbf{k}_n}$.  Finally one needs to verify that the band velocities are nonzero at the FDIR so that the valence and conduction branches are not degenerate away from the FDIR.

\begin{figure}[t]
\vspace{1cm}
\includegraphics[width=8cm]{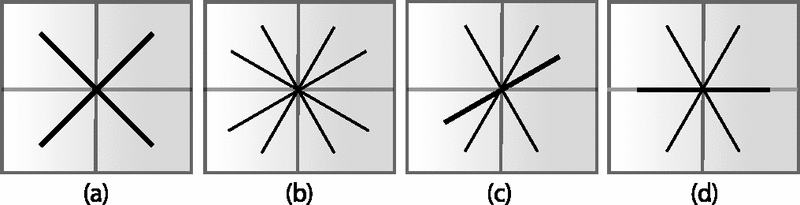}
\caption{Possible linear dispersions of energy (vertical) versus momentum (horizontal) near a symmetry-enforced Dirac point at the zone boundary of a lattice with a nonsymmorphic space group.  In (a) the FDIR occurs at a TRIM and the fourfold degeneracy is lifted to form two two fold degenerate branches (bold). In (b) inversion symmetry is absent and four linearly dispersing branches merge at a FDIR. In (c) and (d) the degeneracy of the FDIR is lifted to form a two fold degenerate and two nondegenerate branches. In (c) the FDIR does not occur at a TRIM while in (d) the two fold degenerate branch has zero slope indicating quadratic dispersion the FDIR occurs at a TRIM. Adapted from  \onlinecite{Young2012}.} \label{ZaheerFig4}
\end{figure}

A table of possible space groups and locations of their Brillouin zone boundary points that host FDIR's has been compiled by  \onlinecite{Zaheer2014}.  One finds that 99 of the 230 space groups have double groups that satisfy the first two symmetry conditions. Approximately one-third of these candidate FDIR's are ``false positives" because they lie along a threefold symmetry axis and describe multi-Weyl points instead of Dirac points.  Fig. \ref{ZaheerFig4} shows the possible decompositions of an FDIR into linearly dispersing branches in the vicinity of the degeneracy: (a) $4 \rightarrow 2+2$, (b) $4 \rightarrow 1+1+1+1$, (c,d) $4 \rightarrow 2 + 1 + 1$. Case (a) is the generic dispersion of a fourfold Dirac point splitting into a pair of two fold degenerate branches as required for ${\cal T}$ and ${\cal P}$ symmetric material. Case (b) occurs when the FDIR occurs in a system that lacks inversion symmetry. Cases (c) and (d) are inversion broken spectra that are distinguished by whether the band crossing occurs at a time reversal invariant momentum (TRIM) ${\mathbf{k}}_n$ [Case (d)] or not [Case (c)].  In the former situation, the FDIR occurs at a TRIM the spectrum must be an even function of ${\mathbf{k}} - {\mathbf{k}}_n$ and the single two fold degenerate branch in the spectrum has zero velocity at the TRIM.

Material realizations of symmetry-enforced DSMs need to satisfy three design criteria: (1) The lattice structure must be stable in one of the active nonsymmorphic space groups that support FDIR's on a zone boundary. (2) Ideally the Dirac points should be spectrally isolated to avoid overlapping Fermi surfaces from other nontopological bands. (3) The stoichiometry must give a band filling for which the Fermi energy is located at (or near) its Dirac points.  These considerations often conflict with each other. For example, the band filling constraint (3) requires an even number of electrons per primitive cell but for a nonsymmorphic space group with sublattice symmetry this may translate into an {\it odd} number of electrons per formula unit. This presents an example of a ``filling enforced semimetal" \cite{Parameswaran2013,Parameswaran2015a,AshvinNonsymmorphic2015} where the zone boundary degeneracy requires that  electron count for a filled band actually results in a gapless state where the conduction and valence bands contact each other. This can conflict with stability requirement (1) since such a structure can be susceptible to a symmetry lowering reconstruction that that produces a gapped spectrum with completely filled bands.  Note that since three dimensional DSM's have point Fermi surfaces they can be perturbatively stable with respect to this kind of reconstruction. However one concludes that material realizations of symmetry-enforced DSMs can generally involve interactions that can coax elements into non-optimized oxidation states.

\begin{table*}
\begin{tabular}{c c c c c c c c c c c c c c }
\hline
\hline
$C_{n}$ & & $|P|$& &$(u_{A,\uparrow},u_{B,\uparrow})$ & & f($k_{\pm}$, $k_{z}$) & & g($k_{\pm}$, $k_{z}$)  & & 2D topological invariant & & $H_{\text{Dirac}}(\textbf{q})$ \\
\hline
\hline
$C_{2}$ & & $\tau_{z}$ & & $-$ & & $-$ & & $-$ & & $-$  & & Not allowed \\
$C_{2}$ & & $\tau_{0}$ & & $-$ & & $-$ & & $-$ & & $-$  & & Not allowed \\
\hline
\hline
$C_{3}$ & & $\tau_{z}$ & &$(e^{i\pi},e^{i\frac{\pi}{3}})$ & & $\beta k_{+}$ & & $\gamma k_{-}$  & & $\nu_{2D}=1$ & & Linear Dirac  \\
$C_{3}$ & & $\tau_{0}$ & &$(e^{i\pi},e^{i\frac{\pi}{3}})$ & & $\beta k_{z}k_{+}+\gamma k_{-}^{2}$ & & $\eta k_{z}k_{-}+\xi k_{+}^{2}$  & &$\nu_{2D}=0$ & & Linear Dirac\\
\hline
\hline
$C_{4}$ & & $\tau_{z}$ & &$(e^{i\frac{3\pi}{4}},e^{i\frac{\pi}{4}})$ & & $\eta k_{+}$ & & $\beta k_{z}k_{+}^{2}+\gamma k_{z}k_{-}^{2}$  & & $n_{M}=\pm1$ & & Linear Dirac
\\
$C_{4}$ & & $\tau_{0}$ & &$(e^{i\frac{3\pi}{4}},e^{i\frac{\pi}{4}})$ & & $\eta k_{z}k_{+}$ & & $\beta k_{+}^{2}+\gamma k_{-}^{2}$  & &
$n_{M}=2\text{sgn}(|\beta|-|\gamma|)$ & & Linear Dirac \\
\hline
\hline
$C_{6}$ & & $\tau_{z}$ & &$(e^{i\frac{\pi}{2}},e^{i\frac{\pi}{6}})$ & & $\beta k_{+}$ & & $\gamma k_{z}k_{+}^{2}$ & & $n_{M}=\pm1$ & & Linear Dirac \\
$C_{6}$ & & $\tau_{0}$ & &$(e^{i\frac{\pi}{2}},e^{i\frac{\pi}{6}})$ & & $\beta k_{z}k_{+}$ & & $\gamma k_{+}^{2}$ & & $n_{M}=\pm2$ & & Linear Dirac \\
\hline
$C_{6}$ & & $\tau_{z}$ & &$(e^{i\frac{5\pi}{6}},e^{i\frac{\pi}{2}})$ & & $\beta k_{+}$ & & $\gamma k_{z}k_{-}^{2}$ & & $n_{M}=\pm1$ & & Linear Dirac \\
$C_{6}$ & & $\tau_{0}$ & &$(e^{i\frac{5\pi}{6}},e^{i\frac{\pi}{2}})$ & & $\beta k_{z}k_{+}$ & & $\gamma k_{-}^{2}$ & & $n_{M}=\pm2$ & & Linear Dirac \\
\hline
$C_{6}$ & & $\tau_{z}$ & &$(e^{i\frac{5\pi}{6}},e^{i\frac{\pi}{6}})$ & & $\eta k_{z}k_{+}^{2}$ & &$\beta k_{+}^{3}+\gamma k_{-}^{3}$ & &
$n_{M}=3\text{sgn}(|\beta|-|\gamma|$) & & Quadratic Dirac \\
$C_{6}$ & & $\tau_{0}$ & &$(e^{i\frac{5\pi}{6}},e^{i\frac{\pi}{6}})$ & & $\eta k_{+}^{2}$ & &$\beta k_{z}k_{+}^{3}+\gamma k_{z}k_{-}^{3}$ & & $n_{M}=\pm2$
& & Quadratic Dirac \\
\hline \hline
\end{tabular}
\caption{\label{DSMI}\noindent {\bf Classification table for Class I 3D Dirac semimetals.}
{
Classification table for 3D topological Dirac semimetals
obtained by an accidental band crossing in systems having $C_{n}$ rotational
symmetry with respect to the $z$ axis (adapted from \onlinecite{YangNagaosa2014}). Here
$C_{n}=\text{diag}[u_{A,\uparrow},u_{B,\uparrow},u_{A,\downarrow}=u^{*}_{A,\uparrow},u_{B,\downarrow}=u^{*}_{B,\uparrow}]$
and $\beta$, $\gamma$, $\eta$, $\xi$ are complex numbers.
For compact presentation, $u_{A,\uparrow}$ and $u_{B,\uparrow}$
are arranged in a way that $0<\text{arg}(u_{B,\uparrow})<\text{arg}(u_{A,\uparrow})\leq\pi$.
$\nu_{2D}$ ($n_{M}$) indicates the 2D ${\mathbb{Z}}_2$ invariant (mirror Chern number) defined on
the $k_{z}=0$ plane. ($n_{M}=\nu_{2D}$ mod 2.)
The $2\times 2$ Hamiltonian
$h_{\uparrow\uparrow}(\textbf{k})=f(\textbf{k})\tau_{+}+f^{*}(\textbf{k})\tau_{-}+a_{5}(\textbf{k})\tau_{z}$.
In the case of $h_{\uparrow\downarrow}(\textbf{k})$, $h_{\uparrow\downarrow}(\textbf{k})=g(\textbf{k})\tau_{x}$ when $P=\pm\tau_{z}$
while $h_{\uparrow\downarrow}(\textbf{k})=g(\textbf{k})\tau_{y}$ when $P=\pm\tau_{0}$.
The leading order terms of $f(\textbf{k})$ and $g(\textbf{k})$ are shown in the table.
$H_{\text{Dirac}}(\textbf{q})$ describes the effective Hamiltonian near the bulk Dirac point,
which is either $H_{\text{Dirac}}(\textbf{q})=\upsilon_{x}q_{x}\Gamma_{1}+\upsilon_{y}q_{y}\Gamma_{2}+\upsilon_{z}q_{z}\Gamma_{3}$ (linear Dirac)
or $H_{\text{Dirac}}(\textbf{q})=\upsilon_{x}(q^{2}_{x}-q^{2}_{y})\Gamma_{1}+2\upsilon_{y}q_{x}q_{y}\Gamma_{2}+\upsilon_{z}q_{z}\Gamma_{3}$ (quadratic Dirac)
where $\Gamma_{1,2,3}$ are mutually anticommuting $4\times 4$
gamma matrices and $\upsilon_{x,y,z}$ are real constants.
Here the momentum $\textbf{q}$ is measured with respect to the bulk Dirac point.
}
}
\end{table*}


\begin{table*}
\begin{tabular}{c c c c c c c c c c c c }
\hline
\hline
$C_{n}$ & & $|P|$& &$u_{A,\uparrow}$ & & f($k_{\pm}$, $k_{z}$) & & $g_{z}$($k_{\pm}$, $k_{z}$)   & & $H_{\text{Dirac}}(\textbf{q})$ \\
\hline
\hline
$C_{2}$ & & $\tau_{x}$ & & $e^{i\frac{\pi}{2}}$ & & $k_{z}F_{1}^{(1)}(k_{x,y})-iF_{2}^{(1)}(k_{x,y})$
& & $\alpha k_{x}+\beta k_{y}$   & & Linear Dirac \\
\hline
\hline
$C_{3}$ & & $\tau_{x}$ & &$-$ & & $-$ & & $-$   & & Not allowed \\
\hline
\hline
$C_{4}$ & & $\tau_{x}$ & &$e^{\pm i\frac{\pi}{4}}$ & & $F_{1}^{(2)}(k_{x,y})-ik_{z}F_{2}^{(2)}(k_{x,y})$
& & $\alpha k_{\pm}$   \\
\hline
\hline
$C_{6}$ & & $\tau_{x}$ & &$e^{\pm i\frac{\pi}{6}}$ & & $k_{z}F_{1}^{(3)}(k_{x,y})
+iF_{2}^{(3)}(k_{x,y})$ & & $\alpha k_{\pm}$  & & Linear Dirac \\
\hline
$C_{6}$ & & $\tau_{x}$ & &$e^{i\frac{3\pi}{6}}$ & & $k_{z}F_{1}^{(3)}(k_{x,y})
+iF_{2}^{(3)}(k_{x,y})$ & &$F_{3}^{(3)}(k_{x,y})
+iF_{4}^{(3)}(k_{x,y})$
& & cubic Dirac \\
\hline \hline
\end{tabular}
\caption{\label{DSMII}\noindent {\bf Classification table for Class II 3D Dirac semimetals.}
{
Classification table for 3D topological Dirac semimetals in systems having $C_{n}$ rotational
symmetry with respect to the $z$ axis when $P=\pm\tau_{x}$ (adapted from \cite{YangNagaosa2014}.
In this Dirac semimetal phase, the location of the 3D Dirac point is fixed either at the center or the edge of the rotation axis,
i.e., at a TRIM on the rotation axis.
Here $C_{n}=\text{diag}[u_{A,\uparrow},u_{B,\uparrow},u_{A,\downarrow},u_{B,\downarrow}]$
$=\text{diag}[u_{A,\uparrow},-u_{A,\uparrow},u^{*}_{A,\uparrow},-u^{*}_{A,\uparrow}]$
and $\alpha$, $\beta$ are complex numbers.
For compact presentation, $\text{arg}(u_{A,\uparrow})$ is fixed to be
$-\frac{\pi}{2}\leq\text{arg}(u_{A,\uparrow})\leq\frac{\pi}{2}$.
But the same result holds even if $\text{arg}(u_{A,\uparrow})$ is shifted by $\pi$.
The real functions $F^{(1,2,3)}$ are given by
$F_{i=1,2}^{(1)}=c^{(1)}_{i}k_{x}+d^{(1)}_{i}k_{y}$,
$F_{i=1,2}^{(2)}=c^{(2)}_{i}(k_{x}^{2}+k_{y}^{2})+d^{(2)}_{i}k_{x}k_{y}$,
$F_{i=1,2,3,4}^{(3)}=c^{(3)}_{i}(k_{+}^{3}+k_{-}^{3})+id^{(3)}_{i}(k_{+}^{3}-k_{-}^{3})$
where $c^{(1,2,3)}_{i}$ and $d^{(1,2,3)}_{i}$ are real constants.
The $2\times 2$ Hamiltonian
$h_{\uparrow\uparrow}(\textbf{k})=f(\textbf{k})\tau_{+}+f^{*}(\textbf{k})\tau_{-}+a_{1}(\textbf{k})\tau_{z}$
where $a_{1}(\textbf{k})=\upsilon k_{z}$ with a real constant $\upsilon$,
and $h_{\uparrow\downarrow}(\textbf{k})=g_{z}(\textbf{k})\tau_{z}$.
The leading order terms of $f(\textbf{k})$ and $g_{z}(\textbf{k})$ are shown in the table.
$H_{\text{Dirac}}(\textbf{q})$ describes the effective Hamiltonian near the bulk Dirac point,
which is either $H_{\text{Dirac}}(\textbf{q})=\upsilon_{x}q_{x}\Gamma_{1}+\upsilon_{y}q_{y}\Gamma_{2}+\upsilon_{z}q_{z}\Gamma_{3}$ (Linear Dirac)
or $H_{\text{Dirac}}(\textbf{q})=\upsilon_{x}(q^{3}_{+}+q^{3}_{-})\Gamma_{1}+i\upsilon_{y}(q_{+}^{3}-q_{-}^{3})\Gamma_{2}+\upsilon_{z}q_{z}\Gamma_{3}$ (cubic Dirac)
where the momentum $\textbf{q}$ is measured with respect to the bulk Dirac point with $q_{\pm}=q_{x}\pm iq_{y}$.
Here $\Gamma_{1,2,3}$ are mutually anticommuting $4\times 4$
gamma matrices and $\upsilon_{x,y,z}$ are real constants.
}
}
\end{table*}


 \subsection{Classification of four band models for Dirac semimetals}

A unified treatment of ``band-inverted" and ``symmetry enforced" DSMs can be developed by studying the combined action of time reversal symmetry ${\cal T}$, uniaxial rotational symmetry ${\cal C}_n$, and inversion symmetry ${\cal P}$ on a minimal four state Hamiltonian that couples two spin and two orbital degrees of freedom \cite{YangNagaosa2014,Gao2016}. For this purpose one examines the Hamiltonian of Eq. \ref{fourstategeneral} which can be written explicitly
 \begin{eqnarray}\label{fourstategeneralv2}
{\cal H}({\mathbf{k}}) = \sum_{ij} \, a_{ij} ({\mathbf{k}}) \, \sigma_i \otimes \tau_j = \left(
\begin{array}{cc}
		h_{\uparrow \uparrow} ({\mathbf{k}}) & h_{\uparrow \downarrow} ({\mathbf{k}})\\
		h_{\downarrow \uparrow} ({\mathbf{k}}) & h_{\downarrow \downarrow} ({\mathbf{k}})\\
\end{array}
\right) \nonumber\\
\end{eqnarray}
where $h_{\sigma,\sigma'}$ are $2 \times 2$ matrix-valued operators expanded in the basis $\{\boldsymbol\tau\}$ which act on the orbital degrees of freedom. In this basis the time reversal operator $\Theta = i \sigma_y K$ where $K$ is complex conjugation.  ${\cal T}$ symmetry and the use of $\Theta$ allows one to express ${\cal H}({\mathbf{k}})$ as

\begin{eqnarray}\label{hamtrs}
{\cal H}({\mathbf{k}}) =  \left(\begin{array}{cc}
                                                                                             h_{\uparrow \uparrow} ({\mathbf{k}}) & h_{\uparrow \downarrow} ({\mathbf{k}})\\
                                                                                             -h^*_{\uparrow \downarrow} ({\mathbf{-k}}) & h^*_{\uparrow \uparrow} ({\mathbf{-k}})\\
                                                                                           \end{array}
                                                                                         \right).
\end{eqnarray}

 Dirac points occur at accidental band crossings between pairs of two fold degenerate branches in the spectrum of ${\cal H}({\mathbf{k}})$.  This is forbidden for a general wavector ${\mathbf{k}}$ but it {\it can} occur along high symmetry lines or points. The presence of a $n$-fold uniaxial rotational symmetry ${\cal C}_n$ along a symmetry line allows one to label the energy eigenstates by their rotational eigenvalues $J_z=\{u_{A \uparrow},u_{B \uparrow},u_{A \downarrow},u_{B \downarrow} \}$. Furthermore,  inversion symmetry relates
 \begin{eqnarray}\label{inversion}
 {\cal H}(- {\mathbf{k}}) = {\cal P} {\cal H(\mathbf{k}}) {\cal P}^{-1}.
 \end{eqnarray}
 Here one can distinguish between two situations based on the allowed matrix representation of the  parity operator ${\cal P}$ \cite{YangNagaosa2014}. When ${\cal P}$ has a diagonal form ($\pm \tau_0, \pm \tau_z$), band crossings, if present, occur in pairs along the symmetry line. When ${\cal P}$ is off-diagonal ($ \pm \tau_x$) inversion interchanges the orbital degrees of freedom and then a band touching point is possible only at a (single) high symmetry point on a zone face along the symmetry line. ``Band inversion" DSMs (Class I) are members of the first class and ``symmetry-enforced" (Class II) DSMs are members of the second class. In the former case one finds that rotational eigenvalues are paired in two fold-degenerate branches with the combinations $\{u_{A \uparrow},u_A{\downarrow}\}, \, \{u_{B \uparrow},u_B{\downarrow}\}$ while in the latter case they are exchanged and paired $\{u_{A \uparrow},u_B{\downarrow}\}, \, \{u_{B \uparrow},u_A{\downarrow}\}$.  These four eigenvalues are not independent.  In both cases because of ${\cal T}$ symmetry they occur in complex conjugate pairs (i.e. $u_{A,\downarrow}  = u^*_{A,\uparrow}$) and for the Class II DSMs one has an additional constraint $u_{B \uparrow} = - u_{A \uparrow}$.  An accidental band crossing can occur only if these groups contain no common eigenvalues, e.g.
 \begin{eqnarray}\label{nooverlap}
 {\rm Class \, I}: \,\, \{u_{A \uparrow},u_A{\downarrow}\} \cap \{u_{B \uparrow},u_B{\downarrow}\}=0 \nonumber\\
 {\rm Class \, II}: \,\, \{u_{A \uparrow},u_B{\downarrow}\} \cap \{u_{B \uparrow},u_A{\downarrow}\}=0
 \end{eqnarray}

This representation allows one to understand how the various properties of DSMs are controlled by the type of rotational symmetry. Tables \ref{DSMI} and \ref{DSMII} catalog results obtained for both classes of DSMs, listed by their rotational symmetry about the $z$ axis and their matrix representations of ${\cal P}$. For each entry the tables list the  rotational eigenvalues (two independent eigenvalues for Class I and one for Class II) and expressions from which the low energy Hamiltonian near the Dirac point can be reconstructed.  An important characteristic of  {\rm Class I} DSMs is that they support quantized topological invariants despite being gapless phases.  Their $k_z=0$ plane is a time reversal invariant plane on which a 2D ${\mathbb{Z}}_2$ invariant can be defined.  When $n $ is even (and greater than 2) it is also a mirror plane and a mirror Chern number can also be defined.  Table \ref{DSMI} for  {\rm Class I} DSMs  lists the relevant topological invariants e.g. the ${\mathbb{Z}}_2$ invariant $\nu_{2D}$ on the $k_z=0$ plane for $n=3$ and the mirror Chern number $n_M$  for $n=4,6$.

For Class I DSMs, the Hamiltonian is an even function of $k_z$ along the symmetry axis and the DP's therefore appear in pairs symmetrically displaced about its center.  This allows one to continuously tune the locations of the band crossings as a function of Hamiltonian control parameters. For example, in the band inversion mechanism in Section \ref{BIDSM} the inversion parameter $m$ provides one such degree of freedom that can be used to shift or even to pairwise eliminate these points of intersection.

 Class I accidental band crossings cannot occur at all as protected degeneracies on a two fold symmetry axis since the condition in Eq. \ref{nooverlap} can not be satisfied. For higher rotational symmetries Class I DSMs can exist and and in fact they can occur both for band inversions between states of the same (${\cal P} = \pm \tau_0$) and of opposite (${\cal P} = \pm \tau_z$) parities.  These two situations  are not distinguished by the rotational eigenvalue criterion in Section \ref{BIDSM} but they can be physically distinguished by 2D topological invariants on their $k_z=0$ symmetry planes which in turn determine the number of topologically protected {\it surface} modes that appear on surfaces parallel to the $z$ axis. The case of an inversion between states of opposite parity is operative in the case of  ${\rm Na_3Bi}$.  Class I DSMs on a sixfold symmetry axis can also support a more exotic bulk Dirac point (labelled ``quadratic Dirac" in Table \ref{DSMI}) with linear dispersion along the $z$ axis but {\it quadratic} dispersion in the two transverse directions. These can be regarded as the stable merger of ``double Weyl" points that each carry Chern number $\pm 2$.

For Class II DSMs the Hamiltonian along the symmetry line is instead an odd function of $k_z$ (Table \ref{DSMII}) and a {\it single} DP occurs at a TRIM.  Since the possibility of a symmetry-enforced Class II Dirac point can be excluded at the $\Gamma$ point\footnote{FDIR's occur only as projective representations of space groups which using Eq. \ref{factorsystem} cannot occur at ${\mathbf{k}}=0$.} Class II DP's must be pinned to a zone boundary face, edge, or corner and cannot be eliminated without changing the lattice symmetry.  Note that a fourfold degeneracy at $\Gamma$ that can occur at a band inversion transition is not symmetry protected in this manner since it does not describe a stable phase, but requires fine tuning parameters to a quantum critical point.

Class II DP's are allowed for a TRIM on a two fold symmetry axis and in fact the singularity predicted at the $T$ point in the body centered orthorhombic ${\rm BiZnSiO_4}$ (space group 74 ($Imma$)) is an just such an example \cite{Steinberg2014}. By contrast DP's for TRIMs on a threefold symmetry axis are forbidden since they cannot have a nonzero Chern number. Interestingly, a Class II DSM with an FDIR on a sixfold axis can occur with a linear Dirac node or a cubic Dirac node. The latter describes a stable merger of two multi-Weyl nodes that carry Chern numbers $\pm 3$ (``triple-Weyl" points) \cite{Ahn16a}.   In this situation the low energy dispersion near the DP is linear in $q_z$ along the symmetry axis and cubic in the two transverse directions with a sixfold symmetry around the rotation axis.

The classification scheme presented in Table \ref{DSMII} applies to inversion symmetric lattices where the matrix representation of the parity operator (${\cal P} = \pm \tau_x$) can be associated with a sublattice exchange produced by a nonprimitive translation in a nonsymmorphic space group. A related classification scheme can be developed in the case of an antiunitary representation of the inversion operator (${\cal P} = \pm i \tau_y$) \cite{Gao2016} and gives the same constraints on the rotational symmetries that can host a DSM. Note also that nonsymmorphic space groups can support FDIR's on TRIMs where the sublattice exchange is generated by a screw or glide plane symmetry in the space group  \cite{BJYangDSMrotationsymmetry2015}.  Creation and annihilation rules for Dirac points stabilized by rotational symmetries in both Class I and Class II DSMs have been identified \cite{Koshino2014}. The ten-fold symmetry classification of Hamiltonians based on their global time reversal, particle-hole and chiral symmetries can be extended to treat gapless systems protected by reflection symmetry \cite{Schnyderreflectionprotection2014}.

\subsection{Phenomena of Dirac semimetals}
There are both similarities and differences in the phenomena exhibited in DSMs vs. WSMs.   Certain of these effects which depend on aspects like the 3D linear dispersion can be imported directly to Dirac case.  Other effects like that of surface states and tranport features like the chiral anomaly need to be considered more carefully.
\subsubsection{Fermi arcs in Dirac semimetals}
\label{DiracFermiArcs}
\begin{figure*}
\includegraphics[width=15cm]{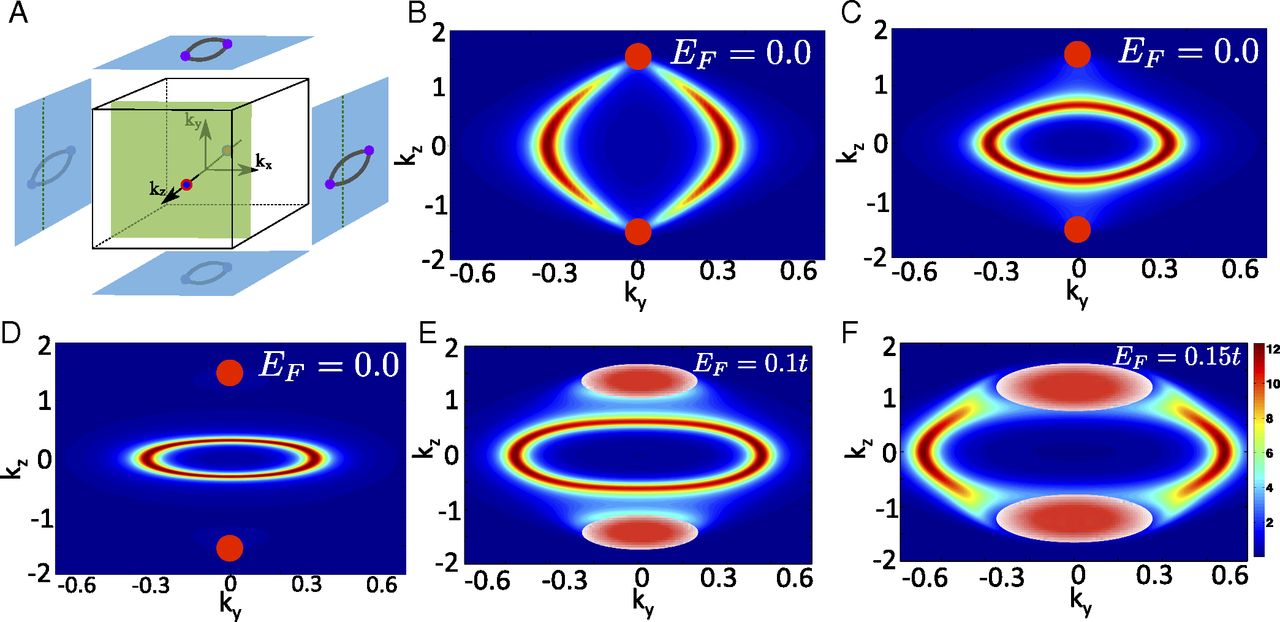}
\caption{Fermi arcs on the surface of DSMs. (a) A schematic of a DSM showing Dirac nodes along the $k_z$ axis in the bulk BZ and double Fermi arcs on the surface BZs. Note that surfaces perpendicular to the $z$ axis have no arcs.  A 2D slice of the bulk BZ perpendicular to the $k_z$ axis is shown as a green (shaded) plane, which projects to the green (dashed) line on the surface BZ.  (b-d).  A symmetry allowed mass term at the surface admits backscattering between these branches at the contact point which dissociates the surface band from the projected Dirac point.  These surface branches can be deformed but not removed from the time reversal symmetric plane at $k_z=0$. If the chemical potential is not aligned with the bulk Dirac points the surface Fermi arcs disappear by merging with the bulk continuum (e,f). Adapted from \onlinecite{Kargarian2015}.}\label{FermiArcs}
\end{figure*}

As discussed above, the Fermi arc on the surface of a Weyl semimetal is a striking manifestation of the topological singularities in its bulk band structure. These boundary states connect the surface ${\mathbf{k}}$-space projections of two bulk Weyl nodes of opposite handedness and are unremovable from any surface where these bulk nodes do not project onto the same surface momentum. The possibility of Fermi arcs at the surface of a Dirac semimetal is more subtle because the analogous bulk node is fourfold degenerate and carries Chern number zero so it is not similarly topologically protected. It can be regarded as the stable merger of two compensated Weyl points that project to the {\it same} surface momentum.  Nonetheless, as detailed in Table \ref{DSMI}, the bulk Hamiltonian for a Class I DSM on the $k_z=0$ plane can support a gapped 2D state with a nontrivial topology which requires boundary edge modes to occur on this symmetry plane \cite{YangNagaosa2014}.  A key issue is therefore whether these edge modes actually reside on open or closed constant energy contours.  The former lines can be combined to form double Fermi arcs, i.e. they are members of a doubled Weyl system where pairs of protected surface modes connect the Weyl points embedded in partner Dirac nodes \cite{WangA3Bi2012,Wang2013} as illustrated in Fig. \ref{FermiArcs}(a).  By contrast in the latter case these branches are not tied to singularities in the bulk bandstructure and therefore can be deformed and even collapsed to a point on the symmetry plane by a continuous change of the Hamiltonian (Fig. \ref{FermiArcs}(b,c)).  ARPES measurements  detect double Fermi arcs on side surfaces of the prototypical Class I DSMs ${\rm Na_3Bi}$ \cite{Xu15a} and ${\rm Cd_3As_2}$ \cite{Yi2014} and have been interpreted as evidence for the former scenario.  Theory further suggests that measurement of ARPES intensities from these modes in a nonequilibrium state with applied fields ${\mathbf{E}} \cdot {\mathbf{B}} \neq 0$ could be used in principle to visualize an induced chiral current in both the WSM or DSM states\footnote{Note that this proposal would require an ARPES setup for a sample immersed in a magnetic field, which would destroy angular resolution.} \cite{BardarsonWeylArpes2016}.

Recent work has focused attention on the fragility Fermi arcs on the boundaries of DSMs \cite{Potter2014,Kargarian2015,Fang16}. The possibility of having double Fermi arcs on a DSM that are pinned to its Dirac nodes is problematic because these nodes are protected only by a bulk spatial (rotational) symmetry\footnote{This issue is entirely avoided in certain glide symmetry protected Dirac nodes which are compatible with surfaces hosting Fermi arcs as discussed at the end of the section.}. One may argue that close to the contact point these surface modes penetrate deeply into the bulk so their properties are controlled by the bulk symmetries. This suggests that the contact points, although not topologically protected, may  be  perturbatively stable with respect to symmetry lowering at the boundary \cite{Potter2014}. However this intuitive picture is not supported by symmetry analysis within a four band model that allows for additional bulk perturbations that anticommute with the DSM Hamiltonian  and allow one to continuously deform the Hamiltonians for $k_z$ Fermi-arc and non Fermi arc sectors into each other \cite{Kargarian2015}. Thus the pinning of the topological band to the Dirac point is not symmetry protected and the Fermi contour may dissociate from the underlying projected Dirac points (Fig. \ref{FermiArcs} (b-d)).  In this interpretation the double Fermi arcs observed in ${\rm Na_3Bi}$ \cite{Xu15a} and ${\rm Cd_3As_2}$ \cite{Yi2014} are to be regarded as slightly displaced from the underlying Dirac singularities possibly manifesting the smallness of the allowed residual mass term. Although the surface Fermi line is unpinned from the projected Dirac points it cannot be removed completely since the plane $k_z=0$ has a higher symmetry which requires an edge mode on the symmetry plane. In principle the surface Fermi line can be shrunk down to a point on this symmetry plane while maintaining all spatial symmetries and producing a new Dirac cone in the surface spectrum. In material realizations where the chemical potential is not aligned with the bulk Dirac points, the occupied bulk states project to a finite area disk on the surface into which the surface mode can disappear (Fig. \ref{FermiArcs} (e,f)) preempting a possible intersection with the Dirac point.

The topological stability (fragility) of Fermi arcs on a Weyl (Dirac) semimetals can also be demonstrated by modeling their dispersions as helicoidal surfaces in the complex $\tilde q = q_x +i q_y$ plane \cite{Fang16}.  This mapping takes its simplest form near a Weyl point with topological charge $\pm 1$ where the energy dispersion on the surface can be mapped to the phase of a holomorphic function of $\tilde q$
\begin{eqnarray}
E(\tilde q) \sim {\rm Im}[\log(\tilde q^{\pm 1})].
\end{eqnarray}
A pair of compensated Weyl points is a ``Weyl dipole" where a single Fermi arc is launched and terminated on the two branch points of the function $ \log[(\tilde q - \tilde K_+)/(\tilde q - \tilde K_-)]$. A Dirac point, regarded as a stable merger of two Weyl points produces
  double Fermi arcs  corresponding to the analogous mapping of the  holomorphic function
\begin{eqnarray}\label{holomorphicDirac}
E(\tilde q) \sim {\rm Im}[\log(\tilde q + \tilde q^{-1} \pm \sqrt{(\tilde q - \tilde q^{-1})^2})]
\end{eqnarray}
Eq. \ref{holomorphicDirac} describes the phase evolution on a double helicoid (an overlapping helicoid/antihelicoid pair) and illustrates an essential fragility of the state. In the double helicoid the two branches of Eq. \ref{holomorphicDirac} intersect on a line where they can be gapped out by symmetry allowed momentum conserving terms in the Hamiltonian.  For energies outside the gap the system retains double Fermi arcs while inside the gap they disappear.  The hybridization of these branches on their line of intersection can be prevented in special two dimensional space groups that support a glide reflection symmetry on the surface.  This occurs in four of the seventeen wallpaper groups which define the lattice systems that host unremovable Fermi arcs for the DSM \cite{Fang16}.

\subsubsection{The chiral anomaly in Dirac semimetals}
\label{DiracChiral}
We have seen in Sec. \ref{ChiralAnomaly} that the defining properties of the chiral anomaly in WSMs in transport require that the scattering rate between nodes of opposite chirality $1/ \tau_a$ is small compared to the current relaxation rate, which determines the conductivity. In WSMs this is achieved by separating Weyl nodes in momentum space so that they cannot be connected by small momentum transfer scattering. However, in Dirac semimetals, opposite chiralities coexist at the same crystal momentum, and are protected instead by symmetries involving rotations and reflections. Thus the nodes of opposite chirality are now protected not by translation symmetry, but by the very same point group elements that forbids Dirac mass terms.  It has been stated that in a magnetic field, the individual Weyl nodes composing a Dirac point will be pulled apart in the Brillouin zone separating the nodes in momentum space.   Although in principle one can make a distinction now between inter and intravalley scattering, for typical physical parameters, this momentum separation is expected to be very small due to the tremendous disparity in energy scales between Zeeman splitting and a typical band dispersion.  Instead if the relaxation is to be small between states of opposite chirality at the same $\k_0$ in a DSM \cite{burkov2015negative}, this is presumably because scattering is suppressed by the same symmetries that protect the Dirac node itself.  Impurities can scatter between these states and even spherically symmetric impurities induce some degree of mixing \cite{Parameswaran14a}, but ab-initio and other studies of the role of impurities should be further explored.  If the $g$ factors are large and bands weakly dispersing, or the effects of a magnetic field are amplified by the presence of magnetic ions, a splitting of the bands to make a Weyl state may lead to observable effects as discussed in the somewhat different context of quadratic band touching systems \cite{Hirschberger16,Cano16a}

Transport in Dirac semimetals with an approximately conserved spin are expected to have additional consequences as discussed in \onlinecite{Burkov16b} due to the DSM  $\mathbb{Z}_2$ topological invariant discussed above that gives a $\mathbb{Z}_2$ topological charge.  The expectation is that such systems could have a spin current proportional to  $\bf{B}$, a spin Hall effect, and inverse spin Hall effect.   These effects may also influence the conventional chiral anomaly to give a stronger angular dependence than $\bf{E \cdot B}$ and which has been proposed to be the source of the observed deviations from the expected angular dependence.

\section{Materials considerations}
\label{MaterialsSection}

Given the above theoretical considerations one may look for actual real materials that exhibit WSM and DSM phases.   In this regard, \textit{ab initio} calculations have proven extremely powerful in identifying real materials that exhibit not just these, but many topological phases \cite{Bansil16a}.  Still various more empirical materials science considerations can be brought to bear in the search for these states of matter.  At the most basic level, the design considerations are similar to those for topological insulators.   One is looking for materials with the appropriate crystal structures, with the heavy elements and the required energetics of the valence and conduction orbitals that give overlapping bands (in the case of WSM and DSM) or bands gaps with inversion (in the case of topological insulators).

Generally the band gap's magnitude and sign depends on both the atomic number ($Z$) (which changes the ordering of bands via spin-orbit coupling) and the electronegativity difference ($\Delta E_n$) of constituents.  Small electronegativity difference tends to decrease band gaps because for small $\Delta E_n$, orbital overlap increases.  Materials with strongly electronegative atomic bonds are associated with wide bandgaps.    Therefore $Z/\Delta E_n$ is perhaps a good figure of merit for band overlap between valence and conduction bands as large $Z$  and small $\Delta E_n$ increases the tendency for bands to be inverted.  One can see the role of electronegativity difference for instance in a comparison between Na$_3$Sb and Na$_3$Bi.   Both have closed shell configurations with six valence electrons, but the electronegativity of Bi is smaller than Sb, which in part makes Na$_3$Bi a DSM rather than a semiconductor like Na$_3$Sb \cite{Ettema00a}.  It is also beneficial (but not required) to have direct gaps at an odd number of points in the Brillouin zone, such as at the $L$ points in Bi$_{1-x}$Sb$_x$ \cite{Fu07b}, or at the $\Gamma$ point (particularly in inversion symmetric systems).

\begin{figure}[htp]
\includegraphics[width=0.97\columnwidth]{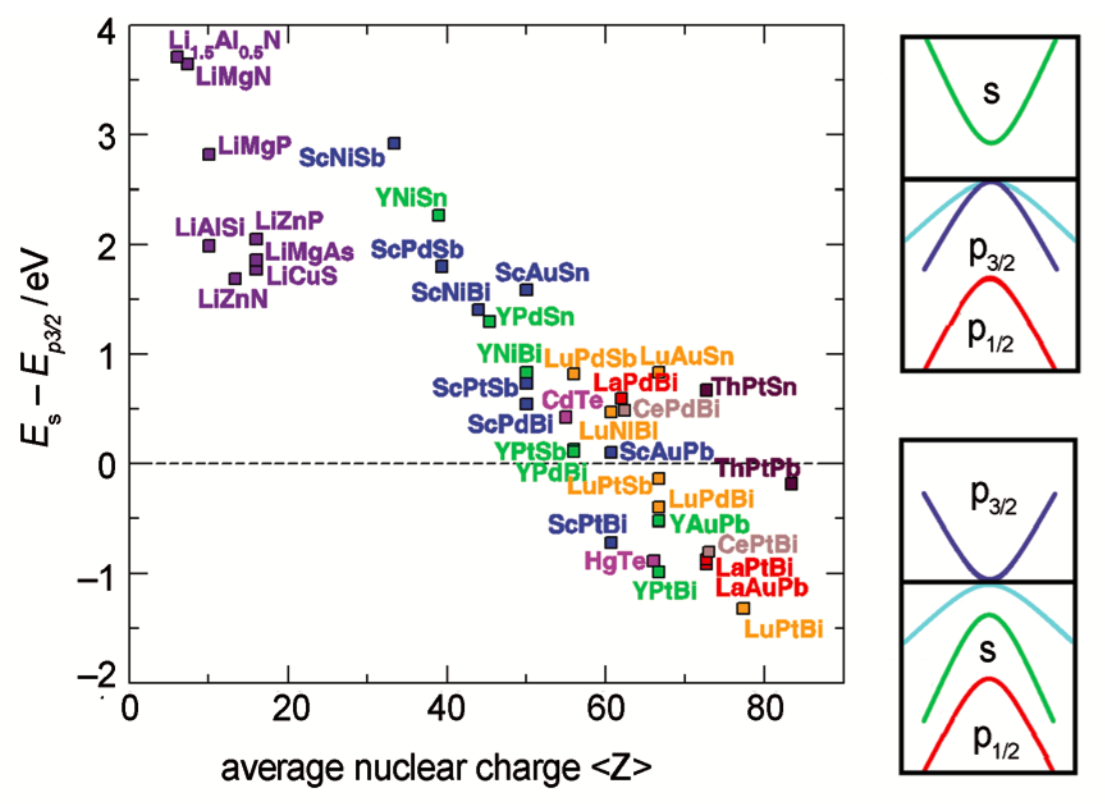}
\caption{(Left) Calculated band gaps as a function of the average nuclear charge $<Z>$ for various half-Heusler phases.  (Right) Schematic band ordering of normal and inverted bands in cubic semiconductors.  From Ref. \cite{Muchler12a}. }
 \label{HalfHeusler}
\end{figure}

The general ideas are nicely illustrated in the XYZ half-Heusler class of compounds.   These are materials that may be described as a tetrahedral zinc blende-like YZ$^{-n}$ structure, the charge of which is compensated by a slightly ionic X$^{+n}$ species, giving three interpenetrating fcc lattices \cite{Jung00a}.  For instance, the 8 valence electron compounds such as LiMgN can be written as Li$^+$ + (MgN)$^-$.  The (MgN)$^-$ forms a zinc blende lattice and is isoelectronic to Si$_2$.  Due to their ternary nature, these materials are an extremely tunable class of materials that show a vast array of interesting semi-conducting and semimetal (and even superconducting \cite{Goll08a}) behavior with broad applications potential \cite{Chen13Heusler,Casper12a}.  Typically the least electronegative element  (usually a main group element, a transition metal, or a rare-earth) is listed first as X.  The half-Heusler phases generally crystallize in a non-centrosymmetric structure corresponding to the space group $F\bar{4}3m$.   Their band gaps can be tuned over a wide energy range by choosing different XYZ combinations and have been proposed to host a variety of topological states \cite{Chadov10a,Lin10a,Xiao10a,al2010topological}.

In the LiYZ compounds, the calculated gap size is larger for compounds with a large Pauli electronegativity difference of the Y and Z species  \cite{Kandpal05a}.  Similar trends are seen in other series.  Fig. \ref{HalfHeusler} shows the calculated energy difference $E_{\Gamma6} - E_{\Gamma8}$ ($E_{s} - E_{p3/2}$) between the  $s-$ and $p-$ symmetry bands as a function of the average nuclear charge $Z$ for many different half-Heuslers.   For  $Z \approx 65 $  the bands are predicted to be inverted from the conventional (e.g. CdTe and GaAs) ordering.  Note that even with band inversion, a TI state is only formed in these materials by breaking the cubic symmetry of these compound by e.g. putting them in a quantum-well structure, under strain, or in a film geometry.   Otherwise at charge neutrality the chemical potential sits inside the $p_{3/2}$ manifold forming a zero gap system with a quadratic band touching.   Systems that could be tuned to precisely the band inversion point (possibly by substitution and mixing variants on either side of the inversion) are described by the Kane model, which can be related to the Dirac equation \cite{Orlita14a,Kane57a}\footnote{The Kane model was originally used to describe tetrahedrally bonded cubic semiconductors and describes at the transition point linearly dispersing Dirac-like bands that are bisected by a quadratic band.}.  Note that in these half-Heusler systems rare-earth metals can be readily introduced as the $f$ states of the rare-earth are strongly localized and do not change the gross scheme of the electronic structure.  Half-Heuslers with rare earths readily show magnetic effects \cite{Canfield91a}.  For instance, GdPtBi shows a Neel transition around 9 K \cite{Suzuki16a}. With the quadratic band touching of the inverted system, it is believed that GdPtBi can be tuned into a Weyl state under applied magnetic field \cite{Hirschberger16a,Suzuki16a,Cano16a} (possible enhanced by the effects of the Gd moments \cite{Shekhar16a}).

\subsection{Weyl semimetals}

\subsubsection{Non-centrosymmetric Weyl semimetals}   Based on the above general ideas, one can set out some general design considerations when considering materials that may exhibit a WSM phase.   As discussed above, the appearance of a WSM phase is possible only if the product of parity and time reversal is not a symmetry.  One wants a material that is close to a band inversion transition and which breaks either $\mathcal{T}$ or $\mathcal{P}$ symmetry.   However, unlike the case of some Dirac systems the existence of Weyl nodes is accidental which can make a systematic search for them challenging.   Moreover, because the band touchings can occur at a generic momentum positions they can be over looked in band structure calculations.

As discussed above, a particularly straightforward mechanism for creation of a WSM phase occurs generically in the band inversion transition between a trivial and topological insulator if the material's space group breaks inversion symmetry  \cite{Murakami07a,Murakami16a}.  On the approach to the band gap inversion transition, the material becomes either a (i) Weyl semimetal or a (ii) nodal-line semimetal for an extended region of parameter space, but there is no direct transition between the two insulating states.   The symmetry of the space group and the wavevector where the gap closes uniquely determine which possibility occurs \cite{Murakami16a}.  In  case (i), the number of Weyl node pairs produced at the band inversion ranges from one to six depending on symmetry.  In (ii) (as discussed below) the nodal line is protected by a mirror symmetry.   \onlinecite{liu2014weyl} proposed to realize a WSM in this fashion in LaBi$_{1-x}$Sb$_x$Te$_3$ and LuBi$_{1-x}$Sb$_x$Te$_3$  by doping close to the band inversion transition for a range of dopings near $x \sim 38 - 45 \%$.

\begin{figure}[htp]
\includegraphics[width=0.82\columnwidth]{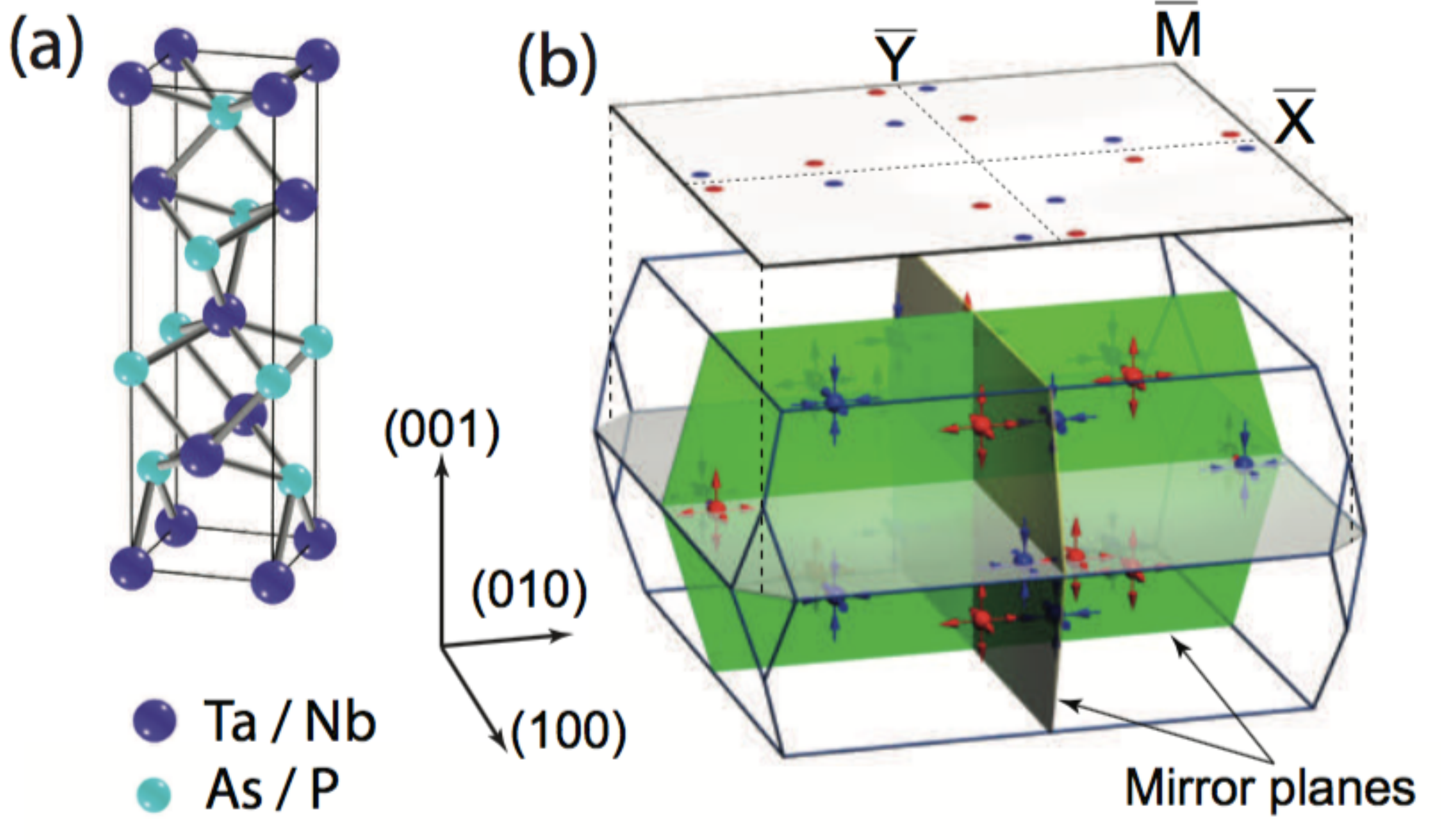}
\caption{(a)  Crystal structure (space group I4$_1md$, No. 109) of the non-centrosymmetric lattice in the TaAs-family of compounds.  It is a body-centered tetragonal structure consisting of interpenetrating Ta and As sublattices, where the two are shifted by $(a/2,a/2,\sim c/12)$.  For TaAs the lattice constants are $a = 3.437$ $\AA$ and $c = 11.656$ $\AA$.
(b) The first Brillouin zone showing twelve pairs of Weyl points.  The red and blue spheres (identified by arrows in black and white) represent the Weyl points with $C = \pm 1 $ chirality.  Note that this pattern of node chiralities may represent the situation more appropriate for the Nb compounds \cite{Belopolski16a,Huang15a,lee2015fermi}.   See discussion in text below.  From \cite{Yan16a}.}
 \label{TaAsxtal}
\end{figure}

This inversion symmetry breaking mechanism may be seen nicely in the pressure tuned transition in Pb$_{1-x}$Sn$_x$Te.  The inversion symmetry broken Pb-based rocksalts have been identified as topological crystalline insulators with surface states protected by mirror symmetry \cite{Ando15a}.  Pb$_{1-x}$Sn$_x$Te has an insulator-to-metal transition at approximately 12 kbar that is believed to be a band closing transition occurring at the L points of the Brillouin Zone.   The metallic phase is stable until about 25 kbar and is reasonably interpreted as an intermediate Weyl phase occurring between topological and trivial regimes \cite{Liang16a}.

\begin{figure}[htp]
\includegraphics[width=0.7\columnwidth]{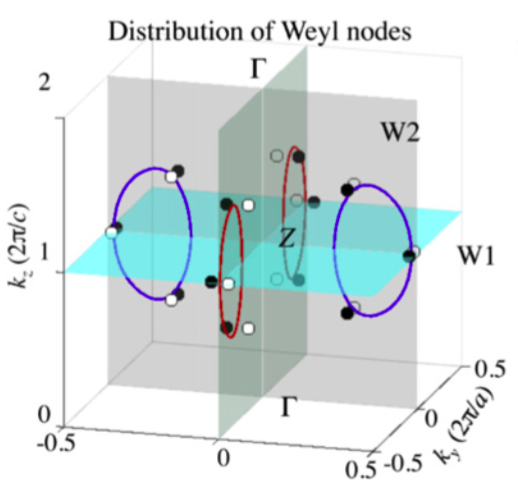}
\caption{Nodal structure of TaAs.   Without SOC, the conduction and valence bands would intersect on indicated 4 closed nodal lines on the $k_x = 0$ (online: red) and $k_y = 0$ (on line: blue) mirror planes.  SOC has the effect of gapping the bands on the mirror planes, but giving nodes slightly displaced from the planes. Black and white denote chiralities of the nodes.  From Ref. \cite{Hasan15a}.}
 \label{TaAsMirrors}
\end{figure}

\begin{figure}[htp]
\includegraphics[width=0.7\columnwidth]{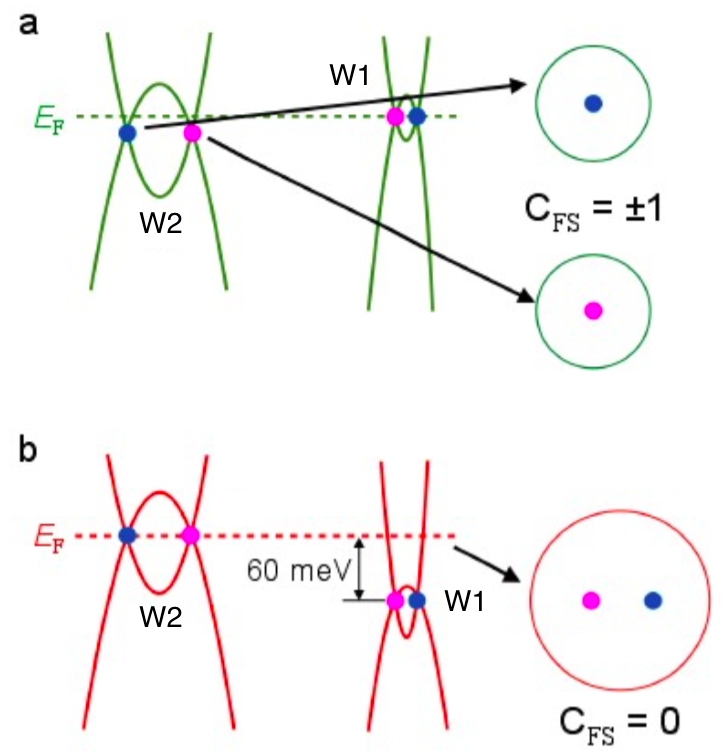}
\caption{Energy dispersions for W1 and W2.  a.)  For TaAs the Fermi surfaces enclose single chiral Weyl nodes and the Fermi Chern numbers (C$_{FS}$) are $\pm1$. b.)  For TaP, one Fermi surface is believed to enclose two W1 Weyl nodes with opposite chirality and hence the C$_{FS}$ is zero. \cite{Xu15c}.  Note that the labeling of W1 and W2 in this figure has been changed with respect to how it appeared originally in the literature so as to make it consistent with the convention used elsewhere in the literature and in the text.}
 \label{TaP}
\end{figure}

A large number of materials that are WSMs through the inversion symmetry breaking mechanism have recently been predicted and discovered.   \onlinecite{Weng2015a,Huang15a} predicted that TaAs,TaP, NbAs and NbP are materials in the Type I class (as discussed above) of WSM.  Although such systems are predicted to have 24 Weyl points, this family of materials are completely stoichiometric without any additional doping, external strain or pressure needed to fine tune the state.   Signatures of the Weyl state were seen by ARPES in TaAs and related materials soon afterwards \cite{Xu15a,Lv15a,Lv15b,Yang15a,Xu15b,Xu15c,xu2015experimental}.   These experiments will be discussed in more detail below.  Symmetry provides a helpful route in organizing and understanding the origin of nodes in WSMs.  There are 24 Weyl nodes in the bandstructure of the TaAs class of compounds \cite{Yan16a}.  The TaAs structure (Fig. \ref{TaAsxtal}) has two mirror planes M$_x$ and M$_y$, $\mathcal{T}$, and a (non-symmorphic) C$_4$ rotation symmetry.   Considering for instance, a Weyl point at ($k_x$,$k_y$,$k_z$) with chirality $C = 1 $.   Each mirror operation taken by itself reverses the chirality and gives  Weyl points with  $C = -1 $ at (-$k_x$,$k_y$,$k_z$)  and ($k_x$,-$k_y$,$k_z$).   Performing two simultaneous mirror operations gives a Weyl point at (-$k_x$, -$k_y$,$k_z$) with $C = 1 $.  $\mathcal{T}$ preserves chirality and gives the four time reversed partners of these 4 Weyl points at  ($\pm k_x$,$\pm k_y$,$\pm k_z$).  Finally these eight Weyl points at ($\pm k_x$,$\pm k_y$,$\pm k_z$) produce another eight partners at ($\pm k_x$,$\pm k_y$,$\pm k_z$) when considering the C$_4$ rotation, which also maintains chirality.  Band structure calculations show that there are two groups of points, one (labeled W1) that is in either the $k_z$ = 0 plane or on the BZ face (depending on the size of the nodal loop, see below) and the other at a non-symmetric intermediate $k_z$ (labeled W2)\footnote{Note that the literature is inconsistent with regards to which sets of Weyl points in the TaAs class are labeled W1 and which are labeled W2.   We have chosen and used figures that has used the convention that the poorly-separated points with $k_z = 0, 2\pi/c$ are labeled W1.  In this regard, we have changed the labeling of Fig. \ref{TaP} from how it originally appeared in the literature to be consistent with this scheme.}.

The Weyl points form through the interplay of mirror symmetries and SOC.   As shown Fig. \ref{TaAsMirrors}, in the absence of SOC, the conduction and valence bands would intersect on 4 closed nodal lines in the $k_x = 0$ and $k_y = 0$ mirror planes.   The addition of SOC gaps the bands on the mirror planes, but creates degeneracies at points slightly displaced from the planes.   

With the above considerations one gets eight W1 Weyl points and sixteen W2 points.   Although all materials in this class have the same general band structure, their different energetics, lattice constants, and SOC can lead to differences in their topological strutures.   For instance, compared to TaAs \cite{Arnold16b}, in TaP \cite{Arnold16a,Xu15c} the bulk pairs of the W2 Weyl nodes, which are well-separated in momentum space are located near the chemical potential while the poorly-separated ones W1 are 60 meV below the chemical potential and so are enclosed by a single Fermi surface.   This gives for the W1 points a Fermi Chern number (the net topological charge enclosed by a Fermi surface) for TaP of zero in a manner shown in Fig. \ref{TaP} \cite{xu2015experimental,Xu15c}.  There may also be differences in the $k_z$ position of W1 between the Nb and Ta compounds.   In some calculations \cite{Belopolski16a,Huang15a,lee2015fermi}, the nodal loops of the Ta compounds are predicted to be smaller making them not extend from one BZ to the next.   This means that for the Ta compounds the W1 point should be on the z axis BZ face (see Fig. \ref{TaAsMirrors}), whereas for the Nb materials the W1 point is on the $k_z = 0$ plane (as in Fig. \ref{TaAsxtal}).   Such an effect will result in a ``chirality switching" of the W1 points in the BZ when comparing Ta and As compounds.  Again, compare Figs. \ref{TaAsxtal} and \ref{TaAsMirrors}.   However, different calculations show different results in this regard \cite{Weng2015a,Huang15a,Yan16a} and there is no resolution on this point.  There may be an extreme sensitivity to the lattice constants used.  Moreover, note that although ARPES has in general been very powerful in finding topological band structures, it may not generally have the energy and momentum resolution to see make well defined statements about chirality of Fermi surfaces.   As discussed below, quantum oscillation experiments can be very useful in this regard \cite{Arnold16b}.

Despite these successes, the search for a more ideal family of WSM materials continues.  As noted, the 24 Weyl nodes in the TaAs family of compounds comes in two non-symmetry equivalent sets \cite{Weng16a}, which with their large number and possible energy offsets give rise to potentially complicated transport and spectroscopic properties.   Moreover, in the TaAs material class, all Weyl physics occur in a narrow range of energies. This requires careful material preparation to ensure the Fermi level falls in this range.  In this regard, Weyl semimetals with larger characteristic energy scales are desirable.  There has been an ongoing search for simpler Weyl semimetals with the minimum 4 Weyl nodes at the Fermi level for inversion breaking systems that have preferably large momentum and energy separations from other bands.  It has been proposed that HgTe and half-Heusler compounds under compressive strain will realize a near ideal Weyl semimetal with four \textit{pairs} of Weyl nodes \cite{ruan2016symmetry} near the Fermi energy.

More recently, Type II WSMs which have strongly tilted Weyl cones have been proposed to exist in the layered transition-metal dichalcogenides WTe$_2$ \cite{Soluyanov15a}, MoTe$_2$ \cite{Wang16b,Sun15b} and their alloys Mo$_x$W$_{1-x}$Te$_2$ \cite{chang2016prediction}.  Such materials are believed to have band structures very sensitive to strain and pressure, which may make experimental identification difficult.   \onlinecite{Sun15b} predicted four pairs of Weyl points in the k$_z = 0$ plane.   Calculations using only slightly smaller lattice parameters showed that two pairs were annihilated by merging along the $\Gamma- X$ line leaving only two pairs of Weyl points \cite{Sun15b}.  Experiments on these materials will be discussed below.    There have also been proposals and experimental claims for Type II WSMs in LaAlGe \cite{chang2016theoretical,xu2016discovery}.

\subsubsection{Magnetic Weyl semimetals} An ongoing search has been for materials that are good examples of a WSM through the $\mathcal{T}$ breaking mechanism.   The first proposed WSM was of this class in the magnetic pyrochlores A$_2$Ir$_2$O$_7$ \cite{Wan11a} (where A =Y or a rare earth element Eu, Nd, Sm, or Pr).  Ab initio (LDA+U) calculations predicted an `all-in, all-out' (AIAO) magnetic structure, which is an unusual Ising ordering that preserves cubic symmetry but breaks $\mathcal{T}$. Depending on the strength of correlations $U$, a Mott insulating phase with this magnetic structure or Weyl semimetal phase with 24 Weyl nodes (all at the Fermi energy) were predicted \cite{Wan11a}. The role of the chemical species A was related to the strength of correlations \cite{WeylResistivityMaeno}, with larger ionic radii (such as A=Pr) implying weaker correlations.  An extra complication is that in addition to the electrons on the Ir sites, local moments on the rare earth atoms order but at a lower temperature. In this regard the A=Y, Eu are the simplest in not having $f$ shell moments.

Experimentally, several pyrochlore iridates are observed to undergo magnetic ordering, and experiments have now confirmed that the order is of the predicted AIAO form \cite{Sagayama13,Disseler,Guo16,Donnerer16}. The magnetically ordered pyrochlores, such as A=Eu, Nd are seen, in clean samples, to be insulating at low temperatures \cite{Ueda12,Tian16,EuIridateExperiments,Ishikawa12,ueda2015magnetic}.  The metallic phase above the magnetic ordering temperature is expected have quadratic doubly degenerate bands that touch at the $\Gamma$ point \cite{Krempa12,Kondo15a,Nakayama16}, a state we term a \textit{Luttinger} semimetal \cite{Abrikosov71a,Moon13a}.  In A=Pr, which remains a nonmagnetic metal down to low temperatures \cite{machida2010time,nakatsuji2006metallic}, a quadratic band touching has been found recent experiments \cite{Kondo15a} (See Sec. \ref{Quad} for further discussion of the Luttiinger semimetal). No direct evidence for Weyl nodes has been found so far in stoichiometric pyrochlore iridates.  However, it is expected that a weak AIAO order imposed on the Luttinger semimetal would lead to Weyl nodes that move through the BZ as the order parameter increases, eventually annihilating leading to an insulating state. Given the continuous magnetic ordering transition observed in the pyrochlore iridates, an intervening Weyl phase should occur just below the ordering temperature.  The temperatures involved are relatively low, compared to the band structure scales, so one should still be able to distinguish a Weyl semimetal band structure. This WSM remains to be found experimentally \cite{Nakayama16} and some numerical studies indicate a direct first order transition between metallic and insulating states \cite{Shinaoka15}.  Experimental evidence for first-order-like behavior was reported in \onlinecite{UedaPressure15a}.  However, proximity of a Weyl phase is indicated in other experiments including A=Eu under pressure \cite{EuIridateExperiments}, and alloying with Rhodium (that substitutes for Iridium) leads to a state where linear in frequency optical conductivity is observed as expected in a  WSM \cite{Ueda12}. A recent comparative study of optical conductivity of pyrochlore iridates with different $A$ ions, is contained in Ref. \cite{Ueda16}.    It has also been shown in mixed Nd-Pr materials that the application of a [001] magnetic field decreases the resistivity and produces a unique Hall response and for field parallel to [111] the resistivity exhibits saturation at a relatively high value typical of a semimetal \cite{ueda2017magnetic}.  Due to uniaxial magnetic anisotropies of these cubic materials, different field directions can drive different magnetically ordered states (all-in all-out to 2-in 2-out for a [100] field and three-in three-out for [111]).  The observed resistivity changes have been interpreted as the emergence of different WSMs with varying numbers of Weyl points and line nodes in respective spin configurations.

The only metallic member in the pyrochlore iridate,  Pr$_2$Ir$_2$O$_7$, is a semi-metallic cousin of the quantum spin ice candidate Pr$_2$Zr$_2$O$_7$ \cite{kimura2013quantum} and is believed to have a low temperature chiral spin liquid phase, where a spontaneous Hall effect is observed \cite{machida2010time}  Its quadratic band touching has been proposed to convert to a magnetic Weyl semimetal phase due to other unconventional broken symmetry states at low temperatures such as spin ice-like  \cite{goswami2017competing} or quadrupolar order \cite{Lee2013a}.

The presence of Weyl nodes in the bulk implies Fermi arc surface states not only on surfaces, but also on magnetic domain walls across which the chirality of Weyl nodes switch. The domain walls are therefore expected to be conducting - and it was argued in \onlinecite{Yamaji14} that the metallic character of domain walls may survive even when the Weyl nodes have annihilated to form a bulk insulator. Indeed such metallic domain wall conduction was reported in A=Nd pyrochlore iridates \cite{UedaAnomalous14a,Fujita16,Ma15} indicating proximity to a Weyl semimetal. 
 
In another family of materials, it was proposed that YbMnBi$_2$ may be an example \cite{Borisenko15a,Chinotti16a} of a magnetic Weyl semimetal.  Materials in this AMnBi$_2$ family are expected to host highly anisotropic Dirac dispersions with a finite gap at the Dirac point due to SOC from first principles DFT band calculations.  Unfortunately, it appears that the canting of the magnetic moment (10$^\circ$) from the $c$ axis that is believed to be required to split the DSM degeneracies, does not exist \cite{Wang16g,Chaudhuri16a}.  Thus YbMnBi$_2$ is likely a DSM (perhaps with a small mass).

Recently, \onlinecite{Wang16c,chang2016room} proposed candidates for magnetic Weyl semimetals based on the Co-based magnetic Heusler compounds  XCo$_2$Z (X=V,Zr,Ti,Nb,Hf, Z=Si,Ge,Sn), VCo$_2$Al and VCo$_2$Ga.  For spontaneous magnetization along the [110] direction (confirmed by experiment) they predict only two Weyl nodes that are formed by bands of opposite C$_2$ eigenvalues.  Importantly, these nodes are predicted to be near the Fermi level and separated by distances of order the BZ size.  The antiferromagnetic half Heusler compounds GdPtBi and NdPtBi have been predicted to be magnetic Weyl semimetals \cite{Shekhar16a,Suzuki16a,Hirschberger16a} under applied magnetic field.

 Of particular note is the recent work on Mn$_3$Sn and Mn$_3$Ge, which are \textit{antiferromagnets} (AFs) with a non-collinear 120-degree spin order that exhibit a large anomalous Hall conductivity \cite{Nakatsuji15a,Kiyohara16a}.  They have been predicted \cite{Yang17a,ZhangY17a} to be WSMs with several Weyl points as well as trivial bands near the Fermi level.  As discussed above, systems with a combined $\mathcal{PT}$ symmetry are constrained to have doubly degenerate bands and therefore while  $\mathcal{PT}$ symmetric AFs may be DSMs, they cannot be WSMs.  The Mn$_3$Sn class is nonsymmorphic with either mirror reflection $M_y$ or inversion plus a half-lattice translation being a symmetry and hence like all WSM AFs they break $\mathcal{PT}$ symmetry.  Although there are similarities in the band structures of Mn$_3$Sn and Mn$_3$Ge, it is predicted that Mn$_3$Sn has fewer Weyl points (3 families of symmetry equivalent points vs. 9).   Paradoxically this may be because its stronger SOC leads to many of the Weyl points annihilating each other \cite{Yang17a}.  In accord with what is expected for WSMs with broken ${\mathcal T}$,  these systems show a large anomalous Hall effect \cite{Yang17a,ZhangY17a,kubler2014non} that is naturally explained by the Berry phase mechanism in these systems.   Note that despite the fact that these systems are believed to have a small net magnetic moment of 0.005 $\mu_B$ per unit cell, the small moment has the effect of only moving the position of the Weyl nodes slightly from their positions in the ideal AF structure.   It is not believed to give an appreciable contribution to the anomalous Hall effect.   Also note that these materials exhibit their anomalous Hall effect up to temperatures near Neel transition near 400 K \cite{Nakatsuji15a,Kiyohara16a}.  This suggests that magnetic Weyl fermions can be available at room temperature and beyond and this family of materials may be useful in spin-current conversion and domain wall effects, which are a topical subject in the field of antiferromagnetic spintronics \cite{Smejkal17a}.

With regards to predicting phases, \onlinecite{Turner12,Hughes11a,Wang16c} give a particularly simple diagnostic to determine if an inversion symmetric system with broken $\mathcal{T}$ hosts an odd number of pairs of Weyl nodes.   One takes the product of the inversion eigenvalue $\zeta_n(K_i)$ of all bands $n$ below the Fermi level at all inversion symmetric points $K_i$.  
\begin{equation}
\prod_{\vec{K}_i = - \vec{K}_i} \prod_{\text{n bands with} \; E_n (K_i)< E_f} \zeta_{n}(K_i) 
\end{equation}
If this number $=-1$ then an odd number of pairs of Weyl points must exist in the bulk.  If this number is $1$ then it is not possible with only inversion symmetry to deduce if a nonzero number of pairs exist.

Two other interesting (and thus far unrealized) proposals that are mentioned above are to create Weyl (and presumably Dirac states as well) states via growing heterostructures of magnetically doped 3D TIs and normal insulators \cite{Burkov11a}, inversion symmetry broken heterostructures \cite{Halasz12a}, and superlattices of alternating layers with odd and even parity orbitals \cite{Das13a}.   There have also been proposals to create a WSM in strained Hg$_{1-x-y}$Cd$_x$Mn$_y$Te films \cite{bulmash2014prediction}.   Here a small applied field would align the Mn spins and if the system is Cd doped to be close to the band inversion transition, the system is expected to result in a WSM with two Weyl points near $E_F$.

\subsection{Dirac semimetals}

With regards to DSMs, the most straightforward manifestation of this state will be found at the phase boundary between a topological insulator and a trivial one when the crystal structure preserves both inversion and $\mathcal{T}$.  The expectation is that the alloying of known topological insulators with lighter elements by tuning spin orbit coupling or lattice constant can cause the bulk bandgap to close and invert at a quantum critical point where the topological class changes.  This physics has been investigated in both the TlBiSe$_{2-x}$S$_x$ \cite{Sato11a,Xu11a,Souma12a} and Bi$_{2-x}$In$_x$Se$_3$ \cite{Brahlek12a,Wu13a}, where evidence of topological phase transition as a function of $x$ between two phases have been found.  In each of these cases, one end member of the series is a topological insulator and the other is a trivial insulator.

Although such systems should in principle show all the physics of a DSM, as discussed above, one would like to identify phases that don't require (or are susceptible to) fine tuning.    Moreover, except for materials like SrSn$_2$As$_2$ \cite{Gibson15a} and ZrTe$_5$ \cite{Li16a,Yuan15a} which are materials believed to be exist naturally near the critical point between TI and non-TI, all the above non-stochiometric systems may exhibit additional effects due to disorder.   For instance Bi$_{2-x}$In$_x$Se$_3$ crystals appears to show a tendency for In segregation in single crystal form \cite{Liu13a}, although thin films seem to be largely free of such effect \cite{Brahlek12a,Wu13a}.  However, as discussed in Sec. \ref{DiracSection}, the conditions for a stable 3D DSM as a $phase$ are rather specific.  In order to not form a gap, it is essential that crossing bands belong to different irreducible representations of the double group \cite{Elliott54a} along a line of symmetry.  Therefore all crystal structures are not equally likely to host DSMs \cite{Gibson15a}.  For example, in the tetragonal group I4/mcm, the Hamiltonian at a general wavevector along the $c$ axis has C$_{2v}$  symmetry.  As the  C$_{2v}$ double group only has one irreducible representation, all states have the same symmetry and so such tetragonal crystals cannot host a DSM.  In contrast, in the tetragonal space group P4/mmm,  degeneracies along the $c$ axis are protected by C$_{4}$  symmetry and a Dirac point is allowed.  The $c$ axes in canonical Dirac systems Cd$_3$As$_2$ (I4$_1$/acd) and Na$_3$Bi (P6$_3$/mmc) have C$_{4}$  and C$_{6}$  that support double groups with multiple irreducible representations and hence a DSM is allowed.  ARPES experiments have now confirmed both Na$_3$Bi \cite{Liu14b,Xu15ab} and Cd$_3$As$_2$ \cite{Liu14a,Borisenko14a,Neupane14a} as DSMs.  Due to the single irreducible representations of the double groups in orthorhombic, monoclinic, or triclinic space groups Dirac semimetals are not possible in such crystal systems.

However, as discussed above, except in the case of the symmetry enforced states, symmetry alone does not definitely predict the presence of a DP.   Notably the same minimal models that describe the Dirac point physics in these materials are used to describe topological insulators in layered materials in the Bi$_2$Se$_3$ family.  In that case the $c$ axis has a C$_{3v}$ symmetry, but different symmetries are allowed for the bands.  However for the TI the low energy bands have common eigenvalues under C$_3$ rotations and an avoided crossing occurs forming a topological insulator and not a DSM.   Thus it is hard to make definitive statements, but one wants to look for materials with heavy elements and the required overlap of the valence and conduction bands that have hexagonal, rhombohedral, tetragonal, or cubic symmetry.

\begin{figure}[htp]
\includegraphics[width=0.9\columnwidth]{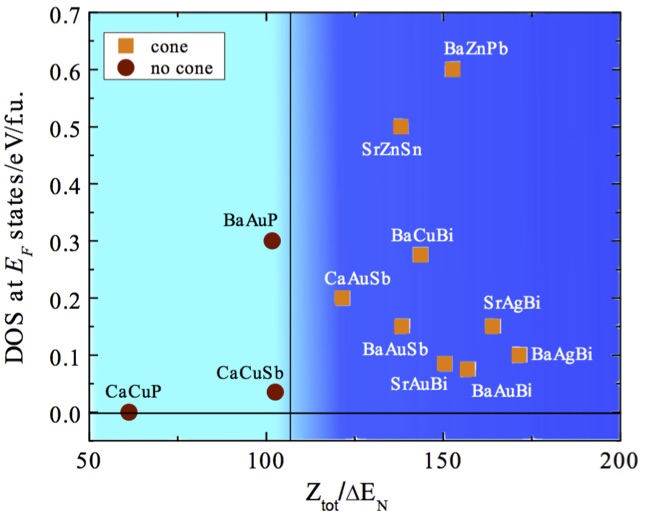}
\caption{The electronic phase diagram of the ZrBeSi family, showing the calculated density of states at E$_F$ as a function of the total $Z$ (atomic number) divided by the electronegativity difference between the large cation (e.g. Ba$^{+2}$) and the average of the anionic honeycomb sublattice (e.g. Ag$^{+1}$Bi$^{-3}$).  The squares (online: orange) represent compounds having a calculated Dirac cone and circles (on line: red) represent compounds with none.  Calculations were performed with a Perdew-Burke-Ernzerhof parameterization of the generalized gradient approximation functional.  From Ref. \cite{Gibson15a}.}
 \label{GibsonPD}
\end{figure}

Within a particular material class trends can be followed that allow one to predict the presence of Dirac semimetals independent of explicit calculations.  Fig.  \ref{GibsonPD}  \cite{Gibson15a} shows how the calculated electronic properties of ZrBeSi-type compounds change as a function of the total atomic number ($Z$) divided by the Pauling electronegativity difference ($\Delta E_n$).  This is the figure of merit for band inversion discussed above.  ZrBeSi-type compounds have a hexagonal (P6$_3$/mmc) space group (the same as Na$_3$Bi) with a very simple crystal structure, with layers of BeSi hexagonal net separated by large cations.  Similar to the above discussed half-Heusler compounds, it is found empirically that once this $Z/\Delta E_n$ metric reaches a certain value, band structure calculations predict that these materials will exhibit a Dirac cone in their band structure structure.   Note that not all these materials are equally good candidates as their near-E$_F$ density of states is predicted to vary widely due to presence of other near-E$_F$ bands.  An ideal DSM would have a near E$_F$ density of states of zero.

\begin{figure*}
\includegraphics[width=1.5\columnwidth]{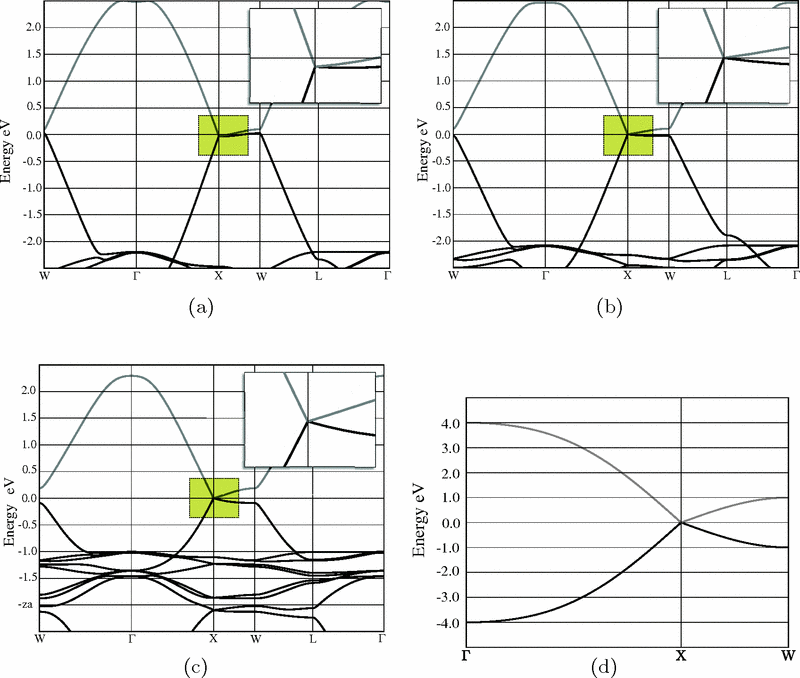}
\caption{Band structures of (a) ${\rm AsO_2}$, (b) ${\rm SbO_2}$, (c) ${\rm BiO_2}$ in the $\beta$-crystobalite structure and (d) the Hamiltonian of Eq. \ref{modelHam} for spin-orbit coupled $s$-states on the diamond lattice. All the spectra feature a symmetry enforced Dirac point in an FDIR at the zone boundary $X$ point.  From Ref. \cite{Young2012}. }\label{Zaheerbands}
\end{figure*}

As discussed in Section \ref{SEDSM} a number of possibly competing conditions must be met in order to achieve symmetry enforced DSMs in a real material.  In addition to the symmetry constraints, one must find compounds with an odd number of electrons per formula unit in orbital states well isolated from other orbitals.  Candidate DSM materials designed to satisfy both the symmetry and the band filling requirements have been proposed for diamond lattice structures (space group 227, $Fd3m$) \cite{Young2012}. This lattice supports a DSM at half filling in the prototype $s$-state Hamiltonian of Eq. \ref{modelHam} which yields the spectrum shown in Fig. \ref{Zaheerbands}(d). Materials that form in this crystal structure are typically $sp$ bonded insulators (e.g. C, Si, ${\rm SiO_2}$, etc.) where the related Dirac singularity occurs ``buried" deep in the occupied spectrum and the band edge valence states derive from bonding combinations of the atomic $p$-orbitals.  Theoretical searches for DSMs in this crystal structure have therefore focused on strategies that decorate the lattice with species that serve the dual roles of (1) boosting the bonding-antibonding splitting in the $p$-manifold to high energy displacing them  away from the Fermi energy and (2) selecting a stoichiometry that fills the topological band precisely to the Dirac point.

Hypothetical group-V oxides ${\rm MO_2}$, ${\rm M=\{As, \, Sb, \, Bi \}}$ in the $\beta$-crystobalite crystal structure where the ${\rm O}$ atoms occupy bridging sites between fourfold-coordinated ${\rm M}$ vertices satisfy the above requirements  as shown in (Fig. \ref{Zaheerbands} (a-c)) \cite{Young2012}.  The symmetry enforced FDIR's occur at the centers of the Brillouin zone faces at the three $X$ points and in ${\rm BiO_2}$ the enhanced spin-orbit scale protects its Dirac-like dispersion near the FDIR on an experimentally useful energy scale $\sim 200 \, {\rm meV}$. Comparison of Fig. \ref{Zaheerbands}(c) and (d) show that the band structure bears a striking resemblance to spectrum of the prototypical Hamiltonian of Eq. \ref{modelHam}. Total energy calculations show that the ${\rm BiO_2}$ is a locally stable structure although it has a substantially higher energy than  the denser oxide ${\rm Bi_2 O_4}$ in the cervantite structure \cite{Young2012}.  Related and possibly more stable forms of ${\rm Bi}$-derived DSMs have also been predicted for quarternary compounds in the family ${\rm BiBSiO_4}$ with ${\rm B=\{Zn, \, Ca, \, Mg\}}$  in a distorted spinel structure \cite{Steinberg2014}. Here the ${\rm Bi}$ species occupy two symmetry related fourfold coordinated sites in the primitive cell as in the diamond lattice, but the lattice breaks the symmetry of the spinel structure (space group 227) to form an orthorhombic body-centered lattice (space group 74 ($Imma$)).  This lower symmetry removes the fourfold degeneracy of the parent spinel lattice at two $X$ points but it retains an FDIR at a single zone boundary $T$ point. Similar to the situation in ${\rm BiO_2}$ the spin-orbit scale provided by ${\rm Bi}$ is robust and is expected to support Dirac physics on a scale of hundreds of meV \cite{Steinberg2014}.

As noted above, the nonsymmorphic character of these DSMs generally leads to situations with an odd number of electrons per formula unit and chemical species in unconventional oxidation states. A possible work around has been to explore low dimensional donor-acceptor structures where a desired band filling can be selected without compromising  structural stability.  Instead of having odd number of electrons per atom, which tends to be either chemically unstable or highly localized, one can use one electron per molecular orbital as occurs in cluster compounds.  Similar approaches have been made recently in attempts to stabilize frustrated magnets and prevent Jahn-Teller instabilities and orbital ordering \cite{Sheckelton12a}.  For example the donor-acceptor family of materials ${\rm AMo_3X_3}$ (${\rm A=\{Na,K,Rb,In,Tl\}}$, $\rm{X=\{Se,Te\}}$)  that exist in structures containing twisted stacks of ${\rm Mo_3}$ clusters with a screw axis symmetry.  Indeed the band structure features quasi-one dimensional dispersive bands along the chain direction with symmetry-enforced zone boundary contact points \cite{Zak1999} and a band filling that can can be selected to some extent by the choice of the ${\rm A}$ site cation \cite{Gibson15a}.  These are DSM candidates, although their one dimensional character inevitably suppresses band velocities in the transverse directions and may leave these materials susceptible to a transition to an insulating state by a Peierls instability. Generalizations of these considerations relating to electron filling and searches of materials' databases in an efficient way has lead to candidate Dirac semimetals \cite{Chen16a}.

In particular, main group elements are not usually found in valence states with odd numbers of electrons.  Configurations such as Bi$^{+4}$ are unstable.   Transition metals are found in odd valence states, however they are prone to various instabilities such as charge density waves in NbSe$_2$ and TaSe$_2$.   Moreover, they are typically found with a number of $d$ orbitals that are close to overlapping each other energetically.  Ir$^{+4}$ in the nonsymmorphic pyrochlore lattice has a half-filled $J_{1/2}$ orbital that is removed in energy from other $d$ orbitals and although it is proposed to be a DSM, it and many other odd electron numbered transition-metal ions  exhibit tendencies towards localized magnetism.

Another possibility to resolve the band filling problem in nonsymmorphic crystals is with intermetallic compounds.  One example is the nonsymmorphic paramagnetic metal Cr$_2$B (space group F$ddd$) that contains interlocking honeycomblike nets of Cr atoms related to each other by glide planes.   Band structure calculations reveal many bands crossing the Fermi energy, with DPs expected to be present.  It is believed that through B deficiency the DP may be brought to E$_F$ \cite{Schoop14a,Gibson15a}.   Through Hall effect experiments on polycrystalline Cr$_2$B it is believed that high mobility $n$-type carriers can be resolved in transport experiments.


\section{Experimental results} 
\label{ExperimentSection}

With the above theoretical and materials considerations, we can now turn to the considerable experimental literature on WSM and DSM systems.   Like other large material classes, characterization has benefited from the application of a  large number of different experimental techniques.   Each have their strengths or limitations in revealing aspects of the underlying physics.

\begin{figure*}[htp]
\includegraphics[width=1.5\columnwidth]{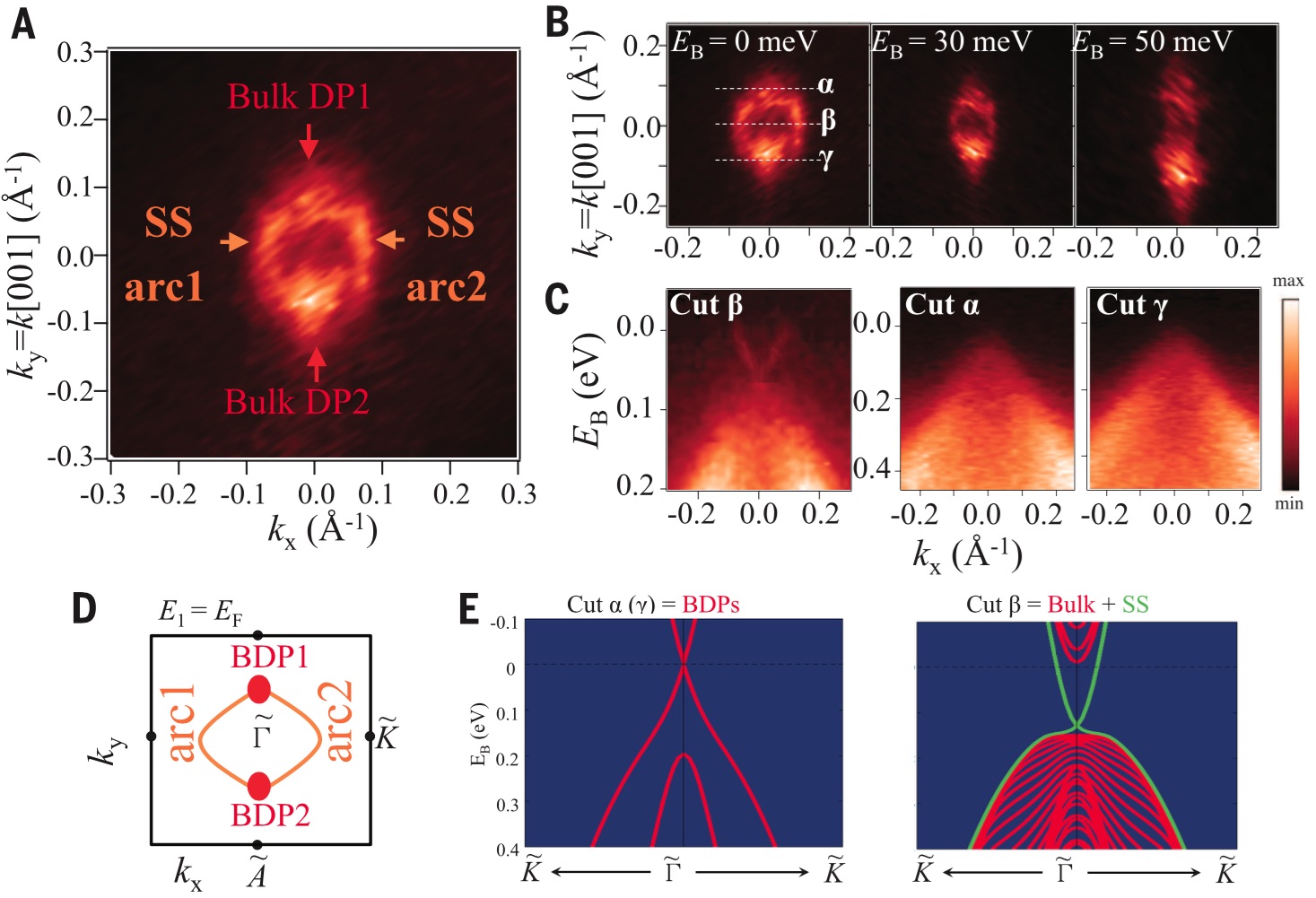}
\caption{(a)  Fermi surface map of the Na$_3$Bi sample on (100) surface at photon energy 55 eV. BDP1 and BDP2 denote the two bulk Dirac points. (b)  Constant energy ARPES spectra as a function of binding energy. (c) ARPES dispersion cuts $\alpha$, $\beta$, and $\gamma$ as defined in (b). (d) Schematic Fermi surface of Na$_3$Bi. The red (shaded) areas and the orange (gray) lines represent the bulk and surface states, respectively.  (e) Calculated band structure along cut $\alpha$ ($\gamma$)  and $\beta$.  Adapted from \cite{Xu15ab}.}
 \label{Na3BiARPES}
\end{figure*}

\begin{figure*}[htp]
\includegraphics[width=1.8\columnwidth]{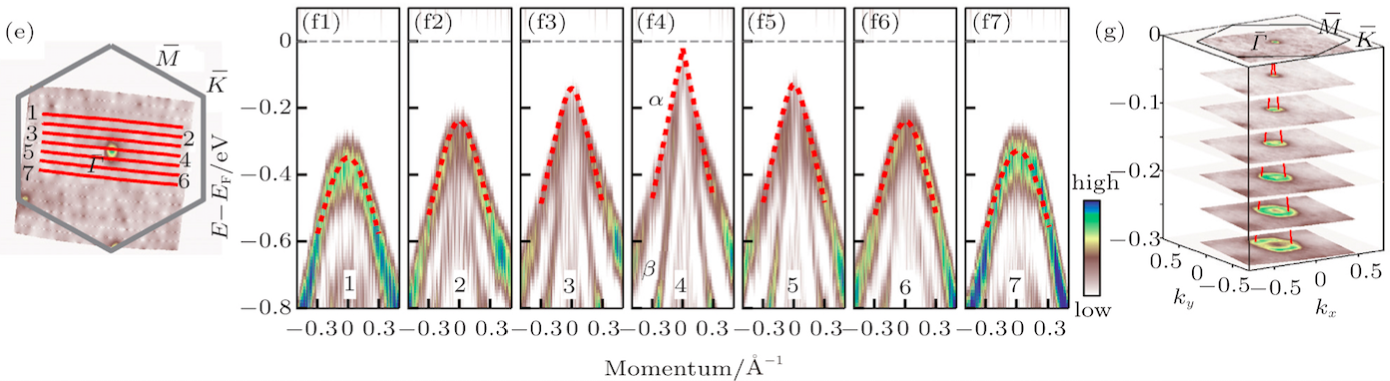}
\caption{ARPES spectra near the $\Gamma$ point on the (001) surface. (e)  Fermi surface mapping. (f) Photoemission spectra measured along the seven momentum cuts labeled as red lines 1 to 7 in panel (e). The red  (gray) dashed lines are curves fitted to the Dirac dispersion expectation. (g) 3D view of the evolution of Fermi surface and constant energy contours at different binding energies. The red (gray) lines are the guide to the eye.  Displayed images are second derivatives of the original data with respect to  energy.  Adapted from Ref. \cite{liang2016electronic}.}
 \label{Na3Bi001}
\end{figure*}

\subsection{Identifying Dirac and Weyl systems through their band structure}
\label{Identifying}

ARPES, STM, magneto-optical transport, and quantum oscillations have all proven to be very useful in determining aspects of the band structures of these materials.  In particular ARPES has emerged as a premier tool for experimentally identifying topological band structures.   Within the framework of certain accepted approximations, this technique measures the single particle spectral function as function of energy and momentum \cite{Damascelli03a}.   For the weakly interacting compounds that comprise most studied topological materials, it gives a direct measure of the band structure.  Due to momentum selection rules in the photoexcitation process, the technique is best suited for 2D materials and has played a central role in the identification of topological insulators surface states \cite{Hasan10} and in probing the physics of quasi-2D superconductors like cuprates \cite{Damascelli03a}.   Its use in 3D materials like most WSM or DSM systems requires more care, as $k_z$ sensitivity is achieved by tuning the incident photon energy.  but it has proved to be equally useful in the identification of material realizations of these states of matter.   It has the constraint of being primarily a surface sensitive probe, but if surfaces have natural cleavage planes then in many circumstances experiments can be done that are reflective of the bulk.  In the case of topological materials spin-resolved measurements have been particularly useful \cite{Hasan10}.

The first relevant ARPES experiments were on the DSM systems Na$_3$Bi \cite{Liu14b,Xu15ab} and Cd$_3$As$_2$ \cite{Liu14a,Borisenko14a,Neupane14a}.   Experiments on (100) oriented Na$_3$Bi single crystals have shown a pair of linearly dispersing three-dimensional Dirac points that are displaced from each other on the $(001)$ line passing through the $\Gamma$-point \cite{Liu14a}. On the (100) surface, the two bulk Dirac nodes are separated from each other in the (001) direction as shown in Fig. \ref{Na3BiARPES}.    At higher binding energies, the two Dirac points were found to enlarge into hole-like contours, whereas the two surface Fermi arcs shrank in an electron-like fashion.   These observations were in accord with the theoretical prediction \cite{WangA3Bi2012}.   Fig. \ref{Na3Bi001} shows ARPES spectra from the (001) surface.   The linear Dirac dispersion is clearly seen, but because the two bulk Dirac nodes project onto the same point in the surface BZ the Fermi arc surface states discussed in Sec. \ref{DiracFermiArcs} are not expected.  

For Cd$_3$As$_2$ there have been discrepancies about the details of the bandstructure.   It is believed that there are a pair of symmetry-protected 3D Dirac nodes near the $\Gamma$ point, but there have disagreements about  location, size, anisotropy, and tilt of these bands.  ARPES studies imply cones extending over a few hundred meV \cite{Borisenko14a,Neupane14a} or even up to an eV \cite{Liu14a}.  However, STM has estimated Dirac cones with energies an order-of-magnitude smaller \cite{Jeon14a}.  Most recently the magnetooptics measurements of \onlinecite{Akrap16a} have demonstrated that the band structure likely includes two types of conical features, one a high energy scales, the second on the small energy scale.  The higher energy structure can be explained within the Kane model that is widely applied to describe the band structure of cubic zinc-blende type semiconductors \cite{Kane57a}.  At low energies this cone ``splits" and Dirac fermions emerge that can be described in the context of the `Bodnar' model \cite{Bodnar77a} that retains the standard Kane parameters (band gap E$_g$, interband matrix element $P$, and spin-orbit coupling $\Delta$), while also introducing a crystal field splitting $\delta$ that in this case reflects the tetragonal symmetry of Cd$_3$As$_2$.

\begin{figure}[htp]
\includegraphics[width=0.95\columnwidth]{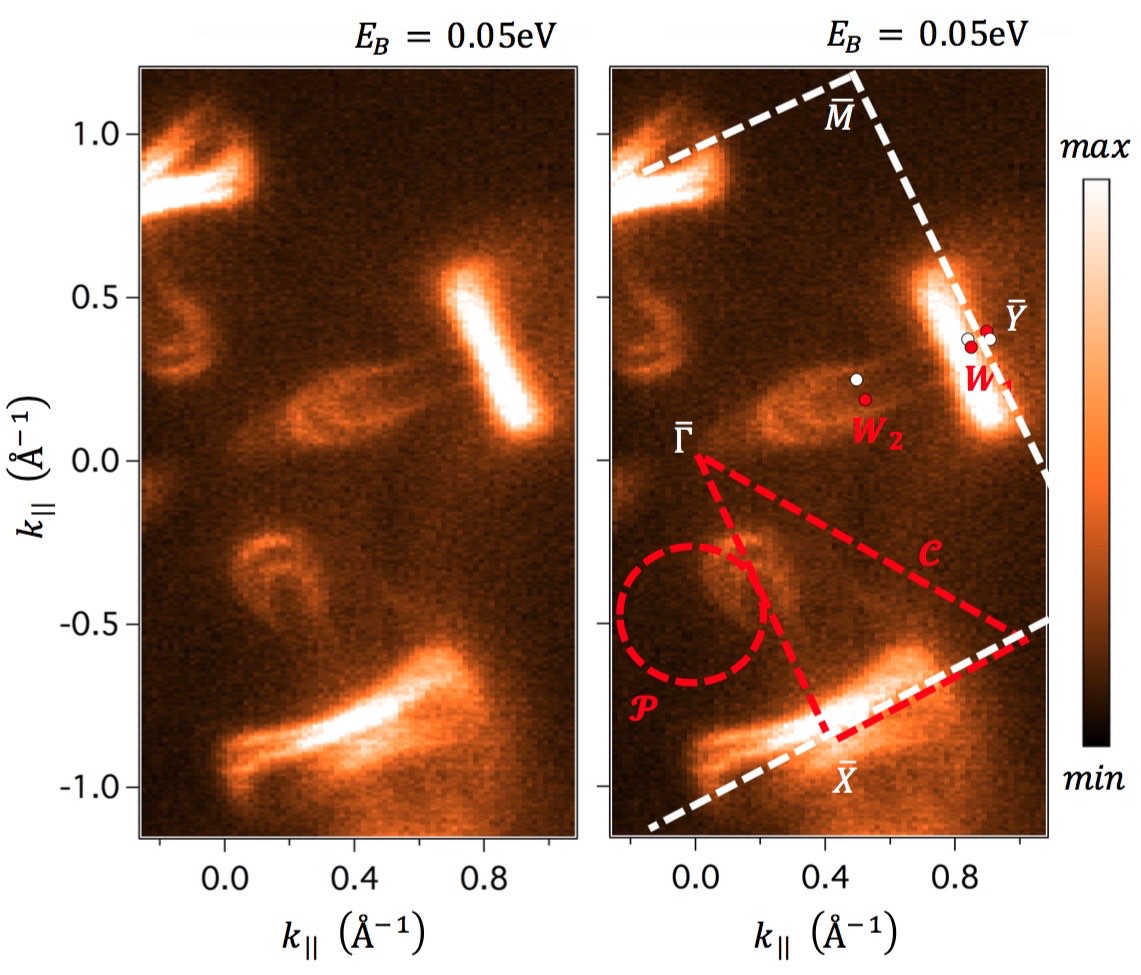}
\includegraphics[width=0.95\columnwidth]{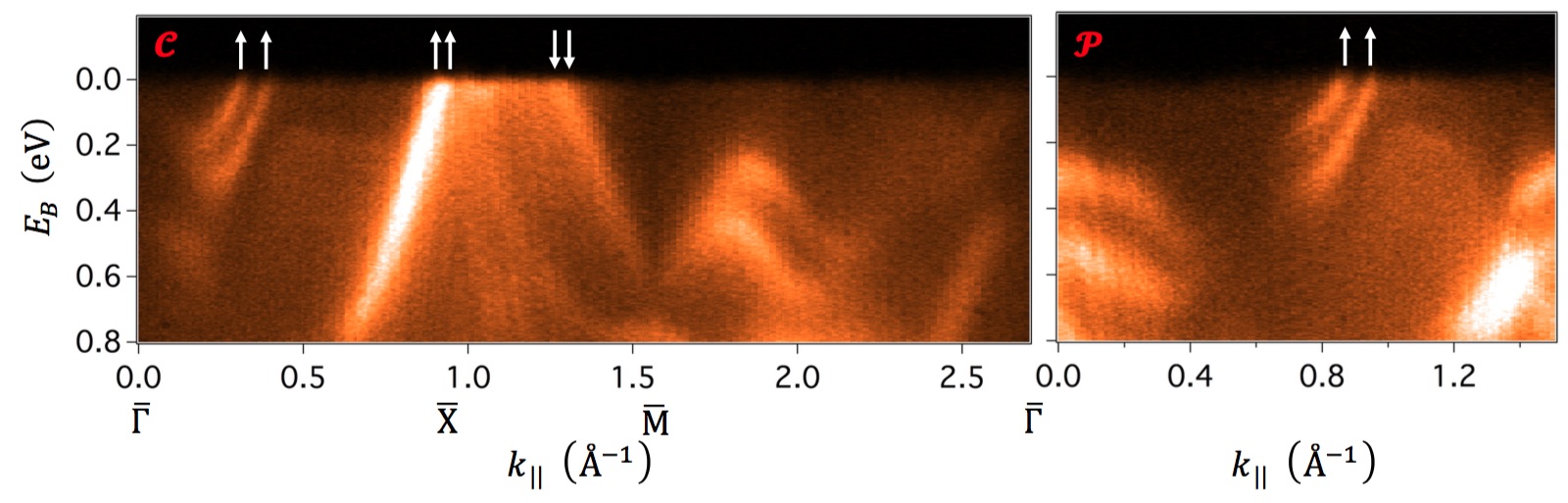}
\caption{Top left) Fermi surface of TaAs  ARPES, at incident photon energy $h \nu = 90$eV, on the (001) surface of TaAs.  Top Right) Again FS of TaAs but with Weyl points indicated and BZ marked.  Two paths in momentum space are denoted as $\mathcal{C}$ and $\mathcal{P}$.  Bottom right) Measured ARPES spectral function  along $\mathcal{C}$, with chiralities of edge modes marked by the arrows. The net Chern number appears to be $+2$, inconsistent with the expectation (Two W2 nodes project to the surface). This can be explained by considering the small separation between the W1 Weyl points. Bottom left)  Measured ARPES spectral function along $\mathcal{P}$. The path encloses only the well-spaced Weyl points and one finds a Chern number $+2$, consistent with expectation.  From \cite{Belopolski16a}.}
 \label{TaAsARPES}
\end{figure}

\begin{figure*}[htp]
\includegraphics[width=2\columnwidth]{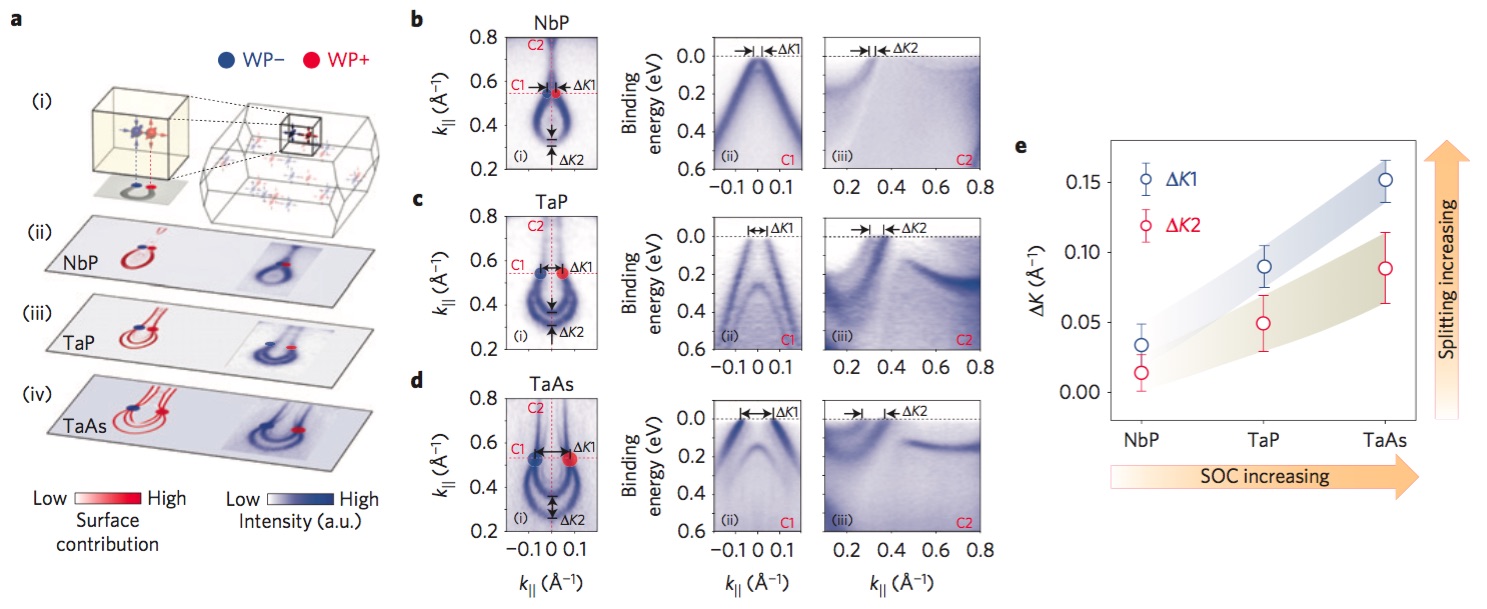}
\caption{Schematic showing the projection of a pair of Weyl points on the (001) surface BZ and the Fermi arc (grey curves) connecting them for materials with increasing SOC strength. (ii-iv) A comparison of the calculated (left) and ARPES measurement (right) of the spoon-like FSs. The red and blue dots denote the chirality of Weyl points. b-d.  ARPES measurements of the spoon-like FS (i) and band dispersions (ii, iii) for NbP, TaP and TaAs, respectively. The positions of the band dispersions presented in (ii, iii) are indicated by the red dotted lines in (i). e. Summary of the extracted $\Delta K1$ and $\Delta K2$ (from b-d) from the three compounds, plotted against a rough measure of the strength of SOC.  $\Delta K1$ and $\Delta K2$ represent the separation between the Weyl points and Fermi arcs, respectively.  From \cite{Liu16a}.}
 \label{ARPEScomparison}
\end{figure*}

\begin{figure}[htp]
\includegraphics[width=0.95\columnwidth]{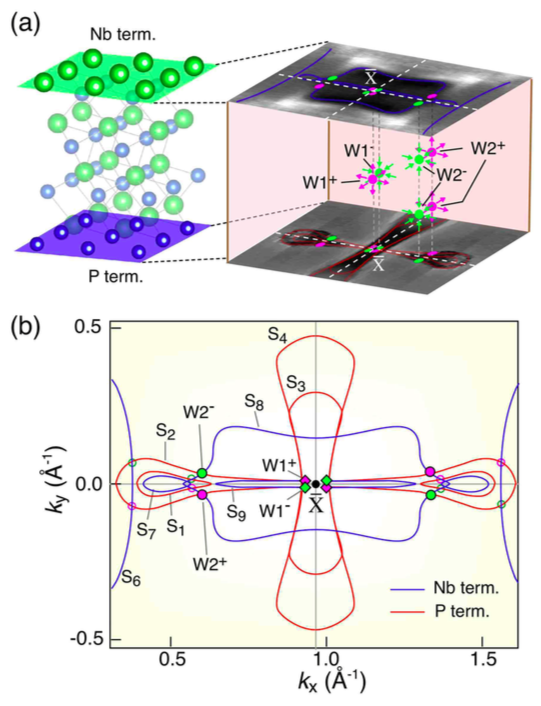}
\caption{(a) (Left) Crystal structure of NbP that has two different surfaces.   Right: Schematics of the experimental ARPES momentum space mapping near the $X$ point for the Nb- and P-terminated surfaces. The bulk Weyl nodes (W1 and W2) projections are illustrated by circles.  (b) A comparison of experimental FS between Nb-terminated (blue, cdntral horizontal loop) and P-terminated (red, central vertical loop) surfaces.  Projection of Weyl nodes W2 at the intersection of FSs for opposite surfaces is shown by filled circles, whereas other intersections are shown by open circles. Weyl nodes W1 are indicated with diamonds.  Fermi surfaces S2, S8, and S9 are believed to be Fermi arcs, whereas S1, S3, S4, S6, and S7 from trivial Fermi surfaces.  From \cite{Souma16a}.  }
 \label{NbPcleaving}
\end{figure}

Despite early predictions, good material realizations and experimental data for a WSM were longer in coming.   After the  prediction \cite{Weng2015a,Huang15a}, a number of groups found evidence in TaAs for a WSM \cite{Lv15a,Lv15b,Xu15a,Yang15a,xu2016spin}.   Subsequently, similar evidence was found for other compounds in this material class \cite{Xu15b,Xu15c,xu2015experimental}.  Shown in Fig. \ref{TaAsARPES}, one can see representative ARPES data for TaAs \cite{Belopolski16a}.   One can see the  characteristic ``lollipop" and ``bowtie" shaped Fermi Surfaces (FSs) of the As terminated surface.   The same general shapes are found in all measured materials in the TaAs material class as shown in Figs. \ref{ARPEScomparison} and \ref{NbPcleaving}.  Although rough agreement of ARPES with band structure calculations has been taken as evidence for a WSM, \onlinecite{Belopolski16a} presented a set of conditions for the ARPES measured surface state band structure to meet to establish the presence of a WSM.  Some of the TaAs class of materials may not rigorously meet this standard due to the energetic position of the nodes in manner of Fig. \ref{TaP}.  To establish a non-zero Chern number, one such condition is to add up the signs of the Fermi velocities of all surface states around a closed loop in the surface BZ in a momentum region where the bulk band structure is everywhere gapped, assigning +1 for right movers and -1 for left movers.  An odd sum, establishes the a Weyl point.  As shown in Fig. \ref{TaAsARPES} for TaAs, one can consider two paths in momentum space that are denoted as $\mathcal{C}$ and $\mathcal{P}$.  However, because two W2 nodes project onto the sBZ's (001) surface, their projection will terminate two Fermi surface arcs.  Therefore around loop $\mathcal{C}$, the expectation is that the net Chern number is $+3$, however it is experimentally observed to be $+2$.  This can be explained by considering the small separation between the W1 Weyl points where the single Fermi surface arc is not resolved.  However, around $\mathcal{P}$, the path encloses only the well-spaced Weyl points W2 and one finds a Chern number $+2$, consistent with expectation of two W2 points projecting to the surface BZ.   This establishes the WSM state.  However, one should take note of the extremely non-trivial shapes of the Fermi arcs in these materials, and compare them to the prediction of the simple model Hamiltonians of Sec. \ref{WSMmodels}, which have a Fermi arc that is a straight line connecting the projection of the Weyl nodes onto the surface.   This shows that although the WSM may be realized in such systems, very careful analysis will generally have to be done to reveal the universal aspects of the WSM state.   Particular experiments may be more sensitive to particular complexities of the band structure.

Although all materials in this class have the same general band structure, their different spin-orbit coupling and other energetics can lead to differences in their topological properties.  One may also study the evolution of the electronic structure with increasing spin-orbit coupling strength \cite{Liu16a}.  As seen in the systematic comparison between three members of the monopnictide family (NbP, TaP, TaAs) in Fig. \ref{ARPEScomparison}, increasing SOC has the effect of pushing the W2 Weyl points with opposite chirality away from the mirror plane.  As seen in Fig. \ref{ARPEScomparison}e, the splitting of these Weyl points ($\Delta K1$) and the splitting of the band dispersions ($\Delta K2$), which causes the splitting of the Fermi-arcs, also increases with the SOC in the various compounds.   Quantum oscillation experiments have also been important in this regard as even aside from the any possible phase offsets discussed below (Sec. \ref{BerryPhase}),  angle-dependent quantum oscillations are a powerful tool to determine the FS topology in such materials.   They have been able to (in conjunction with band structure calculations) establish fine details about the location of the Fermi energy with respect to the Weyl nodes in the different materials and show for instance that despite having the same general band structure but different size of SOCs, NbP \cite{Klotz16a} and TaP \cite{Arnold16a} have FSs that in-circle two Weyl nodes giving zero net chirality to these FSs sections, while TaAs \cite{Arnold16b} has a Fermi energy close enough to both sets of Weyl points to generate chiral particles at $E_F$. 

\begin{figure}[htp]
\includegraphics[width=0.95\columnwidth]{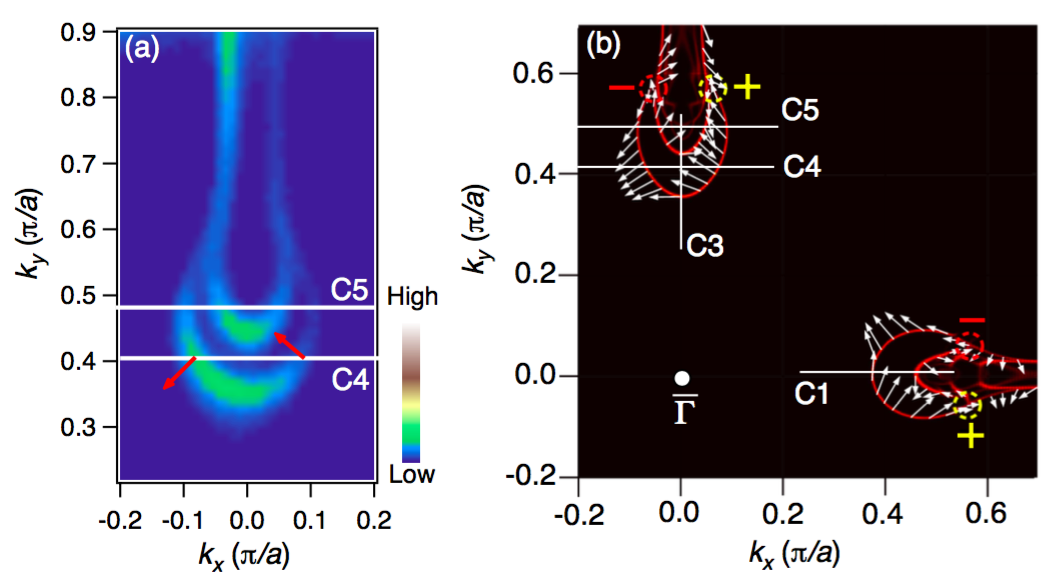}
\caption{(a) Spin-integrated FS map near $\Gamma-Y$ recorded with a spin-resolved ARPES system for TaAs. The red (gray) arrows indicate the direction of measured in-plane spin polarizations of the Fermi arc b2 at C4. (b) Corresponding theoretical spin texture of surface states. Red and yellow dashed circles indicate the Weyl nodes W1 inferred to have negative and positive chirality, respectively.  From \cite{Lv15c}.}
 \label{SpinTexture}
\end{figure}

Like in the case for topological insulators with their well known ``spin-momentum" locking, spin textures are expected and observed (Fig. \ref{SpinTexture}) for the FS arcs in the WSM case \cite{Lv15c,xu2016spin}.  However unlike in the TI case, topological properties cannot be inferred directly from the spin texture of the Fermi arcs.  There is no model-free relationship between the chirality of the Weyl points and the Fermi arc spin texture, other than those mandated by the crystal symmetry itself.  For instance, in TaAs the Fermi arcs that intersect the  $\Gamma - Y$ line have a mirror symmetry ${\mathcal M}_x$ that constrain the spin to be in-plane polarized.  The total spin polarization is as large as 80$\%$ in TaAs \cite{xu2016spin}, which can be compared to  the total spin polarization of Bi$_2$Se$_3$ surface states is only about 40$\%$ \cite{Barriga14a}.  This is because in the Bi$_2$Se$_3$ case the spin textures of the $p$ orbitals interfere partially destructively, whereas in TaAs there is constructive interference. \onlinecite{Lv15c} (Fig. \ref{SpinTexture}) demonstrated consistency between the spin texture and Weyl node chirality as compared to their ARPES data, but it is important to note that the relation between spin texture of surface states and the chirality of Weyl nodes was determined by comparison to experiment in their calculation and not set independently as can be done in TIs.\footnote{We caution that in general great care must be brought to bear in the interpretation of spin resolved photoemission data.  Even in $\mathcal{P}$ and $\mathcal{T}$ symmetric systems that are ensured to have overall two-fold spin degeneracy, pronounced spin polarizations can be observed in spin-ARPES.   This arises from local inversion symmetry breaking within the unit cell \cite{Zhang14b}.  As photoemission is a strongly surface sensitive technique it can preferentially samples a fraction of the unit cell giving the possibility of a spin polarized signal in systems which do not have spin split bands \cite{Bawden16a} or anomalous spin textures that do reflect unit cell averages \cite{Zhu13a}.   This effect interferes with a straightforward interpretation of the spin-resolved ARPES data.}.

The inherent inversion symmetry breaking of the TaAs material class reveals itself in an asymmetry in the photoemission spectra of the top and bottom surfaces as the spectra from the (001) direction is different than the (00$\overline{1}$) direction \cite{Souma16a,Sun15a}.   In the ARPES technique  crystals are typically cleaved in vacuum to reveal a clean surface.  In the case of NbP, this cleaving occurs easier by breaking two Nb-P bonds per unit cell instead of breaking four bonds, (Fig. \ref{NbPcleaving}) so the (001) surface is preferentially Nb terminated and the (00$\overline{1}$)  surface P terminated.  As discussed above there exist two kinds of Weyl points in this material class.  Due to the fact that W2 nodes are located in pairs at positive and negative k$_z$,  their projection onto either surface terminate two Fermi arcs,  whereas the projection of W1 terminates a single Fermi arc.  These broad aspects are independent of the surface termination.  However the particular way in which the Fermi arcs connect the projected Weyl points depends on the surface.  These differences can be seen directly in the photoemission spectra.  As shown in Fig. \ref{NbPcleaving} a number of FSs are seen.  Fermi surfaces S2, S8, and S9 are believed to be Fermi arcs, whereas S1, S3, S4, S6, and S7 from trivial Fermi surfaces.

Scanning tunneling spectroscopy (STS) is a real space surface measurement technique that measures the density of states as a function of position.   However it can provide momentum space information through Fourier transform of the spatial dependence of impurity- or boundary-induced states.  Because it is directly sensitive to scattering it provides information about pseudospin scattering constraints and chirality of quasiparticles even when the pseudo-spin vector is not necessarily associated with the electron spin.   A number of measurements have reported signatures of scattering patterns consistent with Fermi arcs on the surface Weyl semimetals such as TaAs  \cite{Gyenis16a,Inoue16a,Zheng16a,Batabyal16a,chang2016signatures}.   These measurements give evidence not only for the particular spin dependent scattering function indicative of Weyl Fermi arcs, but also the momentum dependent delocalization of the arc states into the bulk of the sample that occur at their projection on the bulk Weyl nodes.

Finally, as mentioned above, it has been proposed that a Weyl state can emerges from the touching of electron and hole pockets in a state that is distinct from the idealized Type I Weyl semimetals with their point-like Fermi surface \cite{Soluyanov15a}.  The Weyl cone in this Type II semimetals is strongly tilted and as a function of chemical potential the Fermi surface undergoes a Lifshitz transition.  Although Type I and Type II WSM's cannot be smoothly deformed into each other, they share electronic behavior that originate in the isolated band contact point in their bulk spectra. They  are anticipated to have a number of different properties including a variant of the chiral anomaly when the magnetic field is well aligned with the tilt direction, have a density of states different than the usual form, possess novel quantum oscillations due to momentum space Klein tunneling, and a modified anomalous Hall conductivity  \cite{Soluyanov15a,OBrien16a,Udagawa16a,Zyuzin16a}.  Evidence for a Type II state has been given in MoTe$_2$ \cite{deng2016experimental,Liang16c,Jiang17a,Tamai16a,Huang16a}, the alloy Mo$_{x}$W$_{1-x}$Te$_2$ \cite{belopolski2016discovery,belopolski2016fermi} and TaIrTe$_4$ \cite{Koepernik16a,Haubold16a,Belopolski16b}.   Detection of the Type II state in the MoTe$_2$ class of materials is challenging.  Although evidence has been claimed for WTe$_2$ in \onlinecite{Wang16e}, it controversial.  As emphasized by \onlinecite{bruno2016observation}, although the bulk band structure is very sensitive to small changes in lattice parameters that push the material in and out of the Weyl state, the feature identified as surface ``Fermi arcs" in WTe$_2$ are largely independent of these changes, and is therefore trivial and cannot be used to show the system is in the non-trivial phase.  The predicted topological Fermi arcs in  WTe$_2$ are predicted to be too small to be observed experimentally.  The situation is somewhat more favorable (although still challenging) in MoTe$_2$.   There is again the same large trivial Fermi arc-like feature, but also small arcs that have been observed, which are consistent with being topological via band structure calculations \cite{Tamai16a,Jiang17a,deng2016experimental}.

\subsection{Semiclassical transport and optics}

\subsubsection{General considerations}
\label{OpticsTheory}

As mentioned above, 3D Dirac and Weyl systems are predicted to have a number of interesting semi-classical transport and optical effects that are diagnostic for this state of matter \cite{Hosur12a,Burkov11a} (Effects related to quantum transport like the chiral anomaly are discussed below in Sec. \ref{QMTransport}).   With small modifications, most of these results apply equally to Weyl and Dirac systems.

In the absence of impurities and interactions the free fermion result for the conductivity in the low energy limit (where quadratic or higher order terms in the dispersion can be neglected) that arises from interband transitions across the Weyl or Dirac node when the chemical potential (E$_F$) is at the Weyl or Dirac point is

\begin{equation}
\sigma_1(\omega) = N \frac{e^2}{12 h} \frac{|\omega|}{v_F}
\label{LinearConduct}
\end{equation}
where $v_F$ is the Fermi velocity and $N$ is number of nodes \cite{Wan11a, Hosur12a,Burkov11a,Hosur13a,Tabert16b}.  This prediction is closely related to the prediction and observation in 2D Dirac system of single layer graphene that for interband transitions that its optical conductance should be  $G_1 (\omega) = \frac{e^2}{\hbar}$, giving a frequency independent transmission that is quantized in terms of the fine structure constant $\alpha$ as $T(\omega) = 1 - \pi \alpha$ \cite{Ando02a,Kuzmenko08a,Nair08a}.  In a 2D material like graphene the Kubo-Greenwood expression for the conductance from interband transitions can be written (for the chemical potential at the Dirac point  and T=0) as $G_1 (\omega) = \frac{\pi e^2}{\omega}  |\textbf{v}(\omega)|^2 D(\omega) $ where $ \textbf{v}(\omega)$ is the velocity matrix element between states with energies   $ \pm \hbar \omega / 2$ and $g(\omega)$ is the 2D joint density of states.  The universal conductance arises because the Fermi velocity factors that come into the matrix element are canceled by their inverse dependence in the density of states.  In these 3D Dirac systems, another factor of $\omega$ comes in in the density of states yielding Eq. \ref{LinearConduct}.

Note that Eq. \ref{LinearConduct} implies a logarithmic divergence of the real part of the dielectric constant through Kramers-Kronig considerations \cite{Rosenstein13a,Jenkins16a}.   The corresponding imaginary conductivity is

\begin{equation}
\sigma_2(\omega) = - \frac{2}{\pi}N \frac{e^2}{12 h} \frac{|\omega|}{v_F}  \mathrm{log} \frac{2 \Lambda v_F}{\omega}.
\label{LinearConductImaginary}
\end{equation}
\noindent where $ \Lambda $ is a UV momentum cutoff.  

Quite generally, in noninteracting electron systems consisting of two symmetric bands that touch each other at the Fermi energy the optical conductivity generically has power-law frequency dependence with exponent $(d-2)/z$ where $d$ is the dimensionality of the system and $z$ is the power-law of the band dispersion \cite{Basci13a}.  Such power-law behavior is a consequence of the scale-free nature of such systems.   It has been argued \cite{Fisher91a} that at a conventional continuous transition the optical conductivity should scale as $(d-z-2)/z$.   Due to their scale free nature one may regard Dirac systems as intrinsically quantum critical with a dynamic exponent equal to the band dispersion power-law.   With the usual substitution for the effective dimensionality of a quantum critical system $d_{eff} = d + z$ the generic power-law expression for the Dirac conductivity follows.

In the presence of impurities or interactions the expectation of Eq. \ref{LinearConduct} is modified.  Impurities in the form of dopants can shift the chemical potential of the system away from the Dirac point, leading to a Pauli-blocked edge at approximately 2$E_F$ below which Eq. \ref{LinearConduct} is not observed \cite{Tabert16b}.  For finite scattering introduced by disorder one will generally find a zero frequency peak which can give a finite dc conductance and additional optical response.   Its scaling form or even the existence of a metallic state at all in the limit of E$_F\rightarrow 0$ is strongly dependent on the dimensionality, the power-law of the dispersion, and the kind of scattering (screened, unscreened, short-ranged) that is being considered \cite{DasSarma15a}.

The effect of weak disorder scattering on the finite-frequency conductivity also depends on whether $\omega \gg T$ or $\omega \ll T$.  Various theoretical approaches (Boltzmann, quantum Boltzmann, and Kubo) are generally in agreement with each other.  Solving the linearized Boltzmann equation for short range disorder with the energy-dependent momentum relaxation rate, the  $\omega \ll T$  Drude-like peak in the optical conductivity has a temperature-dependent spectral weight and width, the latter of which is predicted to scale as T$^2$ with a temperature independent dc limit of $\sigma_{dc} = \frac{e^2 v_F^2}{3 \gamma h}$ \cite{Burkov11a}.  In the low $\omega$ limit a very unusual shape for the conductivity is predicted \cite{Burkov11b}, the real part of which goes as

\begin{equation}
\mathrm{Re} \sigma(\omega) \approx  \frac{e^2 v_F^2}{3 \gamma h} \bigg(   1 - \frac{1}{8} \sqrt{\frac{\omega v_F^3 h^3}{2 \gamma T^2}}      \bigg).
\label{DrudeDisorder}
\end{equation}

\begin{figure}[htp]
\includegraphics[width=0.95\columnwidth]{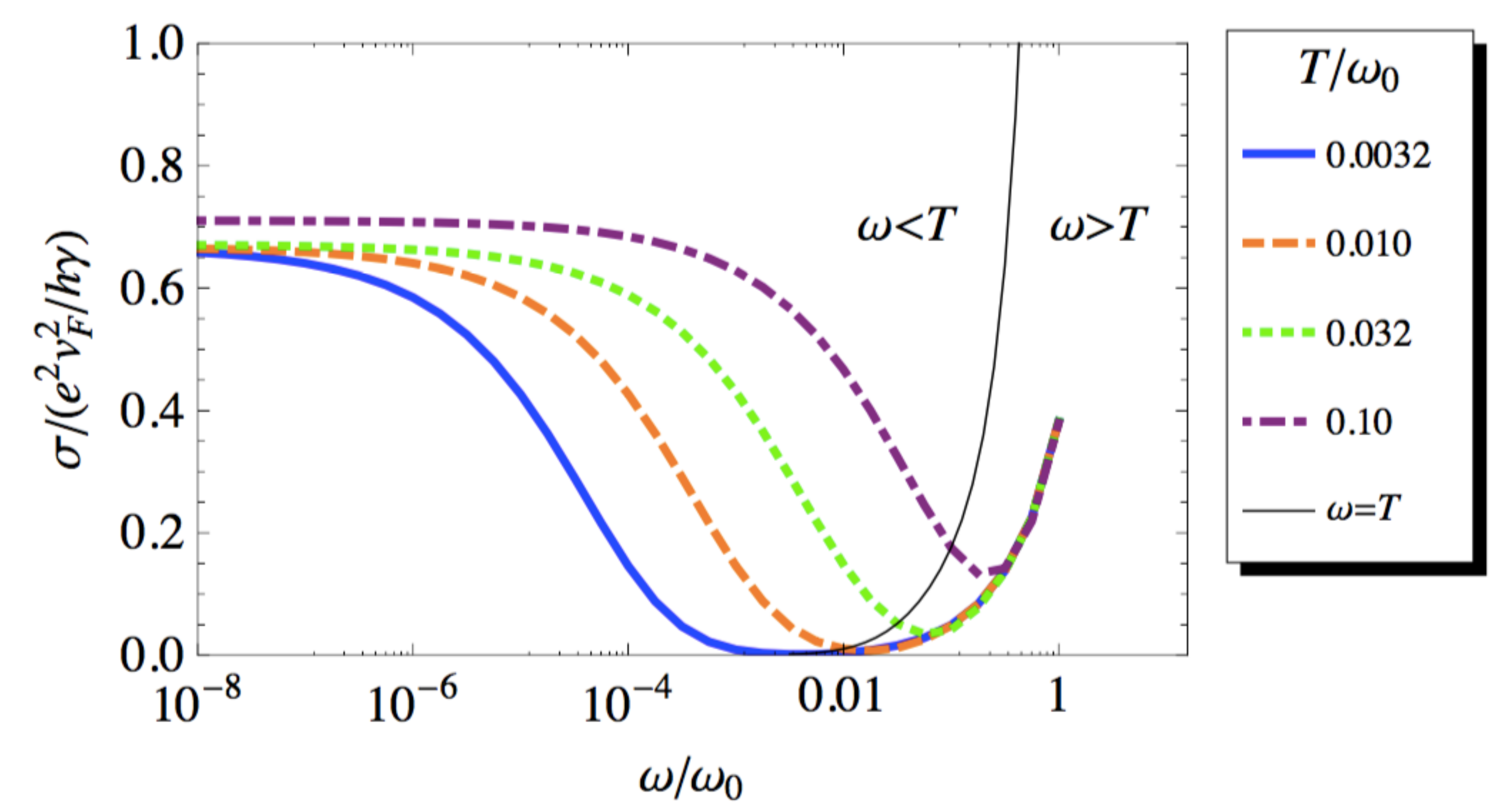}
\caption{Conductivity calculated within the linearized Boltzmann formalism of a single Weyl node with disorder, in units of $e^2v_F^2/h\gamma$ as a function of frequency at different temperatures. Frequencies and temperatures are in units of $\omega_0 =2\pi v_F^3/\gamma$.  One can see that the functional dependence changes at $\omega \sim T$ (given by the thin black line). Note that at the highest temperature plotted deviations from the temperature independent dc result are found.  From \cite{Hosur12a}.}
 \label{TheoryOptics}
\end{figure}

\noindent Note that this is a prediction for a temperature dependence to the \textit{elastic} scattering, which is again quite different from the usual case in metals and arises due to the strongly energy dependent density of states in these systems.  In the opposite $\omega \gg T$ limit the calculation \cite{Hosur12a} gives

\begin{equation}
\mathrm{Re} \sigma(\omega, T) \approx
N\frac{e^2}{12 h} \frac{\omega}{v_F}  \left[1 - \frac{16N \gamma \omega}{15\pi^2 v_F^3}  +\mathcal{O}\left( \left( {\omega}/{\omega_N}\right)^2 \right)\right].
\label{DisorderInterband}
\end{equation}
The leading term is independent of disorder, and is identical to the interband response for the same system in the noninteracting, clean limit.   These functional dependences of the various regimes for the optical conductivity are summarized in Fig. \ref{TheoryOptics}.   Note that $\omega \sim T$ manifests itself as a crossover scale between the behavior of Eq. \ref{DrudeDisorder} and Eq. \ref{DisorderInterband}.   Note that strong disorder in a WSM may necessitate other considerations for the optical response\cite{roy2016universal}.

In the absence of umklapp scattering, electron-electron interactions have a small effect on the optical conductivity of conventional metals as they conserve the total momentum and hence the current.   But in Dirac systems, due to particle-hole symmetry there can be current carrying states with zero total momentum \cite{Fritz08a,Goswami11a}.   This allows interactions to relax current with zero net momentum transfer.  A quantum Boltzmann calculation \cite{Hosur12a} gives the expression for the optical conductivity

\begin{equation}
\sigma(\omega, T) = N\frac{e^{2}}{h}
\left(\frac{k_{B}T}{\hbar v_{F}}\right)
\frac{1.8}{-i\frac{\hbar\omega}{k_{B}T}6.6 + N\alpha^{2}\ln\alpha^{-1}}.
\label{DrudeInteractions}
\end{equation}
Here $v$ and $\alpha$ are the Fermi velocity and fine structure constant renormalized (logarithmically) to energy $k_BT$.  This expression assumes the low frequency limit $\frac{\hbar\omega}{k_{B}T}\ll\alpha^2 $.   The dc limit of this expression can be motivated in a relaxation time approximation via the fact that the temperature dependent density of states goes like $T^2$ whereas on dimensional grounds the transport lifetime is expected to go like $1/\alpha^2 T$.   It is interesting to note the relative temperature dependences of the Drude peak widths that go as $T$ vs. $T^2$ in the interacting vs. disordered cases respectively.

\onlinecite{Lundgren14a} computed various thermal transport coefficients using the semiclassical approach.   With interactions they find that the longitudinal thermal conductivity has a quadratic temperature dependence, in contrast to a linear dependence on the temperature for either charged impurities or short-range disorder (similar to normal metals).  For $k_BT \gg E_F$, both Boltzmann transport  \cite{Lundgren14a} and Kubo formalism \cite{Tabert16b} calculations give a Lorenz number enhanced by interactions from the Fermi liquid value.   \onlinecite{Lundgren14a} also considered the effect of electric and magnetic fields on the thermoelectric coefficients.  With the temperature gradient perpendicular to the magnetic field the transverse thermal conductivity is linear in magnetic field and the longitudinal thermal conductivity has a negative contribution that goes as the square of the magnetic field.  When the temperature gradient is in the direction of the magnetic field there is an increasing quadratic magnetic field dependence for the longitudinal thermal conductivity and zero transverse thermal conductivity.  The presence of finite electric field is predicted to not change these dependences as long as there is no internode scattering.

\subsubsection{Experiments}

The above predictions have been investigated in a number of WSM and DSM systems.  A number of experimental works have reported verification of Eq. \ref{LinearConduct} \cite{Chen15a,Orlita14a,Sushkov15a,Timusk13a}.  ZrTe$_5$ \cite{Li16a,Yuan15a} is a semi-metal with an extremely small and light ellipsoidal Fermi surface that is centered in the bulk BZ.  It is believed to be naturally tuned near a band inversion transition.  A linear optical conductivity consistent with Eq. \ref{LinearConduct} has been found \cite{Chen15a} over a range of 50 -1200 cm$^{-1}$ (6 - 150 meV), albeit with a slope that is some 30 times higher than that expected from the velocities observed in ARPES.  It may be that the quasi-two dimensionality of this material is playing a role.  Linearity over the large range from 50 - 350 meV has also been observed in the zero gapped tuned Hg$_{0.83}$Cd$_{0.17}$Te \cite{Orlita14a}.  This system has been termed as Kane fermion system as its band structure includes an additional nearly flat band contribution from a flat $\Gamma_8$ band that lives between the linearly dispersing bands at the Brillouin zone center.   Much closer agreement between the expected linearly increasing absorption and theory has been obtained in this system.  \onlinecite{Akrap16a} demonstrated linear conductivity over a large energy range in Cd$_3$As$_2$, but this was -- as mentioned in Sec. \ref{Identifying} -- sampling a region of the spectrum describable by the -- also conical -- Kane dispersion and not a Dirac one.  Elsewhere deviations from linearity for Cd$_3$As$_2$ have been reported \cite{Neubauer16a}.

\begin{figure}[htp]
\includegraphics[width=0.95\columnwidth]{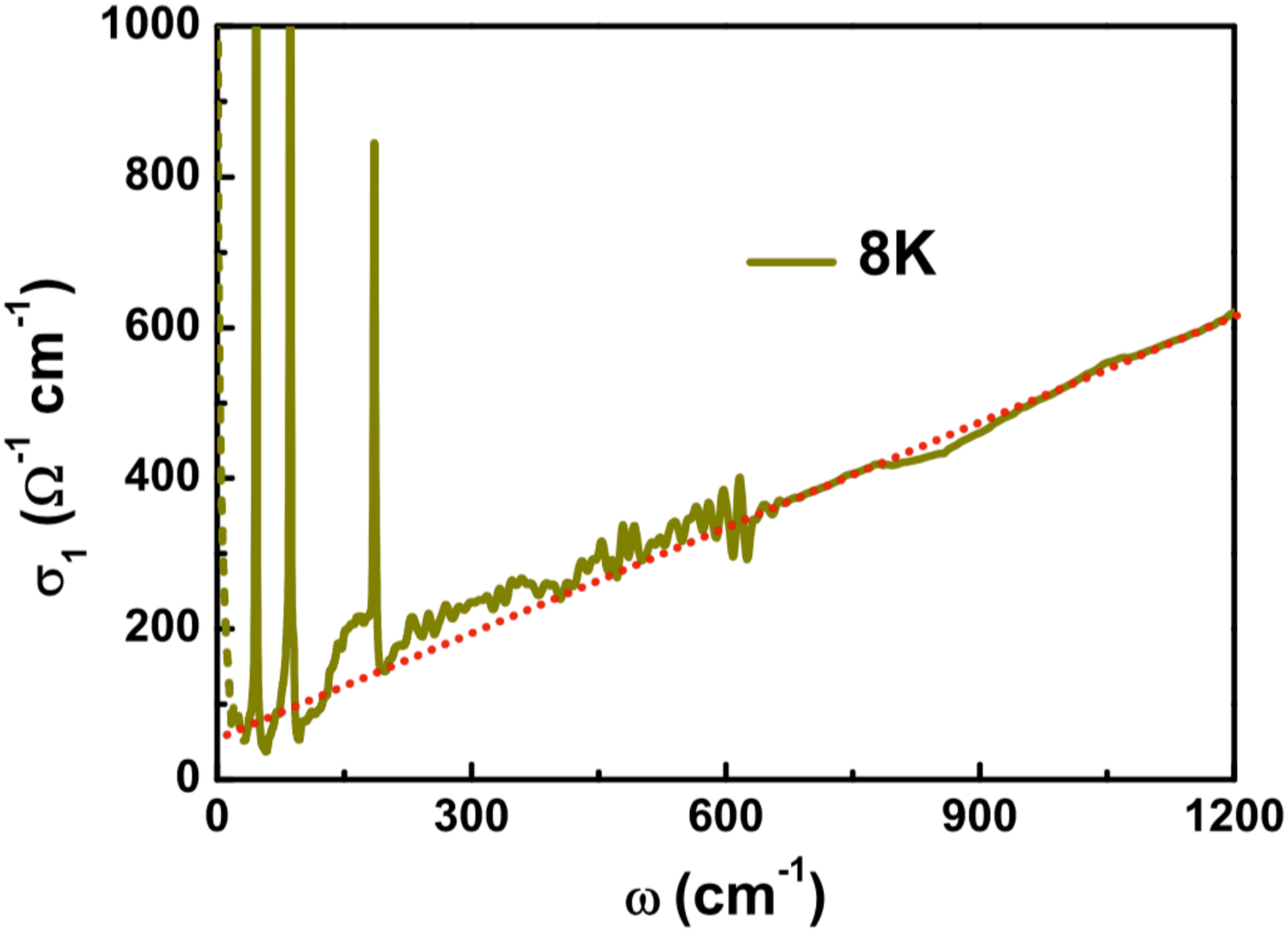}
\includegraphics[width=0.9\columnwidth]{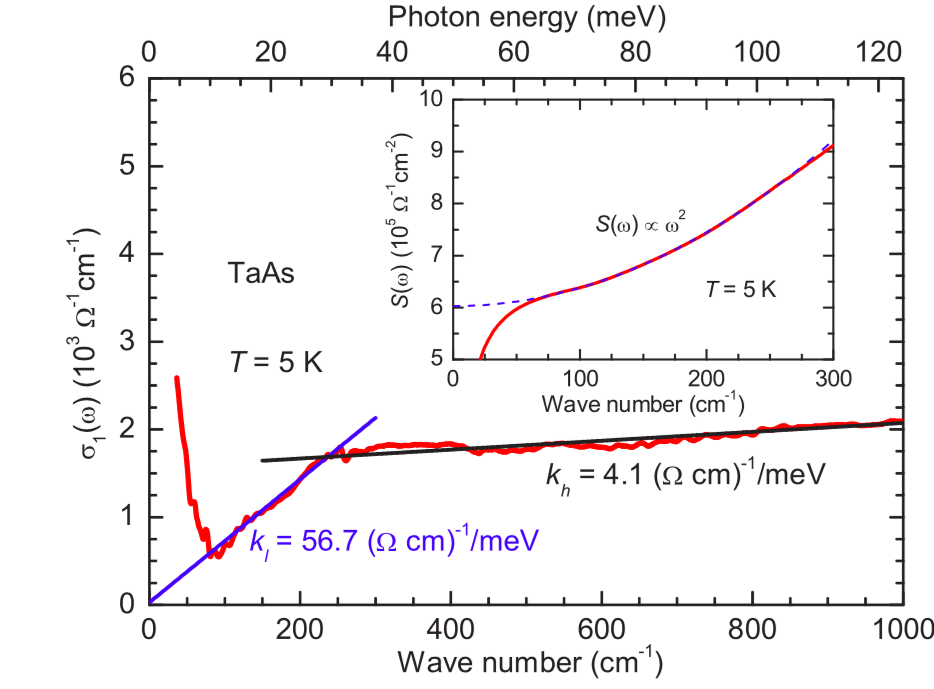}
\caption{(Top) The optical conductivity of ZrTe$_5$ at 8 K at frequencies below 1200 cm$^{-1}$. The red (dotted) line is the linear fitting of $\sigma_1(\omega). $ From Ref. \cite{Chen15a}.  (Bottom) Optical conductivity for TaAs at 5 K. The blue (steep) and black (shallow) solid lines through the data are linear guides to the eye. The blue (steep) line show the Weyl part of the spectrum, while the black (shallow)line comes from higher energy non-Weyl states.  The inset shows the spectral weight as a function of frequency at 5 K (red solid curve), which follows an $\omega^2$ behavior (blue dashed line).   From \cite{Xu16a}.}
 \label{ZrTe5Optics}
\end{figure}

The Weyl semimetal state was first predicted to occur in the antiferromagnetic state of the pyrochlore iridates \cite{Wan11a}.  \onlinecite{Sushkov15a} found that Eu$_2$Ir$_2$O$_7$ has an approximately linear frequency dependence of the optical conductivity down to 3 meV.  Below T$_N$, the Drude spectral weight diminishes consistent with the reduced thermal excitations of a Weyl semimetal.  The data sets can be modeled, assuming a WSM, with 24 Weyl points  and an average Fermi velocity of  $v_F =4 \times 10^7$ cm/s.   A recent optical conductivity study on several pyrochlore iridates \cite{Ueda16} however classifies this material as an insulator. In TaAs, it is claimed \cite{Xu16a} that the low frequency Drude response exhibits a T$^2$ dependence to its width, although neither the strongly temperature dependent spectral weight nor the unusual line shape of Eq. \ref{DrudeDisorder} was observed.  TaAs also shows \cite{Xu16a} a linear in frequency  $\sigma_1$ up to 1000 cm$^{-1}$, which is reasonably associated with the interband transitions associated with the four pairs (of 12) of W1 Weyl points and are predicted to be only 2 meV above the Fermi energy (Fig. \ref{ZrTe5Optics}b).

\begin{figure}[htp]
\includegraphics[width=0.95\columnwidth]{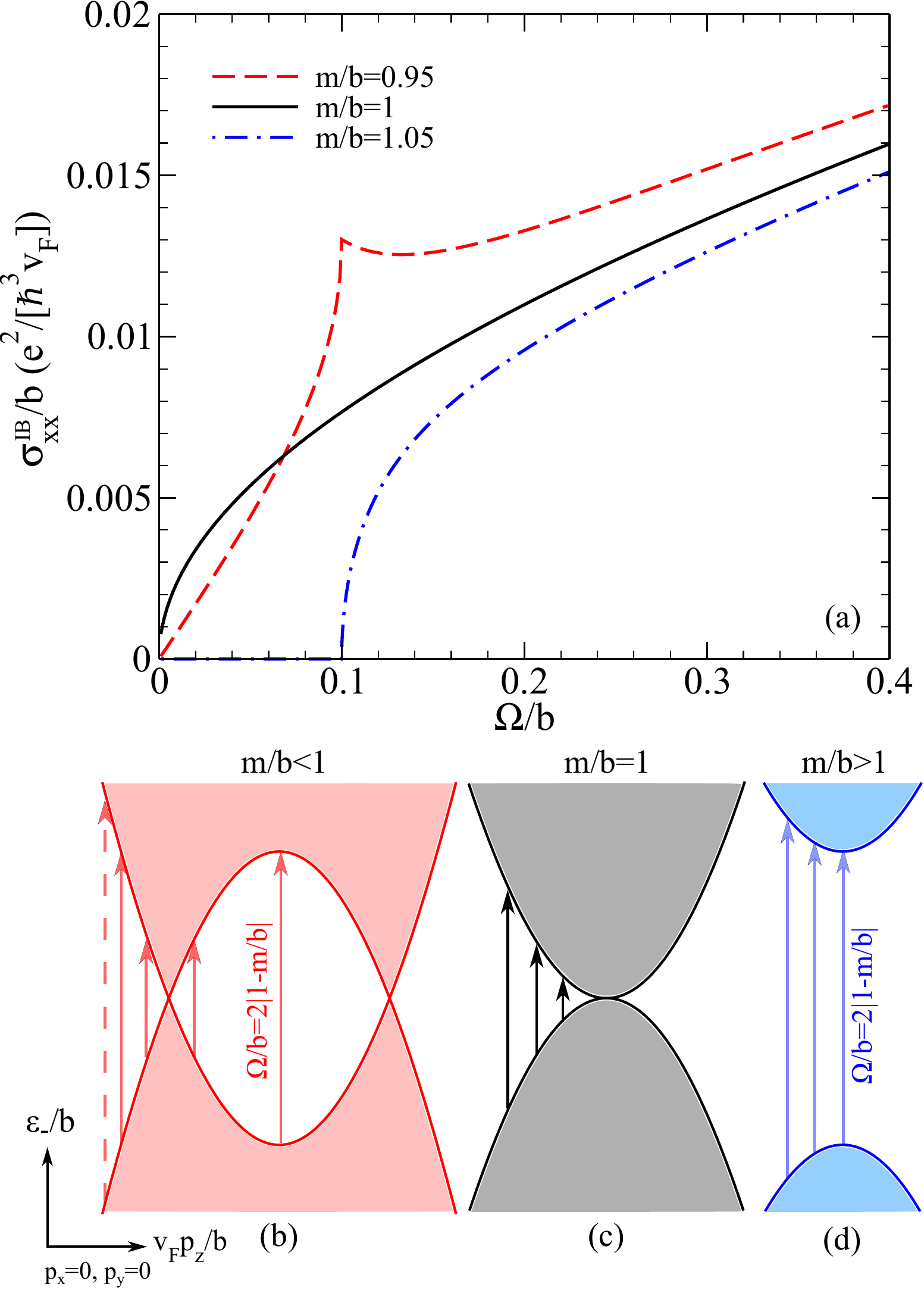}
\caption{Interband optical conductivity from the Weyl semi-metal to gapped semi-metal phase transition.   Here $m$ is a Dirac mass parameter m, and $b$ is an intrinsic Zeeman-field like parameter.  The Hamiltonian is the same as appears in Eq. \ref{4band}.  From Ref. \cite{Tabert16a}.}
 \label{Carbotte}
\end{figure}

It is important to keep in mind that the observability of many of the above predictions rely on some idealities of the band structure that may or may not be present in real materials.  For instance it has been claimed that the lack of a threshold for Dirac cone interband transitions at 2$E_F$ in the purported Dirac system Na$_3$Bi is due to very large Dirac cone anisotropies \cite{Jenkins16a}.   Moreover, it is not clear that even if materials like YbMnBi$_2$ are $\mathcal{T}$ breaking  Weyl semimetals it is possible to see linear in $\omega$ optical conductivity due to certain nonidealities that are certainly present in their bandstructures and the presence of non-toplogical bands \cite{Chinotti16a,Chaudhuri16a}.  Also note that different regions of linear in $\omega$ absorption may be seen that correspond to different linear (or otherwise) parts of the band structure as pointed out by  \onlinecite{Tabert16a} (Fig. \ref{Carbotte}).  This appears to be the case in Cd$_3$As$_2$ where optics sees a linear in $\omega$ conductivity but it is sampling the part of the spectrum that is described by the Kane dispersion, not the lower energy Dirac disperpsion \cite{Akrap16a}.

A very interesting suggestion is that the linear conductivity seen in the optical response of many quasicrystals compounds arises from the fact that these may be realizations of Weyl semimetals \cite{Timusk13a}.   For instance, in the AlCuFe system the conductivity rises linearly with a slope of 5750 cm$^{-1}$/eV.  A comparison to Eq. \ref{LinearConduct} gives -- with the assumption that the Weyl points are located on the faces of the icosahedron gives with spin degeneracy $N=40$ -- the reasonable estimate for the Fermi velocity of 4.3 $\times$ 10$^7$ cm/s.   This idea of WSM state hiding inside quasicrystals deserves further consideration.

Interaction effects may also reveal themselves in an interesting fashion in optical conductivity.  \onlinecite{Jenkins16a} observed the presence of  side-band feature in the optical response of the Dirac semi-metal candidate Na$_3$Bi that they assign to a coupled quasiparticle-plasmon excitation e.g. a $plasmaron$ \cite{Lundqvist67a} that has also been seen in the optical conductivity of the massive Dirac semi-metal elemental bismuth \cite{Tediosi07a,Armitage10a}.   Such coupling is a form of electron-electron interaction,  can cause mass renormalizations, and may be ubiquitous in slightly doped WSM and DSM systems.

\subsection{Quantum mechanical effects in transport}
\label{QMTransport}
\subsubsection{Quantum oscillations}
\label{BerryPhase}

When one puts an electronic system in magnetic field, Landau level (LL) quantization occurs and as field is ramped, the density of states at $E_F$ undergoes quantum oscillations (QO) resulting in the variation of physical quantities as a function of $1/B$.   Measurements of quantities like resistivity (e.g. Shubinikov- de Haas oscillations (SdH)) explicitly measure the rate at which when the LLs are depopulated as the field is increased.  The condition for LL formation is given by a generalized Lifshitz-Onsager quantization expression $A_F  \frac{\hbar}{eB} = 2 \pi (n + \frac{1}{2} + \beta + \delta)$ where  $A_F$ is the cross-sectional area of the Fermi surface normal to the field and $n$ is the Landau level index that depends inversely on $B$.  $ \delta $ is an additional phase shift that results from three dimensional warpings of the Fermi surface that is 0 for a 2D cylindrical FS and $\pm1/8$ for a 3D FS \cite{shoenberg2009magnetic,Murakawa13a}.   

The additional phase shift $\beta$ equals zero in conventional parabolic bands.  However, it was shown by \onlinecite{roth1966semiclassical} that for arbitrary dispersions other values can occur, which can be calculated with knowledge of the Bloch functions.  \onlinecite{mikitik1999manifestation} showed that the expressions of \onlinecite{roth1966semiclassical} could be recast in a form such that $\beta$ can be shown to be equivalent to the Berry's phase experienced by an electron as it travels around a closed loop in momentum space \cite{mikitik1999manifestation}.  One of the distinguishing features of Dirac electrons is this nontrivial Berry's phase, which in principle can be revealed by QO experiments.

Such Shubnikov-de Haas (SdH) oscillation experiments have been extensively in studies of 2D materials like graphene \cite{Novoselov05a,Zhang05a} and topological insulator surface states \cite{Qu10a,Analytis10a,Sacepe11a,Taskin11a}.   For a 2D gapless Dirac system like graphene, the Dirac point plays the role of an infinitely thin solenoid in momentum space with a fictitious effective magnetic field, which is the Berry curvature.   In related massive Dirac systems (for instance boron nitride) the Berry curvature is spread over a region in momentum space near the band minimum.  The effect is still felt in regions of zero Berry curvature through an effective vector potential -- the Berry connection \cite{xiao2007valley}.   Therefore a closed path that encircles a region of net Berry curvature picks up a Berry's phase that can be identified as $\beta$.   In 3D, it is in a similar sense that Weyl points can be considered as sources of Berry curvature as discussed above.

In graphene \cite{Zhang05a}, one locates the peaks and valleys of the SdH oscillations as a function of $1/B$, and plots them against  Landau index $n$ in a ``fan" diagram.  In the ideal case this results in straight lines, with a slope that is the SdH oscillation frequency (which gives the FS area) and an intercept with the $n$ axis that gives the Berry's phase $\beta$ in units of $\pi$.   In practice, curvature of the bands can lead to deviations from linearity.  In graphene, this is a minor effect, but obviously the intercept is most accurately quantified if low LL indices are measured.   For this high fields and low carrier densities are required.   Generally, the latter is more achievable in 2D systems, which can be gated.

In 3D materials, it is generally less straightforward to determine the Landau indices without a detailed analysis.    First, generally densities are high enough that it is difficult to access low LL indices.   Second,  many materials of interest have complex band structures with both linear and quadratic dispersions in the relevant energy range.  This is particularly true in the complex band structures of real materials like WSM and DSM candidates.  Non-idealities in various models and the effect of gap opening terms (e.g. Zeeman fields) have been discussed in Refs. \cite{wright2013quantum,wang2016anomalous,wang2016anomalous}.  It is only in certain limits that the a clear picture can be obtained.

\begin{figure}[htp]
\includegraphics[width=0.95\columnwidth]{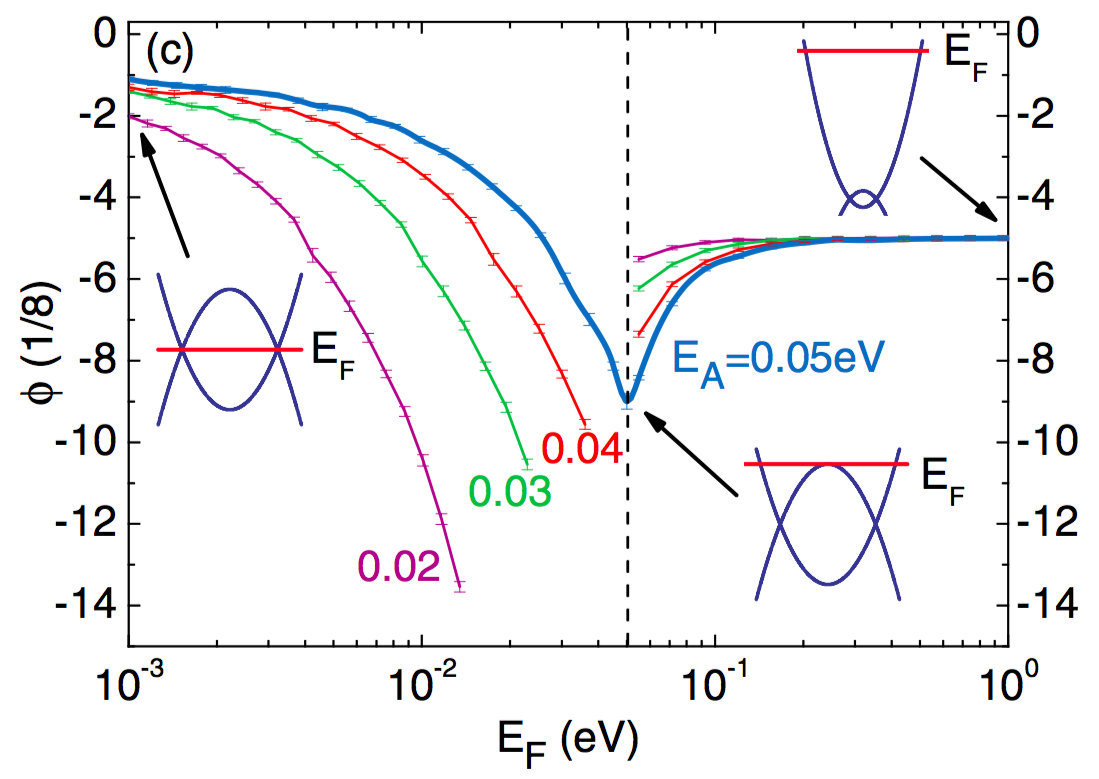}
\caption{ The quantum oscillation phase shift $\beta$ (labeled $\phi$ and given in units of $\frac{1}{8}$) vs. $E_F$ for different relative strengths of linear ($E_A$) and quadratic ($E_M = 0.05$ eV) terms in the energy spectra.   See \cite{wang2016anomalous} for details of the model.  The curves break because  $\beta$ cannot be fit in the parameter range where beats form.  The insets indicate the location of Fermi energy with respect to the model band structure.  The vertical dashed line marks the Lifshitz transition where the system goes from two Weyl pockets to a single larger one \cite{wang2016anomalous}.}
 \label{QOsimulations}
\end{figure}

\begin{figure}[htp]
\includegraphics[width=0.89\columnwidth]{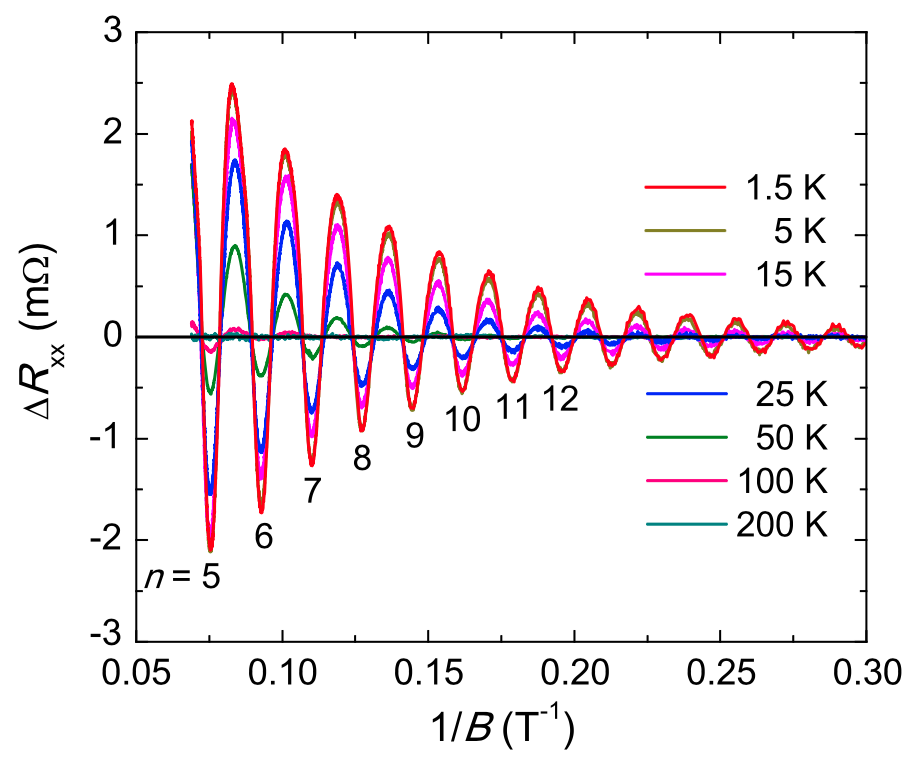}
\includegraphics[width=0.91\columnwidth]{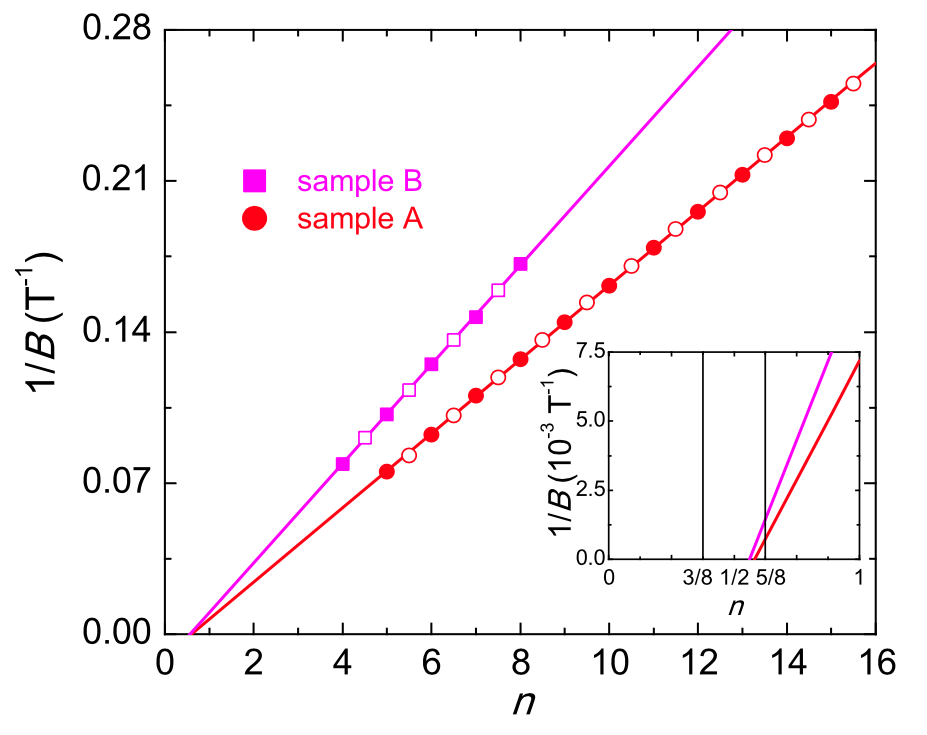}
\caption{ (top) The oscillatory component of the  resistance R$_{xx}$ of Cd$_3$As$_2$ as a function of $1/B$ extracted from R$_{xx}$.  A smooth background has been subtracted.   (bottom)  Landau index $n$ plotted against $1/B$. The closed circles denote the integer index (R$_{xx}$ valley), and the open circles indicate the half integer index (R$_{xx}$ peak) for two different samples. The index plot can be linearly fitted for both samples measured, giving intercepts of 0.56 and 0.58.  Adapted from Ref. \cite{He14a}. }
 \label{QO}
\end{figure}

Even for idealized band structures, interpretation may be complicated.  For instance, \onlinecite{wang2016anomalous} considered a simple model of a WSM in which one can explicitly tune the relative contribution of linear ($E_A$) and quadratic ($E_M$) terms to the energy spectrrum.  The presence of a quadratic contribution arising in this fashion is more appropriate in modeling quantum oscillations experiments than the continuum Dirac model considered in Sec. \ref{WSMmodels}.  A number of the complications pointed out above can be seen explicitly even in this relatively simple band structure.  From the numerical simulations shown in Fig. \ref{QOsimulations} for a $\mathcal{T}$ breaking WSM, for the energy range where $E_A < E_M$ the phase shift  $\beta$ is an very strong function of $E_F$.  Moreover, even for small values of $E_F$, the phase shift is a strong function of the band parameters and is not simply related to the number of Weyl nodes encircled by a FS contour.  It is only in the $E_F \rightarrow 0$ and $\infty$  limits does the numerical simulation recover the simplest analytic results of $ \beta = -1/8$ and $-5/8$ phase shifts for 3D linear or quadratic bands respectively.  The phase shift even becomes non-monotonic in the region where $E_F$ is near the Lifshitz transition.  Near the Lifshitz transition for $E_A \approx E_M$ pronounced ``beats" could be found in the spectra making the whole scheme break down altogether.   Such simulations show that quantum oscillation experiments described by even such a simple band structure must be interpreted very carefully.

A number of additional complications associated with the measurements themselves need to be addressed.  As LLs form, the density of states at $E_F$ changes giving oscillations in various quantities as a function of $1/B$.   In resistivity measurements it has been debated whether one wants to fit minima or maxima in the resistivity.  This depends on a number of issues, including whether or not $\rho_{xx}$ is greater or lesser than $\rho_{xy}$ \cite{Hu16a}.  It was argued in Ref. \cite{wang2016anomalous} that even for the longitudinal geometry the maximum resistivity would be found when $E_F$ is near the Landau level edge due to the vanishing velocity there and hence resistivity maximum should be used.   Moreover, even more complications arise in materials with multiple bands that cross $E_F$.  In addition to simply just complicating the oscillation pattern with additional backgrounds and oscillations, as LL depopulate with increasing field, charges may move between bands to lower the total energy.  This has been documented long ago  \cite{woollam1971graphite} (and more recently \cite{schneider2009consistent}) in graphite where the relative movement of the Fermi energy between bands can be considerable as the quantum limit is approached giving oscillations that are not periodic in $1/B$.   This obviously interferes with a straightforward extrapolation of $1/B$ to infinite field.

Despite all of this, quantum oscillation have been measured in various WSMs and DSMs, and although the interpretation is challenging there is some evidence for non-trivial Berry's phase effects.  See  \cite{wright2013quantum} or the SI of \cite{wang2016anomalous} for a review of experimentally observed phase offsets.  Representative data for quantum oscillations in a DSM can be seen in Fig. \ref{QO} for Cd$_3$As$_2$   The index plot can be linearly fitted for both samples measured, giving intercepts of 0.56 and 0.58 giving evidence for a Berry phase's offset of order $\pi$ \cite{He14a}.  Similar data can be found elsewhere \cite{Narayanan15a,Desrat15a}.  Some evidence exists for a crossover to a trivial Berry phase regime at higher field under high magnetic fields that are directed away from the 001 direction \cite{Cao15a}.   Such a field can cause a gap to form as it breaks the rotational symmetry that protects the Dirac point \cite{WangA3Bi2012}.   Evidence for a non-trivial phase also exists in the quasi 2D system purported Dirac system ZrTe$_5$ \cite{Yuan15a}, although it is still unclear if this system has small band gap and is in fact trivial \cite{Zhang17a}.  This system is believed to be close to a band touching transition \cite{Weng14a} and may be very sensitive to materials preparation.

Interpretation of similar experiments on the TaAs class of materials are much more complicated.   For instance, de Haas-van Alphen measurements \cite{Sergelius16a} on NbP show signs of multiple bands some of which are Weyl FS candidates with low cyclotron masses and a non-trivial Berry phase, and some of which are parabolic with a higher effective mass and close to trivial Berry phase.   For fields applied in the [100] and [010] directions the ``$\beta$" band is identified as a Weyl FS with a non-trivial Berry phase's offset of 0.48$\pi$.   This band was believed to come the W2 Weyl pocket.   Howeiver, showing the complexity of the interpretation of such experiments, an unidentified ``$\theta$" band with a Berry phase of 0.54$\pi$ was found from fields applied in the [001] direction.   This feature is not straightforwardly assigned to the W1 Weyl node as it is expected to be $\sim$ 60 meV below E$_F$ in a manner shown in Fig. \ref{TaP} showing the limitations inherent in using this technique for a definitive topological characterization in a multiband system.  A number of other bands were found with intercepts of less than 0.25 $\pi$ that are likely deriving from conventional parabolic bands.  Similar results have been obtained using Shubnikov-de Haas oscillations \cite{Hu16a}.  The situations seems to be simpler in TaAs due to the more favorable positioning of $E_F$, but even there there are three different types of Fermi surface pockets oscillations found in magnetization, magnetic torque, and magnetoresistance measurements \cite{Arnold16b}.  From a comparison to band structure calculations, two appear to be topologically non-trivial electron pockets around the W1 and W2 points and one is a trivial hole pocket.

In principle magneto-optical experiments have the possibility of revealing the Berry's phase in semimetals \cite{Illes15a,Malcolm15a}, but such experiments and analysis have not been attempted in WSM or DSM candidates.  They have given important such information in graphene \cite{Orlita10a}.

\begin{figure}[htp]
\includegraphics[width=0.95\columnwidth]{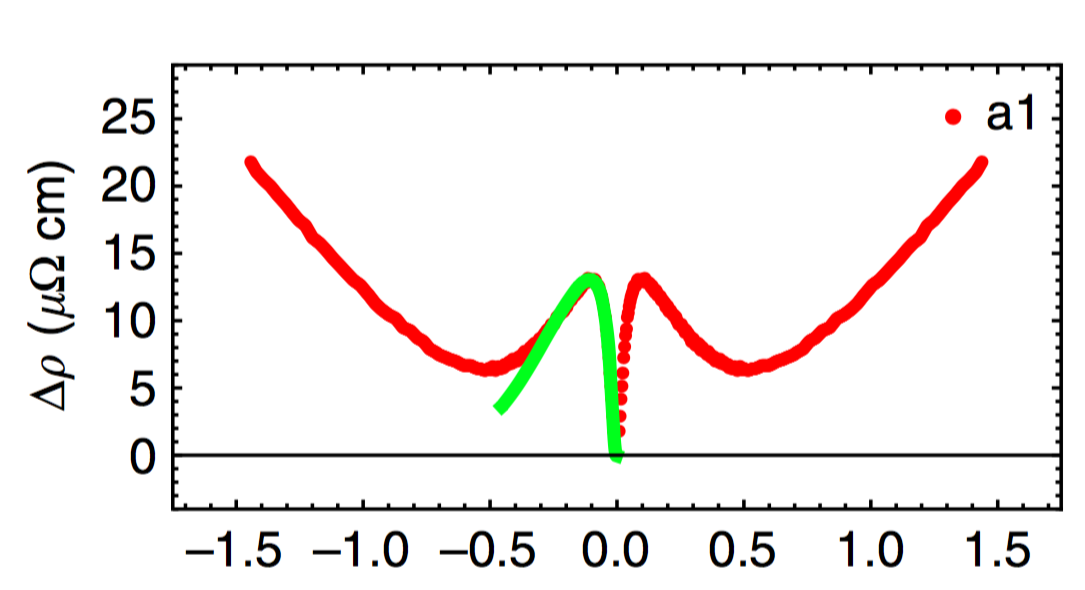}
\includegraphics[width=0.95\columnwidth]{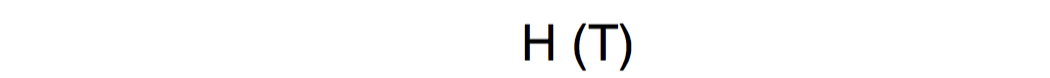}
\includegraphics[width=0.95\columnwidth]{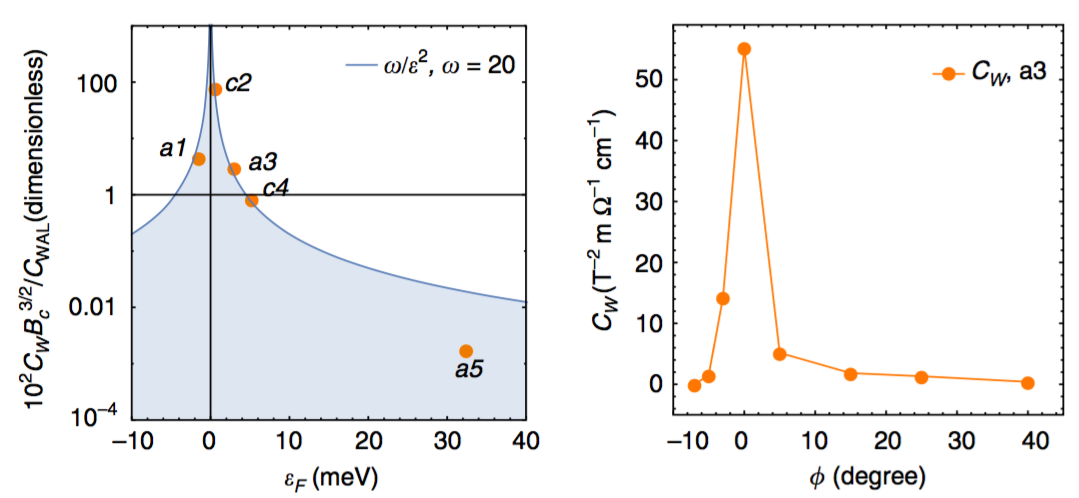}
\caption{ (Top) Inferred longitudinal MR for TaAs samples at 2K for $E$ and $B$ in the $a$ crystallographic direction. The green curve is a fit to the longitudinal MR data in the semiclassical regime based on a chiral anomaly model.  (bottom left)   Chemical potential  dependence of the chiral coefficient $C_W$.   (bottom right) Angular dependence of the chiral coefficient $C_W$.  Adapted from \cite{Zhang16a}. }
 \label{WeylMR}
\end{figure}

\begin{figure*}[htp]
\includegraphics[width=0.8\columnwidth]{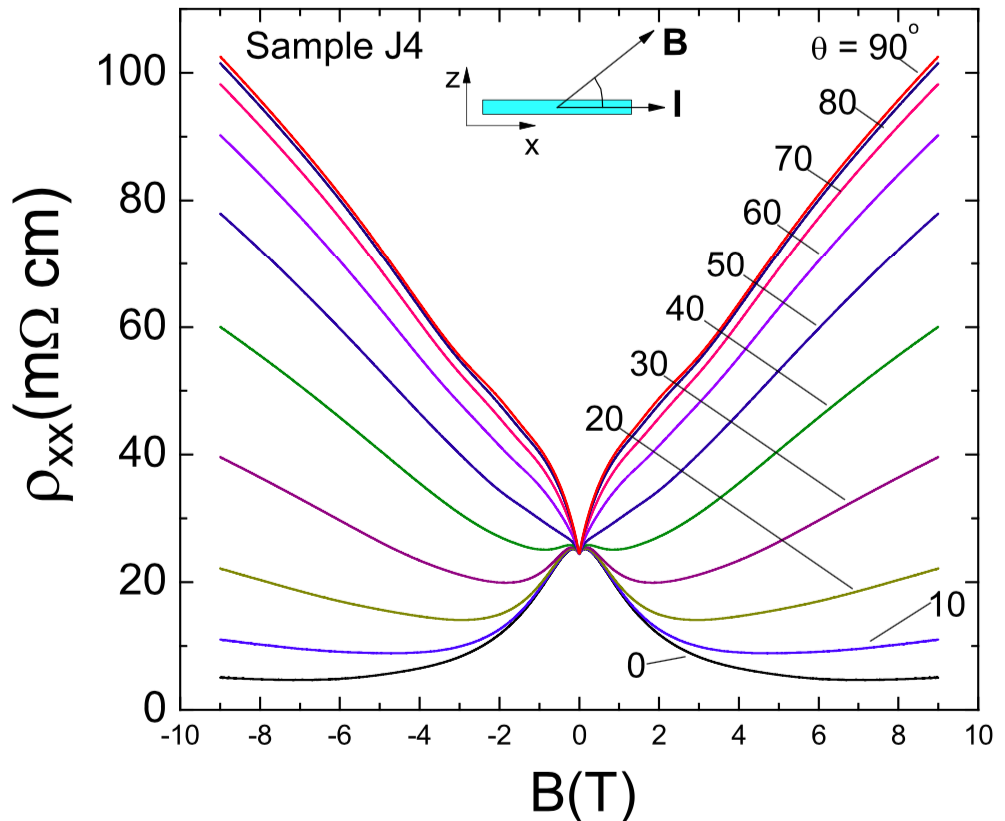}
\includegraphics[width=0.8\columnwidth]{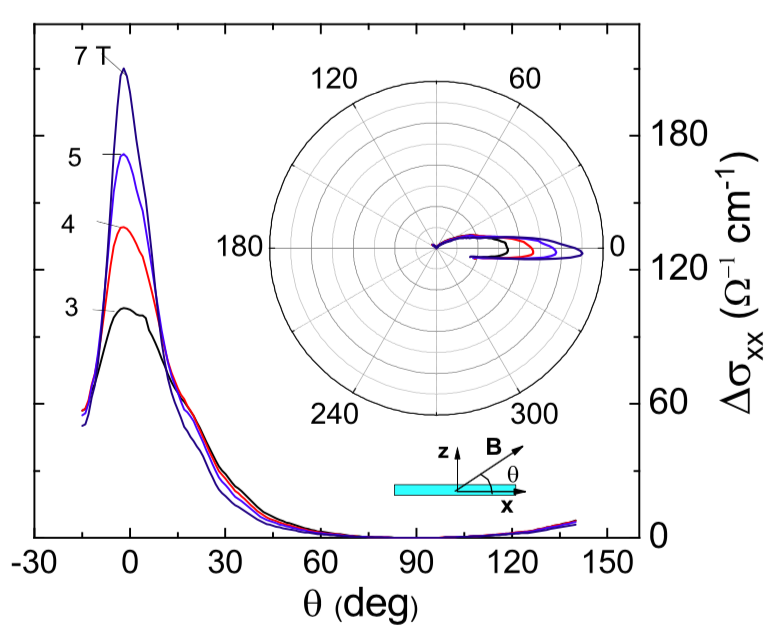}
\caption{ (left)  Inferred magnetoresistance of Na$_3$Bi when  $\textbf{B}$ lies in the $x-z$ plane at an angle $\theta$ with respect to  $\textbf{E}$ that points in the $x$ direction. Plotted as function of field for different angles.  (right) Magneto-conductance plotted as a function of angle for different fields.  Adapted from \cite{Xiong15a}. }
 \label{CurrentPlume}
\end{figure*}

\subsubsection{The chiral anomaly}

As discussed in Sec. \ref{WeylSection}, one of the much touted properties of the WSM and DSM systems has been that of the chiral magnetoresistance effect (CME).  This is an effect that derives from a non-zero $\textbf{E} \cdot \textbf{B}$ that can pump charges from one Weyl cone to the other through their band structure connection below $E_F$.  The axial charge pumping creates an out-of-equilibrium distribution of charges between the Weyl nodes.   In steady state, the charge pumping is relaxed  via through internode scattering.   The chiral anomaly in a WSM will lead to a negative magnetoresistance when the magnetic field is parallel to the current.  In contrast in metals or conventional semiconductors, the longitudinal magnetoresistance (MR) is typically weak, positive, and usually not very sensitive to the magnetic field direction. Therefore, a negative longitudinal MR which depends on the relative orientation of $\textbf{E}$ and  $\textbf{B}$ has been regarded as the most prominent signature in transport of the existence of 3D Weyl points.  A strong negative angular longitudinal MR has been reported for TaAs, NbAs, TaP, NbP and interpreted as this chiral magnetic effect \cite{Huang15c,Zhang16a,Yang15b,Du16a,Wang16f,Shekhar15a}.   Shown in Fig. \ref{WeylMR} is some representative data from \onlinecite{Zhang16a}.  Three principle regions are seen as a function of the magnetic field.  At fields close to zero, an initial sharp increase in the resistance is seen.  Although the low field MR has a shape that corresponds to the low temperature $-\sqrt{B}$ contribution to the magnetoconductance of the weak-anti-localization effect expected in a WSM \cite{lu2015weak}, the fitted coefficient is far larger than expected.  At intermediate fields, the negative longitudinal MR is found.  

At even higher fields, the longitudinal MR starts to increase again.   These features were found to be largely independent of the direction of the applied fields with respect to crystallographic axes as long as  $E$ and $B$ were co-aligned.  As discussed in Sec. \ref{ChiralAnomaly}, in the CME the negative longitudinal MR can be fit with a contribution to the magnetoconductance that goes as $\Delta \sigma_{CME}  = N C_W B^2$, where $N$ is the number of Weyl nodes and $C_W$ is the chiral coefficient, the simplest form of which is $ \frac{e^4 \tau_a}{4 \pi^4 \hbar^4 E_F^2)}$ \cite{SonSpivak,Burkov15a}.   Here $\tau_a$ is the internode relaxation time.    As shown in Fig. \ref{WeylMR} the fitted chiral coefficient was found in Ref. \onlinecite{Zhang16a} to have both a strong dependence on the relative angle between $\textbf{E}$ and $\textbf{B}$ and on the sample's Fermi energy.

Similar behavior has been seen in DSMs like Bi$_{0.97}$Sb$_{0.03}$ \cite{Kim13a}, Na$_3$Bi \cite{Xiong15a}, Cd$_3$As$_2$ \cite{Liang15a,Feng15a,CZhang15,li2015giant,li2016negative}, and ZrTe$_5$ \cite{Li16a}, and the quadratic band touching system \cite{Hirschberger16a}.   In these systems a magnetic field may create Weyl nodes and allows charge to be pumped from one node to the other in a way that is forbidden in a DSM in zero field.   As discussed in Sec. \ref{DiracChiral}, unlike the WSM case, it is not the momentum difference between nodes, but very same symmetry that protects the Dirac node that is expected to suppress the intervalley scattering.  In Fig. \ref{CurrentPlume}, we show some representative data on Na$_3$Bi.   One can see the same characteristic strong dependence of a negative longitudinal MR signal on the relative angle of $\textbf{E}$ and $\textbf{B}$ as in TaAs.   It is interesting to note that this data does not have the large positive longitudinal MR at low and high $\textbf{B}$ characteristic of TaAs and its family members.

Although such data has been extensively interpreted as evidence for the CME, it is not clear that in most cases the materials are in a regime that the effect can be realized easily.  Firstly, in materials like TaP, it appears to be \cite{Arnold16a} that $E_F$ is such that the electron and hole Fermi-surface pockets surrounding the W1 nodes contain a pair of Weyl nodes and hence the total Berry flux through the Fermi surface is zero.   Secondly,  it has been pointed out recently that many of these materials have a large enough transverse magnetoresistance to have the effect corrupted by the classic ``current jetting" phenomenon in compensated semiconductors \cite{Yuan16a,DosReis16a} in which current becomes narrowly directed along the applied field due to a very large field induced transverse resistance.  Historically, the observed effect was known as ``anomalous longitudinal magnetoresistance" in materials like antimony and bismuth \cite{Yoshida76a,Babiskin57a,Steele55a}, but was later shown to arise from current inhomogeneity inside the sample from large transverse magnetoresistance.  It is a strong effect in materials that possess a large field-induced anisotropy of the conductivity, such as almost compensated high-mobility semimetals \cite{Pippard89a}.  In compounds like bismuth the transverse MR can be larger than as $ 10^7$ at 4.2 K in 5 T \cite{Alers53a} causing current to flow in the direction of the applied magnetic field and almost independent of the direction of $\textbf{E}$.  Different mechanisms can give rise to strong transverse MR, but all predict it to be enhanced in low density systems.  Classic two-band magnetotransport predicts a parabolic-field-dependent magnetoresistance with a magnitude that is enhanced with increasing mobility.   For exactly compensated semiconductors, this classic transverse MR should not saturate.   In another mechanism, it was predicted \onlinecite{Abrikosov98a} that a \textit{linear} field dependence of the MR is expected near a linear band touching when the magnetic field is beyond the quantum limit. Spatial mobility fluctuation have been also predicted to cause linear MR anisotropy \cite{Parish03a}.  Also note that the possibility of similar phenomena have been suggested for more generic 3D Fermi surface without the topology of a Weyl node in the presence of parallel electric and magnetic fields \cite{Goswami15a,andreev2017longitudinal}.  Moreover, it has been pointed out that even materials that have their chemical potential above the van Hove point in the Weyl band structure have a nontrivial Berry curvature (despite there being no well defined chirality), which may give a quadratic in field contribution to the magnetoconductivity \cite{Cortijo16a}.


 \begin{figure}[htp]
\includegraphics[width=1\columnwidth]{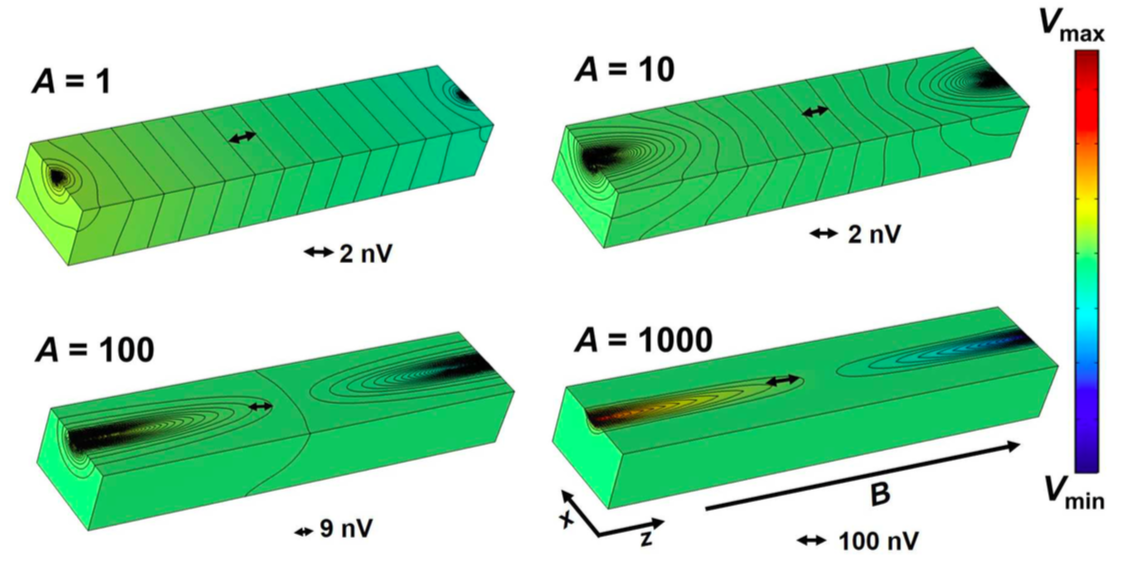}
\caption{ Simulated potential distribution for different conductivity anisotropies $A = \sigma_{zz}/ \sigma_{xx}$ and an geometry of $0.4 \times  0.3 \times  2.0$ mm$^3$ ($w \times t \times l$).  The lines are contour lines of the equipotentials. The increased MR anisotropy strongly distorts the equipotential lines. From Ref. \cite{DosReis16a}.}
 \label{CurrentJetting}
\end{figure}

The possibility of current jetting deserves further discussion.  For samples with a large conductivity anisotropy $A = \sigma_{zz}/ \sigma_{xx}$ current flows predominantly in the high conductivity direction.   Large conductance anisotropies can occur in systems with large transverse MR, such as compensated semi-metals \cite{Pippard89a}.  It is known that for small contacts and samples with a non-ideal aspect ratio, and for cases of a very strong anisotropy current forms a ``jet" between the contacts.  This very non-uniform current distribution inside the sample means that the experimentally measured potential difference between voltage contacts placed between the current contacts is not proportional to the intrinsic resistance.  The inhomogeneous current distribution manifests itself in additional characteristics, such as a strong dependence of the inferred longitudinal MR on the position of the contacts and very strong angular dependences.  Even negative total resistances can be observed if the magnetic field is not extremely well aligned with the current direction \cite{DosReis16a}.  Shown in Fig. \ref{CurrentJetting} is a simulation for the potential distribution for different conductivity anisotropies $A$, which may reflect strong transverse magnetoresistance from a magnetic field in the $\hat{z}$ direction \cite{DosReis16a}.  A large MR anisotropy strongly distorts the equipotential lines.   Even for anisotropies as low as 10, the effect is visible in even this close to ideal geometry with the $l/w$ aspect ratio close to 5.

\begin{figure}[htp]
\includegraphics[width=1\columnwidth]{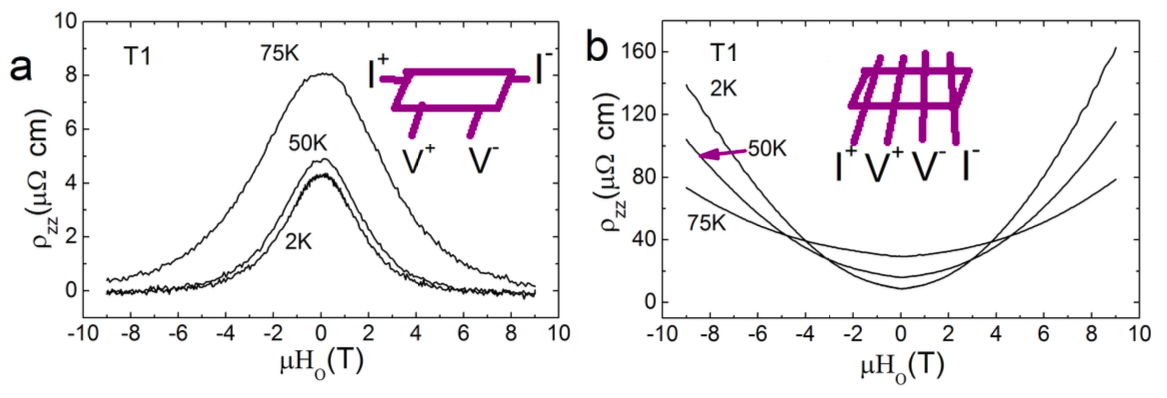}
\caption{ Demonstration of the current jetting effect in the non-WSM or DSM system TaAs$_2$.   (a) Measured apparent  longitudinal MR for when the contacts are not fully crossing the sample.  (b) The \textit{same} sample but for contacts placed such that fully cross the sample.   From Ref. \cite{Yuan16a}.}
 \label{TaAs2MR}
\end{figure}

The TaAs material class has both extremely high mobility ($\sim 10^5$ cm$^2$/V $\cdot$ sec  ) and large transverse magnetoresistance ($\sim 80000 \% $) at low temperature, \cite{Huang15c} in all cases making it likely that current jetting dominates even for a large aspect ratio Hall bar geometry.  In the TaAs case, the negative MR disappeared as the field was rotated only 2$^\circ$ away from the current \cite{Huang15c}.   Moreover, the strong dependence of the observed negative longitudinal MR on $1/E_F$ from Fig. \ref{WeylMR} is naturally explained in terms of the strong transverse magnetoresistance of an almost compensated semiconductor being projected into the longitudinal direction.  Strong negative longitudinal MR has also been inferred recently for non-WSM non-DSM systems like TaAs$_2$ and NbAs$_2$ \cite{Yuan16a,Luo16a}.   Convincing evidence for current jetting in this material class has been shown, that by simply putting  voltage probes across the whole sample  \cite{Yuan16a} the effective negative MR could be made to disappear (Fig. \ref{TaAs2MR}). It is fair to say that at the time of this writing there is no convincing evidence of the chiral anomaly in WSM systems.   All existing experiments appear to be dominated by the current jetting effect.

 \begin{figure}[htp]
\includegraphics[width=0.47\columnwidth]{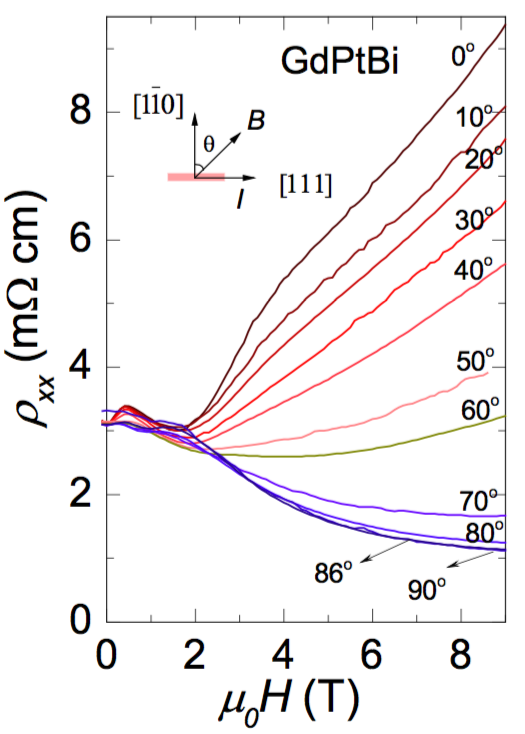}
\includegraphics[width=0.48\columnwidth]{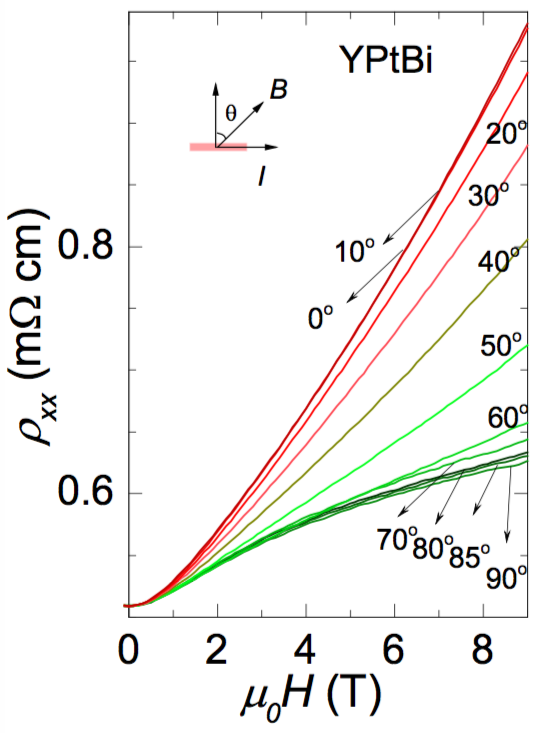}
\caption{Angular dependence of magnetoresistance of GdPtBi and YPtBi.  Adapted from \cite{Shekhar16a}.    }
 \label{GdPtBi}
\end{figure}

Although the situation may be less severe in the DSMs with lower mobilities ($\mu \sim 2,600$ cm$^2$/V $\cdot$ sec in Na$_3$Bi \cite{Xiong15a}, but still as high as 9 $\times 10^6$  cm$^2$/V $\cdot$ sec in Cd$_3$As$_2$ \cite{Liang15a} ) and smaller transverse MRs (approximately 10 at 10 T in Na$_3$Bi \cite{Xiong15a}, but greater than 200 in ZrTe$_5$ \cite{Li16a}, and almost  1000 in Cd$_3$As$_2$ \cite{Liang15a} at approximately the same field) the anisotropy can still be appreciable.   Moreover some of the experiments performed on these materials have been on samples with far less than the ideal shape that would minimize the effects of current jetting.  These issues may be resolved by experiments that vary the sample shape and position of the contacts. Current jetting is expected to be less of a concern in the quadratic band touching system GdPtBi (the transverse MR is only of order 3)  \cite{Suzuki16a,Hirschberger16a,Shekhar16a} and some checks for inhomogeneous current have been performed \cite{Hirschberger16a}.  It was proposed \cite{Shekhar16a} that the mechanism for turning a quadratic band touching into a Weyl system in field was enhanced by the role of Gd moments (even with the large $g$ factor of 40 in GdPtBi, the Zeeman scale is too small to split the nodes appreciably) as no such negative MR was seen in the YPtBi system (Fig. \ref{GdPtBi}).  However, one must bear in mind that as this is a magnetic system and possible alternative origins of the angular dependence is directional magnetoresistance.   In such a mechanism, the spin is assumed to follow the direction of the applied field and spin-orbit coupling gives an anisotropy in the scattering rate when current is aligned along or perpendicular to this direction \cite{Gorkom01a}.

To what extent can current jetting account for the negative longitudinal MR of DSMs and quadratic band touching systems? This is an important open question for future investigation. Since current jetting can manifest even for small anisotropy ratios $A$ when geometries are non-ideal, further experiments are needed to clarify the origin of longitudinal MR.  We also note that in virtually ever experiment to date, the relative angular dependence is much stronger than the $\textbf{E} \cdot \textbf{B}$ form would suggest.   Although there have been proposals about how such deviations may arise intrinsically in a DSM \cite{Burkov16b} (See Sec. \ref{DiracChiral}), as an exceedingly strong angular dependence is the precise expectation from current jetting, any deviations from cos$\theta$ need to be carefully considered.

Given all the above considerations, is it still possible to extract information about the chiral magnetic effect?  One must ensure current homogeneity inside the sample.   As shown in Fig. \ref{TaAs2MR}, one may improve the reliability of the measurement by using both voltage and current probes that reach across the sample, but even then it can still challenging to get a homogenous current.   Even for uniformly applied silver paste contacts applied on the entire end of the sample, current tends to enter where the local contact resistance is minimum. In classic experiments on samples like potassium metal that have a very large transverse magnetoresistance, \onlinecite{Lass70a} used contacts made with liquid Hg that made an amalgamated bond with a long skinny sample, but even there it was judged that the current distribution inside was inhomogeneous.   To do such measurements, one wants to use an ideal geometry with large aspect ratio $l/w$.  The resistance anisotropy can be viewed as an effective changing of the aspect ratio by a factor of $1/ \sqrt{A}$ \cite{Pippard89a}.   Obviously anisotropy ratios of $10^3$ in the WSM like TaAs puts severe constraints on sample geometries.   And with conventional aspect ratios of order $5$, the effects of inhomogeneous current distribution can manifest even with the comparatively low anisotropies found in DSM systems. We believe these issues make it hard to demonstrates conclusive signatures of the CME in the longitudinal MR for high mobility systems.   Other (non-contact) probes of ``dc" transport are essential.  In this regard, the older classic literature may be a guide \cite{Simpson73a}.

One intriguing result that may be evidence for the chiral anomaly is the observation of large magneto-optical Kerr effect in Cd$_3$As$_2$ crystals the size of which is dependent on the applied in-plane $\bf{E \cdot B}$.  Even up to room temperature, \onlinecite{CZhang15} found that the Kerr rotation followed an almost pure cos$\theta$ dependence of $\bf{E \cdot B}$  with a maximum Kerr rotation of over 0.04$^\circ$.    Putting aside its unexplained extremely large value (which is even greater than that found in some ferromagnets \cite{xia2009critical}) this result closely follows the prediction of \onlinecite{Hosur14a}, in which  coaligned electric and magnetic fields will pump charge into one Weyl node from another, allowing the intrinsic gyrotropic coefficient of a node to manifest, giving net optical activity.

\subsubsection{Surface state transport}

As discussed abvoe, the surface of WSM and DSM systems are expected to have unique properties.   Unfortunately, due to the inherently conducting nature of the bulk of these materials, surface transport signatures are hard to isolate.   This is unlike the case of topological insulators, where after sufficiently insulating bulk materials were grown, definitive surface transport signatures soon followed \cite{Wu16a,Xu14a,Checkelsky11a,Analytis10a}.

One of the most promising avenues to isolate the surface transport is through measurements that take advantage of the hybridization between surface and bulk states.   As discussed in Sec. \ref{Sec:AnomalyArcs}, \onlinecite{Potter2014} proposed a unique form of quantum oscillations in WSMs that involves the hybrid motion of electrons through half an orbit on the surface via a Fermi arc, transport through the bulk to the bottom surface, half an orbit on the other surface via the other Fermi arc, and then transport back to the beginning of the arc on the top surface.  The relevant quantization condition is given in Eq. \ref{ZhangEquation}.  The orbit is very different from the typical closed path that electrons take on going around a conventional Fermi surface.  Due to phase factors accumulated in the propagation through the bulk (where no Lorentz force is experienced) there is an explicit dependence to the observed signal on the thickness of the sample and/or the length of the classical trajectory.

\begin{figure}[htp]
\includegraphics[width=0.95\columnwidth]{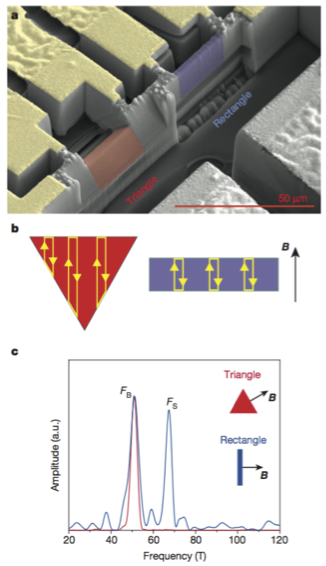}
\caption{a.) Scanning electron microscope image of triangular and rectangular devices used to observe hybrid surface-bulk quantum oscillations.  The rectangular sample is 0.8 $\mu$m wide, 3.2 $\mu$m tall and 5 $\mu$m long. The other device features an equilateral triangular cross-section with a base of $a = 2.7$ $\mu$m. Both devices have a similar cross-sectional area and circumference of the cross-section. The crystallographic direction perpendicular to the surface of the rectangular device is [102] and [010] parallel to the surface.  b.) Sketch of the  hybrid surface-bulk Weyl orbits for rectangular and triangular cross-sections. c.) Frequency spectrum of the triangular and rectangular samples, for field orientations perpendicular to each of the surfaces (0$^\circ$ for the rectangle and 60$^\circ$ for the triangle).
  From Ref. \cite{Moll16a}.}
 \label{SurfaceTransport}
\end{figure}

\onlinecite{Moll16a} investigated the possibilities of these unique orbits in Cd$_3$As$_2$ by performing transport experiments on focused ion beam machined nanostructures of a variety of shapes and size scales ($\sim$150 nm).  As discussed above, Cd$_3$As$_2$ is a DSM and to manifest this physics requires the material functioning as two independent Weyl subsystems overlapping in $k$-space.  Experiments on these samples give evidence for these unconventional orbits in a few different ways.   When field is applied in the long direction of the sample, quantum oscillations are exhibited, the frequency $F_B$ of which that are in good agreement with previous measurements of bulk crystals \cite{He14a}.  However, when the field is applied in the thin direction of the sample, a second oscillation frequency $F_S= 61.5$ T appears that can be distinguished from the higher harmonics of the bulk.   This higher frequency exhibits a distinct $1/\mathrm{cos} \theta$ variation with the angle of applied field that is emblematic of two-dimensional Fermi surfaces. However the frequency spectrum of the additional quantum oscillations is found to be strongly thickness dependent and is only observed in samples where the sample thickness is shorter than the bulk mean-free-path showing the bulk plays an essential role in the fashion anticipated.

Shown in Fig. \ref{SurfaceTransport}a from Ref. \cite{Moll16a} is a scanning electron microscope of two devices made with triangular and rectangular cross sections.  Both devices have a similar cross-sectional area and circumference.   As sketched in Fig. \ref{SurfaceTransport}b, the expectation is that the unconventional orbits for the two devices should be very different.   In the data the rectangular device clearly shows frequencies that reveal the presence of both unconventional orbit and the conventional bulk one, the triangular device shows only the bulk frequency.   Presumably this occurs because in the triangular device, destructive interference results from the sum of oscillations with a random phases, rendering the quantum oscillations unobservable in experiment.  However, given that Cd$_3$As$_2$  is a DSM, it would be important to reproduce this experiment, particularly on a WSM material.

Evidence for surface state transport has also been inferred from transport measurements on low carrier concentration Cd$_3$As$_2$ nanowires that have a large surface-to-volume ratio \cite{wang2016aharonov}.   When large enough field is applied along the length of the nanowire, they find that the conductance oscillates as a function of magnetic flux with peaks at $\Phi = (n + \frac{1}{2}) \frac{h}{e}$ e.g. peaks at odd integers of $h/2e$ with a period of $h/e$.   The $\frac{1}{2}$ is interpreted as it is in the case of topological insulator nanoribbons \cite{peng2010aharonov} as indicative of a $\pi$ Berry's phase \cite{zhang2010anomalous,bardarson2010aharonov}.

\subsubsection{Non-linear probes}

In addition to the considerable contribution that optical techniques in the limit of linear response can make to the study of these systems, it appears that non-linear optical probes may be able to give particular insight.  Generally these effects such as high harmonic generation, photovoltaic effects, shift (dc) currents, and nonlinear Kerr rotations, are related to the Berry connection and Berry curvature \cite{Hosur11a,Moore10a}.  For instance, it can be shown that the second harmonic generation (SHG) signal can be related to the shift vector $\mathbf{R}$, which is a gauge-invariant length formed from the momentum derivative of the phase of the velocity matrix element and the difference in the Berry connection \cite{Morimoto16a}.  It has been predicted \cite{Morimoto16b,Wu16a} that transitions that arise near Weyl nodes between bands with nearly linear dispersion should give a near universal prediction in the low $\omega$ limit for the non-linear susceptibility as

\begin{equation}
\chi^{(2)} = \frac{g(\omega) <v^2\mathbf{R}>}{2 i \omega^3 \epsilon_0}
\end{equation}
With the density of states $g(\omega)$ proportional to $\omega^2$, the SHG signal is predicted to diverge as $1/\omega$.   Although this result is reminiscent of the $\omega$ dependence of the optical conductivity discussed in Sec. \ref{OpticsTheory}, the 1/$\omega$ divergence is a unique signatures for inversion-breaking WSMs in particular because it vanishes in DSMs.  However, similar to the case of the linear in $\omega$ conductivity, the SHG divergence will be cutoff by disorder and nonzero Fermi energy in real materials.   Recently \onlinecite{Wu16a} have found a giant, anisotropic $\chi^{(2)} $ at 800 nm (1.55 eV) in TaAs, TaP, and NbAs, which may be related to this effect.  In the spectral range measured, the effect is of order 7000 $pm/V$, which is an order magnitude larger than in GaAs, which is the material with the next largest coefficient.  In the future it will be important to probe SHG and the shift current at even lower frequency to look for the dependence of $\omega$.  Other effects have been proposed such as a nonlinear Hall effect arising from an effective dipole moment of the Berry curvature in momentum space  \cite{Sodemann15a} and a photoinduced anomalous Hall effect\cite{Chan16b} for WSMs and photogalvanic effects \cite{Cortijo16b} in DSMs.

In DSM systems there is no direct photocurrent without driving electric field \cite{Shao15a}, but because of their spin selective transitions the photoconductivity is anisotropic for polarized radiation. \onlinecite{Chan16a} propose that inversion symmetry breaking WSMs with tilted Weyl cones (Type II most effectively) and doped away from the Weyl point, will be efficient generators of photocurrent and can be used as low frequency IR detectors.  Such a photocurrent has recently been demonstrated via the circular photogalvanic effect (CPGE)  \cite{xie2017direct} in TaAs.  The CPGE is the part of a photocurrent that switches its direction with changes to the handedness of incident circular polarization.   It can be shown \cite{Moore10a} to be sensitive to the anomalous velocity derived by Karplus and Luttinger \cite{Karplus54a} that was later interpreted as a Berry-phase effect \cite{Sundaram99a,Jungwirth02a}.  \onlinecite{xie2017direct} also pointed out that such experiments can also measure uniquely the distribution of Weyl fermion chirality in the BZ.

An interesting proposal of \onlinecite{Juan16a} was that of a quantized response also in the CPGE of inversion symmetry broken WSMs that possess no mirror planes or four-fold improper rotation symmetries (e.g. structurally chiral).  The CPGE usually depends on non-universal material details.  \onlinecite{Juan16a,Morimoto16b} predict that in Weyl semimetals and three-dimensional Rashba materials without inversion and mirror symmetries, that the trace of the CPGE is quantized (modulo multi-band effects that were argued to be small) in units of the fundamental physical constants.  It is proposed that the currents obey the relation

\begin{equation}
\frac{1}{2}\left[{dj_{\circlearrowright} \over dt}- {dj_{\circlearrowleft} \over dt} \right]= C {2\pi e^3 \over h^2 c \epsilon_0} I 
\label{eqinjection}
\end{equation}
where $C$ is the integer-valued topological charge of Weyl point and $I$ is the applied intensity.    Alternatively, the right hand side of the equation can be expressed as $ C {4\pi \alpha e \over h} I $ where $\alpha$ is the fine-structure constant.  In this expression, the currents for left and right circular polarization are perpendicular to the polarization plane.  Alternatively the quantity $\left[j^{\mathrm{sat}}_{\circlearrowleft}-j^{\mathrm{sat}}_{\circlearrowright}\right] $ may be measured if the relaxation time $\tau$ is sufficiently long and known independently.  An attractive property of this response is that it is related to the chiral charge on a single node.  The total node chirality in the BZ must of course be zero, however this does not prevent a CPGE.  In an $\mathcal{P}$ breaking material with no mirror planes, the Weyl nodes of opposite chirality do not need to be at the same energy.   One node can be Pauli blocked rendering it inert and giving a quantized response for some finite range in frequencies.   The proposed double Weyl system SrSi$_2$ \cite{Huang16b} that has no mirror planes (unlike TaAs) or RhSi \cite{Chang2017}  (which is predicted to have  six-fold-degenerate double spin-1 Weyl nodes and a four-fold-degenerate node) may be good candidates for this effect.  Its predicted magnitude is well within the range of current experiments.  Using the universal coefficient $ \frac{e^3}{\hbar^2 c \epsilon_0}$ = 22.2 $\frac{A}{W \cdot ps},$ \onlinecite{Juan16a} predict for $\tau \sim 1$ ps a steady state photocurrent of $\sim 2 \frac{nA}{W/cm^2}$, which is approximately 100 times that found in the topological insulator films \cite{Okada16a}.

It has also been appreciated that the low energy linear dispersion of WSM's and DSM's make them potentially useful for plasmonic applications particularly in the retarded (low frequency) limit \cite{ZhouPlasma15,Hofmann15a,Hofmann15b,Hofmann16a}.  $\mathcal{T}$ broken Weyl phases can support nonreciprocal one-way surface plasmon polaritons \cite{zyuzin2015chiral,Hofmann16a}.

\section{Related States of Matter}
\label{RelatedSection}

The electronic WSM and DSM states of matter discussed here provide inspiration and source for analogies for related realizations in other physical systems.   There may be Weyl or Dirac-like states of matter that do not have have electrons as their fundamental degrees of freedom (for instance photons \cite{Khanikaev13a}, phonons \cite{Xiao15a,Rocklin16a,Yang16a}, or magnons \cite{Li16b,Mook16a}).    Or WSM and DSM systems may serve as platforms for  phenomena like new forms of superconductivity \cite{Meng12a,Yang14a,Li17a}.  It is also the case that some aspects of DSM and WSMs were anticipated in essential aspects of exotic superconductors or superfluids \cite{Volovik87a,VolovikBook,Schnyder15} and important connections can be made here.  Moreover the stable band touchings found in WSMs and DSMs are just two of the possibilities for interesting semimetal states of matter.  Here we discuss some of these related states in detail.

\subsection{Topological Line Nodes} 
In addition to Weyl and Dirac nodes, other zero gap semimetal possibilities exist.  In Weyl and Dirac semimetals bands touch at points in the 3D BZ. If the band touchings occur along lines in the 3D BZ, these states are termed  topological  nodal line semimetals \cite{Burkov11b} and are reviewed in \cite{Fang16}.  Nodal line semimetals are expected to have particular transport properties including a number of different conduction regimes as a function of temperature, doping, and impurity concentration that arises from a Coulomb interaction that falls as $1/r^2$ over a large range in $r$ and a weak-localization correction with a strongly anisotropic dependence on magnetic field \cite{Syzranov16b}.  They may have been observed in ZrSiS \cite{Schoop16a,neupane2016observation}, PbTaSe$_2$ \cite{Bian16a}, HfSiS \cite{Takane16a}, and TlTaSe$_2$ \cite{bian2016drumhead} via ARPES.  Below we discuss some basic aspects of these systems. First, we note that typically the band touching lines are not expected to be at the same energy so the Fermi surface itself is not expected to be a line in the 3D BZ, but rather a collection of 2D surfaces. Under what conditions can line nodes appear? There are two distinct scenarios - depending on whether we neglect or include spin-orbit coupling. 

\subsubsection{Nodal lines in the absence of spin-orbit coupling} In the absence of spin orbit coupling (or in the limit where it is small), the spin SU(2) symmetry is retained and we can effectively ignore this degeneracy of the bands. If in addition we have both time reversal and inversion symmetry ${\mathcal P}$, the combined action $\tilde {\mathcal T} = {\mathcal T}{\mathcal P}$ leaves the crystal momentum unchanged and acts like a local time reversal symmetry in the BZ. Furthermore, we can set ${\mathcal T}^2=+1$ since we ignore the spin degree of freedom. With the same symmetries it is known that the 2D graphene Dirac nodes cannot be gapped, suggesting that it is a 2D \textit{topological} semimetal. By extension, we should expect a {\em line} node in 3D with these symmetries. Indeed as discussed in \cite{kim2015dirac} there is a ${\mathbb{Z}}_2$ index associated with loops in the BZ that encircle line nodes. This ensures the stability of the line nodes and also implies that graphene is a topological semimetal. Surprisingly, there is a second ${\mathbb{Z}}_2$ invariant associated with surfaces in momentum space, much like that used to characterize Weyl nodes \cite{Fang15}. 

To understand these in a unified fashion, let us revisit the procedure to identify topological semimetals, using Weyl semimetals as an example. We first choose a sub-manifold in the $d$-dimensional BZ that has dimension $d'<d$, where there is a band gap. We can then treat the band structure on this sub-manifold like that of a gapped insulator and use the classification scheme for band insulators summarized in Table \ref{8FoldWay}. For the case of Weyl semimetals, $d=3;\,d'=2$ and the absence of symmetry puts us in class A. The existence of an integer ${\mathbb{Z}}$ classification of $d'=2$ insulators by Chern number, implies that if there is a nontrivial Chern number on the surface it leads to Weyl semimetals with point nodes. Extending this to other situations with symmetry, requires the symmetry to be present for any choice of submanifold. The combination $\tilde{\mathcal T} = {\mathcal T}P$ fixes the crystal momentum, and one may then expect semimetals with this symmetry to be classified by the class AI for which $\tilde{\mathcal T}^2=+1$. However, unlike regular time reversal, the crystal momentum remains invariant under this effective time reversal, which does not allow us to read off the answer directly from the table. However, one can make the following adjustment - we can interpret the momentum as a real space coordinate which would naturally be left invariant under time reversal. Then, following \cite{Teo10}, who considered space dependent  band structures $H(k,r)$ to account for topological defects, the space coordinates can be treated as `negative' dimensions. This is meaningful because of the periodicity in dimension of the classification shown in Table \ref{8FoldWay}. Therefore, nodal lines in 3D can be captured by considering a $d'=1$ submanifolds, which can be classified according to $-1+8=7$ dimensional topological insulators in class AI. This has a ${\mathbb{Z}}_2$ classification consistent with the more direct calculation of \cite{Fang15, Morimoto2014}. However, we also notice that  a ${\mathbb{Z}}_2$ index for the $6$ dimensional insulator corresponds to a $d'=2$ submanifold. Indeed, there is an additional invariant for nodal lines in 3D that does not allow us to simply shrink a nodal line to zero \cite{Fang15}. As with Weyl nodes, these nodal `monopoles' must come in pairs to give net zero charge in the BZ.

 \begin{table}
  \centering
\begin{ruledtabular}
\begin{tabular}{c|ccc|cccccccc}
\multicolumn{4}{c|}{Symmetry } & \multicolumn{8}{c}{ $d$} \\
\multicolumn{1}{c}{AZ} &$\hspace{1.5mm}{\mathcal T}\hspace{1.5mm} $ &
                        $\hspace{1.5mm} \Xi\hspace{1.5mm} $ &
                        $\hspace{1.5mm} \Pi\hspace{1.5mm} $ &
 $1$   &  $2$ &  $3$ &  $4$ &  $5$ & $6$ & $7$& $8$ \\
 \hline
A & $0$ & $0$ & $0$  &$0$& $\color{red}\mathbb{Z}$ &$0$& $\mathbb{Z}$ &$0$& $\mathbb{Z}$ &$0$& $\mathbb{Z}$\\
AIII & $0$ & $0$ & $1$ & $\mathbb{Z}$ &$0$& $\mathbb{Z}$ &$0$& $\mathbb{Z}$ &$0$& $\mathbb{Z}$& $0$\\
\hline
 AI & $1$ & $0$ & $0$  &$0$&$0$&$0$&$\mathbb{Z}$&$0$&$\color{blue} \mathbb{Z}_2$&$\color{blue} \mathbb{Z}_2$& $ \mathbb{Z}$ \\
 BDI & $1$ &$1$ &$1$ & $\mathbb{Z}$ &$0$&$0$&$0$&$\mathbb{Z}$&$0$&$\mathbb{Z}_2$& $\mathbb{Z}_2$\\
 D & $0$ &$1$ &$0$ & $\mathbb{Z}_2$& $\mathbb{Z}$ &$0$&$0$&$0$&$\mathbb{Z}$&$0$&$\mathbb{Z}_2$\\
 DIII&$-1$ &$1$ &$1$ &$\mathbb{Z}_2$& $\mathbb{Z}_2$& $\mathbb{Z}$ &$0$&$0$&$0$&$\mathbb{Z}$&$0$\\
 AII & $-1$ & $0$ & $0$ &$0$&$\mathbb{Z}_2$& $\mathbb{Z}_2$& $\mathbb{Z}$ &$0$&$0$& $0$&$\mathbb{Z}$\\
 CII & $-1$ &$-1$ & $1$&$\mathbb{Z}$ & $0$&$\mathbb{Z}_2$& $\mathbb{Z}_2$& $\mathbb{Z}$ &$0$&$0$&$0$ \\
 C & $0$ & $-1$& $0$ & $0$ &$\mathbb{Z}$ &$0$&$\mathbb{Z}_2$& $\mathbb{Z}_2$& $\mathbb{Z}$ &$0$& $0$\\
 CI & $1$ & $-1$ & $1$& $0$ & $0$&$\mathbb{Z}$&$0$&$\mathbb{Z}_2$& $\mathbb{Z}_2$& $\mathbb{Z}$& $0$ \\

\end{tabular}
\end{ruledtabular}
\caption{Periodic table of topological insulators and superconductors.
The 10 symmetry classes are labeled using the
notation of Altland and Zirnbauer (AZ) \cite{PhysRevB.55.1142}  and are specified by presence or absence of $\mathcal T$ symmetry, $\Xi$ particle-hole symmetry and $\Pi ={\mathcal T}\Xi$ chiral symmetry. Here $\pm 1$ and 0 denotes
the presence and absence of symmetry, with $\pm 1$ specifying
the value of ${\mathcal T}^2$ and $\Xi^2$. As a function of symmetry and space
dimensionality, $d$, the topological classifications (${\mathbb{Z}}$, ${\mathbb{Z}}_2$ and 0) show a regular pattern that repeats when $d \rightarrow d + 8$.  From \cite{Ryu10}. Colored entries are referred to in the text. }
\label{8FoldWay}
\end{table}

Materials candidates in this class of nodal semimetals necessarily involve light elements for which the spin orbit interaction is expected to be weak. In graphene, this is believed to be the case, and some of the early proposals of 3D line node semimetals also involved carbon based structures \cite{Weng15,Chen15b}, models based on generalized 3D honeycomb networks in the absence of spin orbit coupling \cite{Ezawa16a}.  Other proposals include a new form of Ca$_3$P$_2$ \cite{Xie15}.  A review of materials candidates is contained in \cite{Fang16,Yu16}.

\subsubsection{Nodal lines in  spin-orbit coupled crystals} When spin-orbit interactions are included, the combination of inversion and time reversal alone are insufficient to protect nodal lines. Instead one necessarily requires a glide or  twofold screw symmetry which can protect a double nodal line, where two sets of doubly degenerate bands cross each other. When two orthogonal glides are simultaneously present, such nodal line band touchings are symmetry enforced \cite{Chen2016b}. An example of such a nodal line semimetal is furnished by SrIrO$_3$ \cite{Fang15,Chen2016b}.  

Generally, topological semimetals are expected to be accompanied by surface states like Fermi arcs in the case of Weyl semimetals. An additional requirement here is that the surfaces preserve the symmetries that protect the semimetal dispersion. In general the nodal lines band touchings are not all at the same energy. However, if particle-hole symmetry were additionally present, which would pin the entire nodal line at the same energy, the associated `drumhead' surface states involve a flat band over the surface BZ enclosed by the projection of the nodal line. In the absence of this additional symmetry, one may still discern nearly flat surface bands \cite{Burkov11b}. However, in superconductors with nodal lines, the additional symmetry is indeed present as we discuss below.

In ether case, when the bulk system is doped so that the Fermi surface surrounds the nodal line its Fermi surface may be susceptible to various interaction driven instabilities. This problem was studied by Nandkishore \cite{nand2016} who suggested that the leading instability on a toroidal Fermi surface in the particle-particle channel would lead to a fully gapped  $\mathcal T$-breaking chiral superconductor.   At lower density,  instabilities in the particle hole channel  lead to gapless states that can break either mirror or rotational symmetries  \cite{sur2016}.

\subsection{Relation to nodal superconductors and superfluids}
While there has been much recent effort dedicated to studying and realizing topological superconductors, with a full gap in the bulk and gapless modes at the edge, there is of course extensive history of work on nodal superconductors and superfluids, where the energy gap closes at points or lines in the BZ \cite{Leggett76,VolovikBook,Hu94}.  These can then be understood in the framework of topological semimetals, albeit with additional symmetries arising from the superconducting nature of the gap.  This can lead to protected  states at the boundary that are precisely at zero energy. Here we will focus on a few examples that are physically relevant, a more general discussion can be found in the reviews \cite{Schnyder15}.

Superfluid He-3 at milliKelvin temperatures forms a paired superfluid. While at ambient pressure a fully gapped topological superfluid is realized (the B-phase), at higher pressures a different `A' phase obtains, and is believed to be described by the Anderson-Brinkman-Morel order parameter. This pairing spontaneously breaks time reversal symmetry and leads to a pair of nodal points where the gap vanishes \cite{Leggett76}. Qualitatively it may be understood as a $p_x+ip_y$ superfluid where spin up fermions pair (spins being defined along the $z$ axis) with the same pairing function between spin down fermions. At the north and south pole of the Fermi surface where $p_x=0,\,p_y=0$ and $p_z=\pm p_F$, the pairing function vanishes and leads to opposite Weyl nodes. These are the superconducting analog of Weyl nodes; their relation to chiral fermions was investigated in \cite{Volovik87a,VolovikBook}. Here the role of charge conservation is played by spin rotation invariance about the $z$ axis, which is respected by the pairing function. A direct consequence is the presence of Fermi arc surface states at the boundaries of the superfluid, as in Weyl semimetals, which will be pinned at zero energy \cite{Heikkila11}. Other nodal superconductors can realize line nodes in the 3D BZ, and realize zero energy Andreev bound states at the surface BZ momenta that  lie within the projections of these line nodes \cite{Schnyder11}. Non-centrosymmetric superconductors such as CePt$_3$Si \cite{Bauer04} have been proposed as candidate material realizing  this physics \cite{Schnyder15}. In a 2D system, the equivalent is a point node in the gap. The best known example of this is the d-wave spin singlet superconducting gap of the cuprates. Here too there is a topological origin for the gap protection which leads to zero energy modes \cite{Hu94, Wang12} along certain edges that have been reported to have been observed in several tunneling experiments \cite{Walsh92}.

A related set of questions involve the properties of Weyl semimetals in the presence of superconductivity. For magnetic Weyl semimetals, the single Fermi arc surface state splits into a pair of chiral Majorana modes, which are attached to different gapless bulk nodes, which can be moved independently of one another \cite{Meng12a}. Turning to pairing in $\mathcal T$ symmetric Weyl semimetals, time reversal invariant pairing is expected to gap the nodes but could potentially lead directly to topological superconductivity as discussed in \cite{Qi10,Hosur14,Li17a}. For a Weyl metal with Fermi surface Chern numbers $C_i$ (the Fermi surfaces being labeled with the index $i$) and time reversal symmetric  superconducting pairing gaps $\Delta_i$ on the different Fermi surfaces, the resulting topological superconductor is labelled by an integer topological index $\nu$ which can be expressed as 
\begin{equation}
\nu = \frac12 \sum_i C_i {\rm sign}(\Delta_i),
\end{equation}
thus for the minimal situation of four Weyl nodes, if the pairing changes sign between the pair with $C_i=1$ and the pair with $C_i=-1$, this leads to a topological superconductor with $\nu=2$.  

The interesting case of pairing between Fermi surfaces surrounding nodes of opposite Chern number was discussed in Ref. \cite{Li17a}.  In this case, the complex-valued gap function cannot be globally well-defined over the entire Fermi surface; its distribution exhibits non-zero total vorticity and hence, the gap function must have nodes.  However the pairing symmetry cannot be described by the usual spherical harmonic functions (or their lattice analogs) since these describe regular functions over the Fermi surface.  It determines a novel topological class characterized by monopole charge rather than the usual unconventional nodal superconductivity which is typically characterized by angular momentum quantum numbers.  Notably, this \textit{monopole} superconductivity is determined by the normal state topology alone, rather than any particular pairing mechanism and thus should be very robust.  Although pairing in $\mathcal T$ broken superconductors takes special consideration,  if it exists in a $\mathcal T$ broken WSM (even through conventional electron-phonon mechanisms or via a proximity effect) a nodal structure is mandated.

\subsection{Quadratic band touchings and the Luttinger semimetal}
\label{Quad}
 
One can generally understand the physics of linear Weyl and Dirac semimetals within the framework of weakly interacting fermion theories.  In contrast one expects a number of important interaction effects to intervene in 3D quadratic band touching (QBT) systems.   Such effects are expected to be more pronounced that in linear band crossing systems due to the scaling of the density of states with energy and thus QBT systems are expected to be strongly interacting.  Seminal work by Abrikosov and Beneslavskii showed that in the vicinity of the band edge, quadratic band touching systems are always strongly interacting and that at  energies well below the exciton scale ($  \frac{ 2 \mu e^4 }{  \varepsilon_{\infty}^2  \hbar^2 } $) the single particle concept is inapplicable \cite{Abrikosov71a,Abrikosov71b}.  Here $\mu$ is the reduced mass of the conduction-valence band system and $\varepsilon_{\infty}$ is a background dielectric constant.  This was a remarkable demonstration almost fifty years ago of a non-Fermi liquid conductor.  Taking advantage of the inherent scale-free criticality in such a system and using an $\epsilon$-expansion about 4 spatial dimensions,  Abrikosov derived scaling relations \cite{Abrikosov74a} and the forms for various observables.  More recently \onlinecite{Moon13a} argue that for the 3D case the Coulomb interactions may stabilize a new stable non-Fermi liquid \textit{phase}, rather than driving the system to an instability.  They show that it can be understood as a balance of the screening of Coulomb interactions by electron-hole pairs and mass enhancement of the quasiparticles dressed by the same virtual pairs.   However, a number of other effects are possible.  Even more recently, it has been argued that in 3D and for the single band touching found in known materials that the quadratic band touchings are unstable at low energies to opening a gapped nematic  \cite{Herbut14a,Janssen15a} or $\mathcal{T}$ breaking phase \cite{Lai14a}.   A nematic phase has not been observed in experiment, although possibly relevant magnetic phases are observed in iridate pyrochlores \cite{Matsuhira07a}.  It has also been argued that because short-range correlated disorder scales identically to Coulomb interactions at tree level, and dominates them in a one-loop RG analysis,  the Abrikosov and Beneslavskii phase is unstable to disorder and may result ultimately in a localized phase \cite{nandkishore2017disorder}.

\begin{figure}[t]
\includegraphics[width=8.5cm, angle=0]{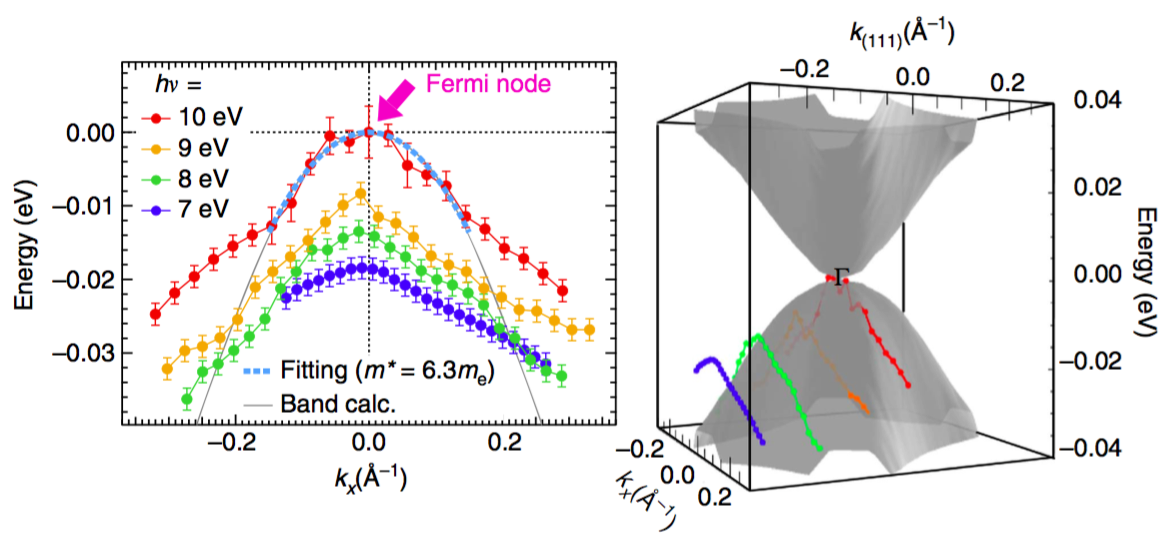}
\caption{(Left) ARPES measured dispersion curves along the k$_x$ direction measured at photon energies of 7, 8, 9 and 10 eV.  (The corresponding momentum cuts are indicated by the colored lines on the right figure).  The dispersion curve obtained by the band calculations is superimposed (grey curve). The data close to E$_F$ is fitted by a parabolic function shown as the light-blue (dotted) curve. The estimated effective mass at the $\Gamma$ point is 6.3 free electron masses. (Right)  The calculated band dispersion in the k$_x - k_{111}$ sheet. The ARPES data from the left panel are overlayed.  From \cite{Kondo15a}. }
\label{Quadratic}
\end{figure}

In the absence of these low temperature phase transitions, quadratic contact points between bands can be protected -- as they are in the 3D massless Dirac case -- by point group and time reversal ($\mathcal{T}$) symmetries that enforce a particular degeneracy.  Their valence and conduction bands belong to the same irreducible representation of the symmetry groups and the 4 fold degeneracy at the touching point cannot not be removed unless the symmetries are broken.  This is the case in well known materials such as $\alpha$-Sn and HgTe.   Of course, these materials have been of recent interest also due to the fact that they are near a topological band inversion transition that can be accessed under uniaxial strain or in a thin film geometry.  In their fully symmetric cubic state they posses a quadratic band touching (QBT) at the zone center.  In the non-interacting, disorder free, case these systems can be described in a minimal band structure by the Luttinger Hamiltonian for inverted gap semiconductors \cite{Luttinger56a} and so we term this phase with a stable QBT a \textit{Luttinger} semimetal.

 Although in principle the strong interactions proposed by  \onlinecite{Abrikosov74a} should exist in the classic HgTe and $\alpha$-Sn systems, such effects were never observed.   One expectation is that due to virtual excitations across the band touching the dielectric constant can become greatly enhanced.  Broad bands decrease the overall scale of the interaction effects and residual doping giving a finite chemical potential $E_F$ was sufficient to cutoff the divergences associated with these zero energy transitions.  In classic QBT systems like $\alpha$-Sn and HgTe the dielectric enhancement was relatively modest ($ \tilde{\varepsilon } \sim$ 3.5 and 7 respectively) \cite{Wagner71,Grynberg74a}.  It has recently been shown by ARPES \cite{Kondo15a} that the pyrochlore oxide Pr$_2$Ir$_2$O$_7$ \cite{balicas2011anisotropic,tokiwa2014quantum,machida2007unconventional} possesses a QBT (Fig. \ref{Quadratic}).   The QBT is believed to be formed between $J=3/2$ bands in the same fashion as the classic systems.   The effective mass of the conduction band was found to be approximately 6.3 m$_e$, which is almost 300 times that of $\alpha$-Sn \cite{Wagner71}.  Larger mass enhances the relative role of interaction and opens the possibility of probing the strongly interacting regime.  Recent optics work on  Pr$_2$Ir$_2$O$_7$ shows that the dielectric constant in this material becomes of order 200 at low temperatures demonstrating that this material is in the strongly interacting regime \cite{cheng2017dielectric}.  Recent magnetotransport measurements show that the purported quadratic band touching system GdPtBi has a mass of about 1.8 m$_e$ and an E$_F$ of about 3.1 meV \cite{Hirschberger16a}, which may also put this material in the strongly interacting regime.
 
Quadratic band touching points that are not topologically protected are generically unstable to cubic symmetry or $\mathcal{T}$  breaking perturbations and in this regard these systems can be viewed as ``parent" states to a number of topological phases \cite{Moon13a,Cano16a}.    For example uniaxial strain induces a gapped 3D TI phase as observed in HgTe \cite{Brune11a}.   Applied magnetic field allows non-degenerate bands to cross along the applied field direction and for fields in the (001) direction, give a pair of double Weyl points, with linear dispersion along the applied field direction axis and quadratic dispersion normal to it  \cite{Moon13a}.  These Weyl points correspond to $\pm$2 monopoles in momentum space.    Although the quadratic dispersion normal to the field is due to symmetry, the touching itself is protected by topology in the standard Weyl fashion.  The low frequency optical response of topologically protected double and triple Weyl nodes has been calculated \cite{Ahn16a} and reveals power law scaling in the parallel and perpendicular components of the low frequency conductivity that can be used to distinguish these states from linear WSMs. Note that creating a double WSM with a magnetic field in this manner is a fine tuned situation. Fields tipped away from the (001) direction will  cause the double Weyl point to split into two single Weyl points.  The consequences of this anisotropy on Landau quantization of the spectrum in strong magnetic fields has also been considered \cite{Li2016a}. An analogous route to realizing Weyl semimetals by straining HgTe was discussed in Ref. \cite{ruan2016symmetry}. 

\subsection{Kramers-Weyl nodes and ``New" Fermions}
As discussed above, WSMs are robust against small perturbations that preserve translational invariance.  However, one may destroy the semimetal state by changing Hamiltonian parameters (such as SOC) to uninvert their bands, as in all previously known WSM states, band inversion is an essential feature for system's realization.  When tuning the SOC, Weyl nodes may move and eventually pairwise annihilate to drive the system into a gapped trivial phase, while preserving all symmetries.  In contrast,  in $\mathcal{T}$-symmetric chiral crystals, Weyl nodes based on Kramers doublets can be locked at time reversal invariant points in the BZ, which makes them stable against annihilation with opposite Weyl nodes.  Space groups in which such ``Kramers Weyl fermions" could appear were analyzed in \cite{Chang16b} and density functional theory calculations were employed to identify materials candidates such as Ag$_3$BO$_3$. However, the splitting between bands arising from breaking of inversion symmetry is typically weak. As a result, although Fermi surfaces with Chern number are expected, they will typically occur in opposite Chern number pairs, closely separated in momentum. Characteristics associated with the Fermi surface Chern number are then observable only if impurity induced mixing between the two Fermi surfaces can be neglected.

One may ask what are other degenerate band touchings protected by crystal symmetry are allowed in addition to the ones discussed in this article (Weyl, Dirac, nodal line, double Weyl, quadratic)?  \onlinecite{Bradlyn16a}  show that in condensed matter systems, the usual field theoretic categorizations of the three kinds of fermions  of free space (Majorana, Weyl, and Dirac) is incomplete and that there are more possibilities that are stabilized by crystal symmetries e.g. ``New" fermions. It was shown that in the presence of $\mathcal{T}$ the only possibilities are 2, 3, 4, 6, and 8 fold degenerate band touchings.  These phases are stabilized as a consequence of non-symmorphic symmetries.  

It has also been shown recently that three-fold degeneracies can also be stabilized by rotation and mirror symmetries  \cite{Zhu16a,Chang16a,Weng16b} even in symmorphic structures.  The band touching is pinned to a high symmetry line that it can move along by tuning a Hamiltonian parameter. These three component fermions are intermediate between two component WSMs and four component DSMs and should have properties that are different from either.  There has been recent claim to have observed such triply degenerate points in MoP \cite{Lv16a} and in tungsten carbide WC \cite{ma2017three}.  
These states are important in the general framework of WSM and DSM phases as in some cases they can be seen as an intermediate phase separating Dirac and Weyl semimetals in materials with a C$_{3v}$-symmetric line.

In contrast to the band crossing induced threefold degeneracies, one may regard the symmetry protected threefold crossings as Spin-1 Weyl points as their Hamiltonian at low energies has form of $H = \textbf{ k} \cdot \textbf{ S }$  where instead of the Spin-1/2 Dirac matrices one has the Spin-1 $\textbf{ S }$  matrices.  Such degeneracies give a natural generalization of ``conventional" Weyl fermions.  To leading order these Spin-1 Weyl points are formed by two linearly dispersing bands bisected by a flat band.  If one computes the Berry curvature in a fashion discussed in Sec. \ref{WeylSection}, one finds monopoles of charge $\pm$ 2 as compared to the usual Weyl case of $\pm$ 1.  In contrast to the band crossing induced three fold degeneracies, the symmetry enforced chiral spin-1 Weyl fermions proposed in \cite{Bradlyn16a} occur at high symmetry points in the Brillouin Zone and can only appear in certain nonsymmorphic cubic space groups. Furthermore, these are necessarily accompanied by topological surface states similar to the Fermi arcs of Weyl semimetals. Obtaining experimental candidates with well isolated spin-1 Weyl fermions is however challenging.  In related space groups the combination of $\mathcal{T}$  and inversion results in a 6 fold degenerate Spin-1 Dirac system that is two copies of a Spin-1 Weyl systems with opposite chirality pinned on top of each other.  These can be seen as three-fold degeneracies that are doubled by the presence of $\mathcal{T}$ symmetry.   These are a symmetry protected version of the 6 fold degenerate Kane fermions that are expected at the fine tuned point of the band inversion transition in Hg$_{1-x}$Cd$_{x}$Te \cite{Orlita14a}.  Ab initio calculations have been carried out and a number of different materials realizations have been made \cite{Bradlyn16a}.   In keeping with this general picture, earlier  \onlinecite{Wieder16a} proposed the existence of ``double-Dirac" semimetals with an 8-fold degenerate touching in crystals with a non-symmorphic space groups.   This double Dirac semimetal can be gapped into a trivial or topological insulator by applying strain.  It is important to note that all these analyses have only been carried out for non-magnetic groups.  In principle the inclusion of broken $\mathcal{T}$ makes the possibilities even richer.  A particularly promising family of materials that is expected to combine multiple types of band touching and long Fermi-arcs on certain surfaces may be found in the AB material class (A = Co, Rh and B=Si, Ge) \cite{Chang2017, Tang2017}.

\subsection{Possible realizations in non-electronic systems}

\begin{figure}[t]
\includegraphics[width=8.5cm, angle=0]{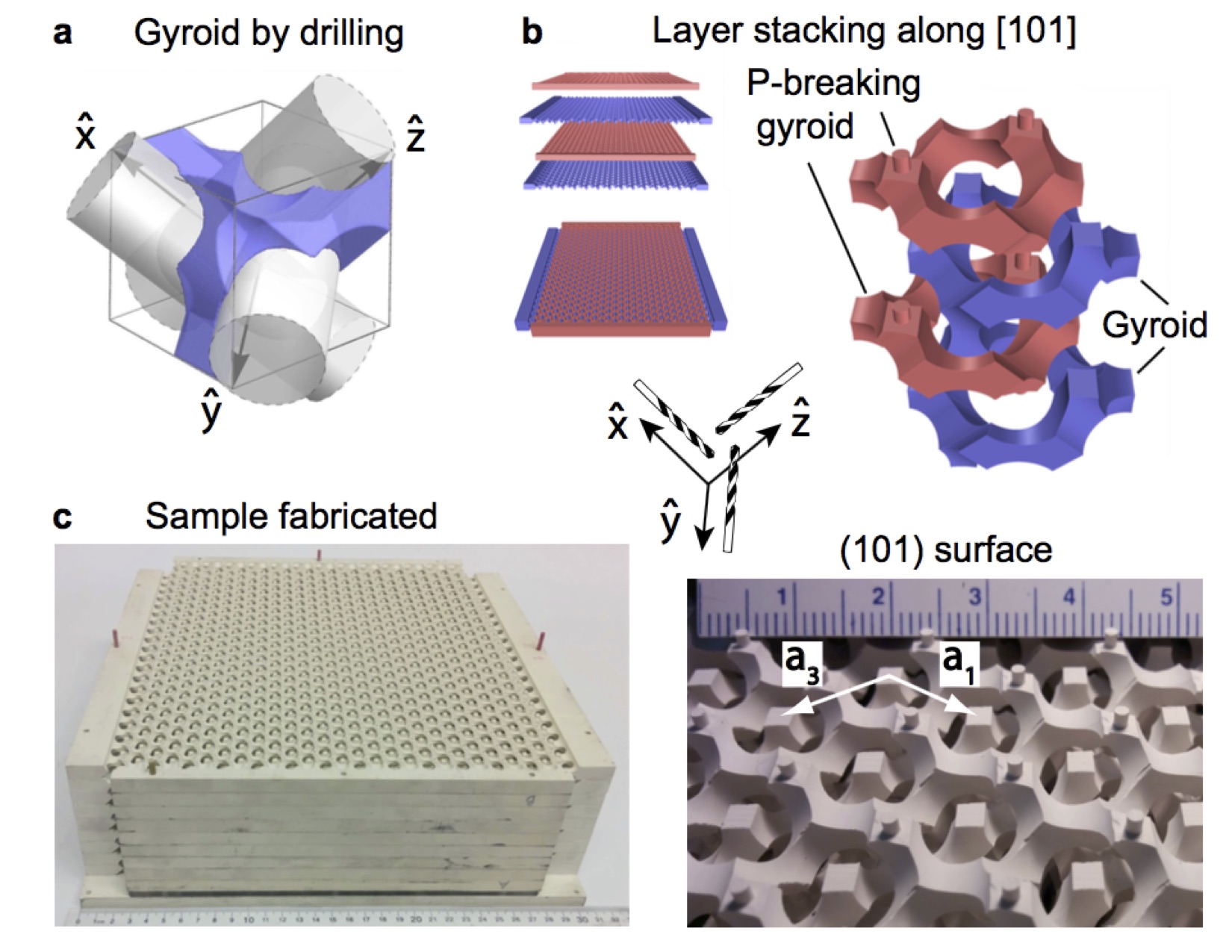}
\caption{ a) Gyroids can be fabricated by drilling holes along $x, y, z$ directions.  Shown is a bcc unit cell in which a single gyroid structure can be approximated by drilling holes. b) The double-gyroid structure is made by stacking layers along the [101] direction.  The structure is made with two inversion counterparts interpenetrating each other.  Inversion symmetry is broken by reducing the vertical connections to the thin cylinders for the red (first and third) gyroids. c) Shown on the left is the assembled structure.   A magnified view from top is shown on the right with a centimeter ruler in the background. From \cite{Lu15a}. }
\label{PhotonicWeyl}
\end{figure}

Motivated by the incredible interest in topological states for electronic systems, a rapidly emerging area is the realization of topological states for photonic systems \cite{Haldane08a,Wang09a,Umucalilar2011a,Hafezi13a,Khanikaev13a,Rechtsman13a,Lu2014a}.  To make a WSM-like system, \onlinecite{Lu15a} fabricated a precise array of holes (Fig. \ref{PhotonicWeyl}) into several ceramic layers, which they stacked together in a interpenetrating double gyroid structure to make a 3D photonic crystal with broken inversion symmetry.  This structure is predicted to host the electromagnetic analog of Type I Weyl nodes, which can be accessed by tuning the frequency of incident microwaves to the frequency where the Weyl node occurs. \onlinecite{Lu15a}  performed angle-resolved microwave transmission measurements and showed that the bulk showed two linear dispersing bands touching at four isolated points in the three-dimensional BZ, indicating the observation of Weyl points.  Using a 3D structure consisting of laser-written waveguides, \onlinecite{Noh16a} have observed photonic type-II Weyl points at optical frequencies, in a 3D photonic crystal structure consisting of evanescently-coupled waveguides.  There are  proposals for realizing Dirac dispersions in hyperbolic photonic \cite{Narimanov15a} and metamaterial crystals \cite{Xiao16a} and nonsymmorphic \cite{Wang16d} and other photonic crystals \cite{WangPhotons16a}.  In general there is interest in studying Weyl and Dirac points in photonic systems as they may potentially be used for large-volume single-mode lasing \cite{Bravo12a}.

In a homogeneously magnetized plasma, the cyclotron frequency can exceed the plasma frequency, which results in crossing points between the helical propagating mode at the plasma frequency and the longitudinal plasmon mode.  These crossing points in the momentum space are Weyl points with a finite Berry curvature and can in principle give non-trivial topological features \cite{Gao15a}.   Such a system is expected to have electromagnetic effects in reflection with no analog in electronic systems.

There have also been proposals to realize topological semimetals systems in ultracold atoms in optical lattices \cite{sun2012topological,jiang2012tunable,Dubvcek15a,Li15a}.  One would use laser-assisted tunneling to engineer the complex tunneling parameters between lattice sites.  For instance, \onlinecite{Syzranov16a} propose that Weyl-like dispersions can emerge  in 3D arrays of dipolar particles in the presence of a weak magnetic field due to dipole interaction induced transitions between internal angular-momentum $J = 0$ and $J = 1$ states.  Of particular interest here is that although the single particle properties of such a system would be expected to be the same as electronic Weyl systems, their many-particle properties are expected to be different, opening up the possibilities for new functionalities and applications beyond those accessible with solid-state systems.

In acoustic system, it has been proposed that inversion symmetry breaking through structurally engineering interlayer couplings can generate an effective gauge field \cite{Xiao15a}.  In 2D this can give an acoustic analog of the topological Haldane model \cite{Haldane88a}.  In 3D these acoustic systems possess topological Weyl points and are thus realization of the Weyl Hamiltonian in sound waves.   Recently it was proposed that a class of noncentrosymmetric crystals will have double Weyl points in both their acoustic and optical phonon spectra \cite{zhang2017double}.

\section{Concluding remarks}

Going forward there are many possible interesting avenues in the field of WSMs and DSMs.  First and foremost, we still need to find a truly ideal WSM realization e.g. the ``graphene" of WSMs.  One would like to find a material in which all Weyl nodes are symmetry related and close to E$_F$ with a large momentum separation and no non-topological bands near in energy.   Indeed, these are properties of the  graphene band structure that have proved useful in isolating the Dirac node physics in 2D. Further, one may require a small number of Weyl nodes, for example just the minimal pair of opposite chirality nodes allowed for magnetic WSMs. Although proposals for such systems exist \cite{Wang16c,Ruan16a}, all known realizations fall short of this ideal.  As discussed above, there have been recent attempts to optimize the screening of materials candidates focusing on promising space groups, supplemented with a filling condition which constrains chemical formulas \cite{Gibson15a,Chen16a}.  These directions combined with traditional materials searches, particularly in magnetically ordered systems which may host magnetic WSMs will hopefully lead to materials that can accurately be termed `3D graphene'.  It is interesting in this context, to note how common topological band touchings are.  First principles calculations have shown that even bcc iron appears to have many such band touchings (some deep in the bandstructure) of the Weyl, double Weyl, and nodal loop variety \cite{Gosalbez15a}.  Two Fermi pockets surround isolated Weyl points and are likely to give a major contribution to its anomalous Hall effect.  Furthermore we note that while there has been rapid theoretical development in classifying topological semimetal phases, and identifying materials candidates, the equally important work of characterizing them by identifying their signature properties and distinct phenomenology awaits a similar degree of development. 
  
The analysis of topological semimetals discussed in this review has been developed using the band theoretic language of free fermion theories on the lattice.  A conceptual frontier is to expand this to better understand the role of interactions.  Treated at the mean field level, interactions are important in driving states that violates $\mathcal{T}$.  This was a key feature in the early development of this subject and in the pyrochlore iridates where the interaction scale is comparable to the spin-orbit scale \cite{Wan11a,Krempa12} this leads to a rich mean-field phase diagram featuring topological insulating and semimetallic phases with transitions driven by the type of magnetic order.  Yet one recognizes that for interacting systems at strong coupling, other gapped ground states are possible that, while lacking a simple band theoretic representation, may nonetheless admit a useful topological classification.  A proof of principle is the demonstration of fermion fractionalization in toy models of two dimensional fractional topological insulators \cite{Levin09a}.   The investigation of similar effects in 3D systems even at the level of model Hamiltonians is in its infancy.  The evolution of toy models into theories of real materials and their ultimate material realizations poses an outstanding challenge to modern condensed matter science.  

There may also be applications potential for these system.   For instance, it has been recently proposed that one may utilize topological electronic states to enhance catalytic activity.  In this regard, it was shown that the combination of topological surface states and large room temperature carrier mobility (both of which originate from bulk bands of the WSM and DSM) may be a recipe for high activity hydrogen evolution reaction catalysts and may be used in solar energy harvesting to produce hydrogen from water \cite{Rajamathi16a}.  In the device domain, it has been proposed that WSMs in thin film form can be used to build a spin-filter transistor with a controllable spin polarized current.  A loop device made of 2D WSM with inserted controllable flux to control the polarized current has been demonstrated \cite{Shi15a}.  The device has good on/off ratios with controllable chemical potential induced by liquid ion gate.  Other possibilities for spintronics exist \cite{Smejkal17a}.   For instance, it has been predicted that in an antiferromagnet DSM, charges can be controlled by the spin-orbit torque reorientation of the N\'eel vector \cite{Smejkal17b}.  Valley degrees of freedoms in WSMs also open up further possibilities for chiral and valleytronics applications in 3D systems \cite{Kharzeev13a,Schaibley16a}.  These systems have a unique coupling to electromagnetic radiation that can be exploited.  For instance, it has been proposed that inversion symmetry breaking WSMs with tilted Weyl cones will be efficient generators of photocurrent and may be used as IR detectors \cite{Chan16a}.  And we have noted above  the WSM systems TaAs, TaP and NbAs have the largest ever recorded SHG $\chi^{(2)}$ coefficient \cite{Wu16b}.  In the photonic Weyl systems discussed above, the number of optical modes has an unusual scaling with the volume of the photonic crystal, which may allow for the construction of large-volume single-mode lasers.

One of the remarkable and continuing themes in physics is that concepts and mathematical structures are repeated in different contexts across vastly different length scales.  The realization of real three dimensional materials described by the Weyl and Dirac equations is a extraordinary part of this particular story that began with Dirac's intellectual leap almost ninety years ago.  Whether nature chooses to repeat itself on this occasion with the realization of Weyl fermions as fundamental particles of the vacuum of free space is a open question.   However, in the meantime we can continue to marvel at the -- thus far virtually limitless -- possibilities and rich phenomena that the different ``vacuums" of solid state systems provide.

\section{Acknowledgements}

We would like to thank F. Baumberger, I. Belopolski, A. Bernevig,  S. Borisenko, C. Butler, A. Burkov, J. Cano, B. Cheng, J. Checkelsky, A. Cortijo, S. Crooker, S. Das Sarma, N. Drichko, C. Felser, M. Franz, N. Gedik, Z. Hasan, E. Hassinger, P. Hosur,  H.-Y. Kee, K. Landsteiner, Y. Li, Z.-M. Liao, H.-Z. Lu, S. Nakatsuji,  B. Spivak, B. Ramshaw,  M. Rechtsman, F. Ronning, A. Taskin, K. Ueda, B. Wieder, L. Wu, and X. Zhou for helpful conversations, correspondences on these topics and/or careful reading of this manuscript.

Our work was supported by the DOE-BES through DE-FG02-08ER46544 and ARO through W911NF-15-1-0560  (NPA),  DOE-BES through DE-FG02-ER45118 (EM), and NSF DMR-1411343, a Simons Investigator grant, and the ARO MURI on topological insulators W911NF- 12-1-0961(AV).

\bibliographystyle{apsrmp}
\bibliography{WeylDiracBib}

\end{document}